\documentclass[11pt]{article}
\pdfoutput=1 

\usepackage{jheppub} 

\usepackage[T1]{fontenc} 

\usepackage{bm}

\usepackage{dsfont}
\def\openone{\mathds{1}}

\def\Tgen{{\mathbb T}}

\def\alphas{\alpha_{\rm s}}
\def\Nc{N_{\rm c}}
\def\Nf{N_{\rm f}}
\def\CA{C_{\rm A}}

\def\qhatA{\hat q_{\rm A}}
\def\Re{\operatorname{Re}}

\def\eps{\epsilon}

\def\b{{\bm b}}
\def\p{{\bm p}}

\def\B{{\bm B}}
\def\C{{\bm C}}
\def\P{{\bm P}}
\def\B{{\bm B}}
\def\S{{\bm S}}
\def\grad{{\bm\nabla}}

\def\ix{{\rm i}}
\def\fx{{\rm f}}
\def\xx{{\rm x}}
\def\xbx{{\bar{\rm x}}}
\def\yx{{\rm y}}
\def\ybx{{\bar{\rm y}}}

\def\seq{{\rm seq}}

\def\uOmega{\underline{\Omega}}

\def\calX{{\cal X}}
\def\Ybar{\overline{Y}}

\def\quote#1{``#1"}
\def\quotea#1{``#1"_{\!\!\rm a}}
\def\quotes#1{``#1"_{\!\!\rm s}}

\newcommand{\mtrx}{\underline}

\def\one{1}
\def\onexp{1^\times_+}
\def\onexm{1^\times_-}
\def\onexpm{1^\times_\pm}
\def\Ap{{\rm A}_+}
\def\Am{{\rm A}_-}
\def\Apm{{\rm A}_\pm}
\def\Ax{{\rm A}^{\!\times}}
\def\sone{|\one\rangle}
\def\sonexp{|\onexp\rangle}
\def\sonexm{|\onexm\rangle}

\def\sAp{|\Ap\rangle}
\def\sAm{|\Am\rangle}
\def\sApm{|\Apm\rangle}
\def\sAx{|\Ax\rangle}
\def\vbxi{\vec{\bm\xi}}
\def\Jvec{\vec{\bm J}}

\def\Ninf{{N=\infty}}

\def\tr{\operatorname{tr}}

\title{\boldmath The LPM Effect in sequential bremsstrahlung:
   $1/\Nc^2$ corrections}

\author{Peter Arnold}
\author{and Omar Elgedawy}
\affiliation{Department of Physics, University of Virginia,
  P.O.\ Box 400714, 
  Charlottesville, VA 22904, U.S.A.}
\emailAdd{parnold@virginia.edu}
\emailAdd{oae2ft@virginia.edu}

\abstract{
  An important question concerning in-medium
  high-energy parton showers in a quark-gluon
  plasma or other QCD medium
  is whether consecutive splittings of the partons in a given shower
  can be treated as quantum mechanically independent, or whether the
  formation times for two consecutive splittings instead have significant
  overlap.  Various previous calculations of the effect of
  overlapping formation times
  have either (i) restricted attention to a soft bremsstrahlung limit,
  or else (ii) used the large-$\Nc$ limit (where $\Nc{=}3$ is the number of
  quark colors).
  In this paper, we make a first study of the accuracy of
  the large-$\Nc$ limit used by those calculations of overlap effects that
  avoid a soft bremsstrahlung approximation.
  Specifically, we calculate the $1/\Nc^2$ correction to previous
  $\Nc{=}\infty$ results for overlap $g \to gg \to ggg$ of
  two consecutive gluon splittings $g \to gg$.
  At order $1/\Nc^2$, there is interesting and non-trivial color dynamics 
  that must be accounted for during the overlap of the formation times.
}

\begin{document} 
\maketitle
\flushbottom

\newpage


\section{Introduction}
\label{sec:intro}

Consider a high-energy gluon showering
as it traverses a QCD medium, such as a quark-gluon
plasma, via splitting processes such as gluon bremsstrahlung
$g \to gg$.
At high energy, the formation time for a bremsstrahlung gluon
becomes large and encompasses multiple scatterings with the medium,
so that one must take into account the Landau-Pomeranchuk-Migdal
(LPM) effect
\cite{LP1,LP2,Migdal,BDMPS1,BDMPS2,BDMPS3,Zakharov1,Zakharov2,Zakharov3}.%
\footnote{
  For English translations of refs.\ \cite{LP1,LP2}, see
  ref.\ \cite{LPenglish}.
}
It is possible for two consecutive splittings in the shower to have
overlapping formations times.  The corrections to an in-medium parton
shower due to overlapping formation times are formally suppressed by
a power of $\alphas$, and it has been of interest for many years to figure
out exactly how significant such corrections are.%
\footnote{
  See, for example, the motivation described in the
  introduction of ref.\ \cite{2brem}.
}

Such corrections were first analyzed in refs.\ \cite{Iancu,Blaizot,Wu}%
\footnote{
  See ref.\ \cite{LMW} for earlier, related work on soft radiative
  corrections to transverse momentum broadening.
}
for the case where one of the two overlapping splittings is
relatively soft (with various other simplifying assumptions we will
review later).  Subsequently, the program of
refs.\ \cite{2brem,seq,dimreg,QEDnf,qedNfstop,qcd} has been
working toward analysis of the more general case where neither
splitting is necessarily soft.  However, the analysis
of this more general case has used the large-$\Nc$
approximation.  The formalism for treating $\Nc{=}3$ is known
in principle \cite{color}%
\footnote{
  For earlier work in a similar but slightly different context,
  see refs.\ \cite{NSZ6j,Zakharov6j}.
}
but may be challenging to implement numerically.

We'd like to know whether or not the $\Nc{=}\infty$
overlapping formation-time calculations in the literature
are a reasonable or poor approximation to the physical case
of $\Nc{=}3$.  In this paper, we investigate that question by
calculating $1/\Nc^2$ corrections to earlier $\Nc{=}\infty$
results \cite{2brem,seq} for the effect of
overlapping formation times on real double splitting
$g \to gg \to ggg$.  Our goal is to see whether those corrections,
when extrapolated to $\Nc{=}3$,
are large, small, or comparable to the purely parametric estimate
$O(1/\Nc^2) \sim 10\%$.

In this paper, other than going beyond the $\Nc{=}\infty$ approximation,
we will make the same sort of simplifying assumptions and approximations
as in the earlier work of refs.\
\cite{2brem,seq,dimreg,QEDnf,qedNfstop,qcd}.
We will assume that the medium is static and homogeneous on scales
of the formation time and corresponding formation length.
We also make the high-energy multiple scattering approximation and
so take interactions with the medium to be described by the
$\hat q$ approximation.

Before proceeding, we should clarify why the first corrections to
$\Nc{=}\infty$ are $O(1/\Nc^2)$ instead of $O(1/\Nc)$.
If one were to think {\it inclusively} about double splitting, then
the $g {\to} gq\bar q$ (pair production overlapping bremsstrahlung)
rate would be an
$O(1/\Nc)$ correction to the purely gluonic
$g {\to} ggg$ rate because of the relative number of quark colors
vs.\ gluon colors.  However, though a calculation of
$g {\to} gq\bar q$
has not yet appeared in the literature without soft approximations, it
may be computed using the same $\Nc{=}\infty$ techniques that
were used to compute $g {\to} ggg$ in refs.\ \cite{2brem,seq}.
So computing $g {\to} gq\bar q$ to leading order in the
large-$\Nc$ limit would give no information on
the size of corrections to $\Nc{=}\infty$ methods.
Instead, we will focus in this paper {\it exclusively}
on purely gluonic (overlapping) $g {\to} ggg$.
In the $\hat q$ approximation, the corrections to $\Nc{=}\infty$ for
the purely gluonic process will be $O(1/\Nc^2)$.%
\footnote{
  In the $\hat q$ approximation, the details of the quark vs.\ gluonic
  content of the medium are swept up into the value of $\hat q$.
  When making $1/\Nc$ expansions in this paper, we treat $\hat q$
  as fixed: we do not expand $\hat q$ in powers of $1/\Nc$.
  Our calculation of overlap effects for
  $g {\to} ggg$ in the $\hat q$ approximation therefore effectively
  involves only gluons.  In standard discussions of large $\Nc$
  for diagrams that involve only gluons, the expansion is
  an expansion in powers of $1/\Nc^2$ \cite{tHooft}.
}


\subsection*{Outline}

In the next section, we first review the interesting,
non-trivial color dynamics
that take place for finite $\Nc$
in calculations of overlapping formation time effects
for $g{\to}ggg$.  We next discuss the $\Nc{\to}\infty$ limit and
isolate the $1/\Nc$ and $1/\Nc^2$ corrections to the effective
``Hamiltonian'' that describes medium-averaged evolution of
high-energy gluons involved in the splitting process.
In section \ref{sec:seq}, we start with what are called
``sequential diagram'' contributions to the $g{\to}ggg$ rate
and show how to obtain analytic integral
expressions for the $1/\Nc^2$ corrections.
The integrals can all be done analytically except for three time integrals,
which later will be performed numerically.
Section \ref{sec:reps} fills in details about
low-level $\Nc{=}\infty$ formulas for applying the $\hat q$
approximation to different color combinations of the high-energy
gluons involved in the $g{\to}ggg$ process.
Section \ref{sec:cross} generalizes the approach for
sequential diagrams in section \ref{sec:seq} to now also cover
what are called ``crossed diagrams.''  That completes the analytic
work, and we move on to numerically evaluate the size of our
$1/\Nc^2$ corrections in section \ref{sec:numerics}.
We also discuss there the relation of our work to earlier
work in a different context (relaxing the
$\Nc{\to}\infty$ limit for {\it single} splitting rates $g{\to}gg$
that have not been integrated over transverse momentum).
Finally, section \ref{sec:conclusion} offers our conclusion.

We should clarify that, for simplicity, we will study $1/\Nc^2$
corrections to only the subset of $g{\to}ggg$ processes
that were studied for $\Nc{=}\infty$ in refs.\ \cite{2brem,seq}.
This leaves out, for example, direct $g{\to}ggg$ through a 4-gluon
vertex, as opposed to a sequence of two 3-gluon vertices with
overlapping formation times.
Such direct 4-gluon processes have been studied in ref.\ \cite{4point}
and found to be numerically small for $\Nc{=}\infty$.
Our study also leaves out effective 4-gluon vertices that appear
in Light Cone Perturbation Theory from integrating out
longitudinally polarized gluons in light-cone gauge.
Their contribution even for $\Nc{=}\infty$ has not yet been completed.
(A  calculation of their contribution in large-$\Nf$ QED is included
in ref.\ \cite{QEDnf}.)


\section {Background: Color dynamics}

\subsection {Warm-up: The BDMPS-Z single splitting rate}

\begin {figure}[t]
\begin {center}
  \includegraphics[scale=0.6]{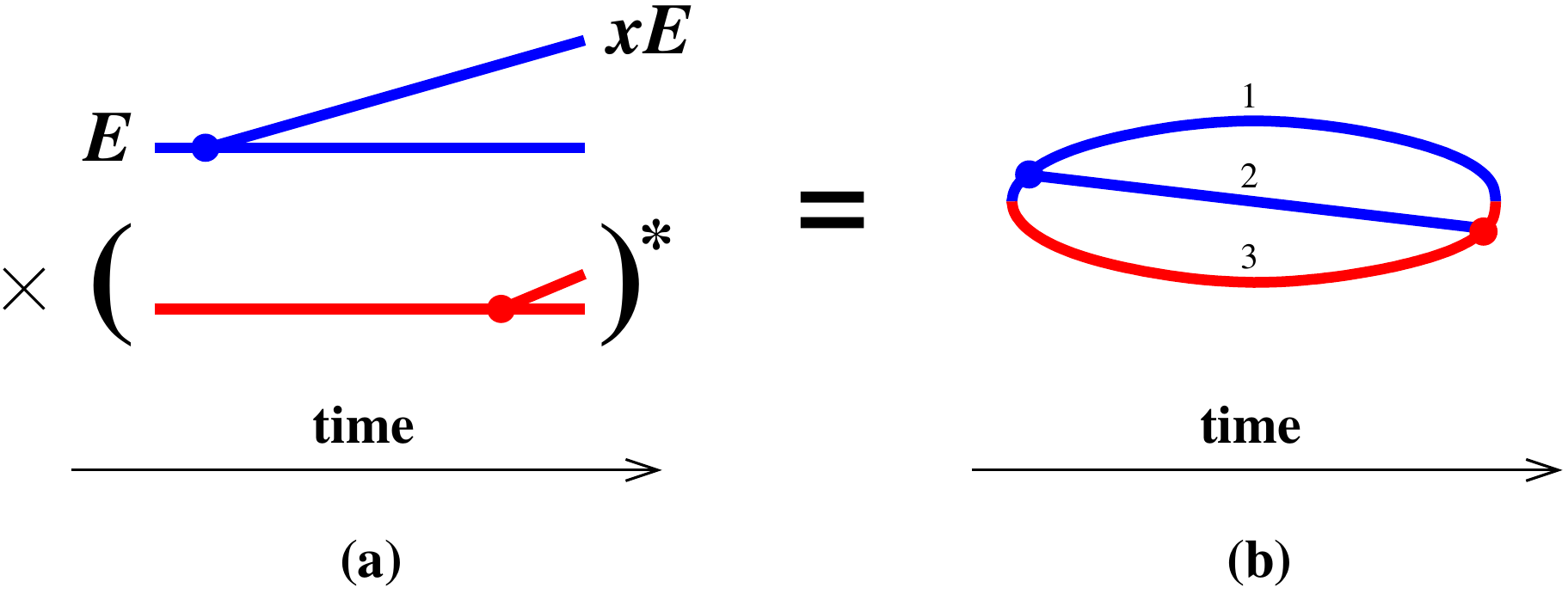}
  \caption{
     \label {fig:split}
     (a) A time-ordered contribution to the rate for single splitting,
     such as $g \to gg$, with
     amplitude in blue and conjugate amplitude in red.
     (b) A single diagram representing this contribution to the rate.
     In both cases, all lines implicitly interact
     with the medium.  We need not follow
     particles after the emission has occurred in both the amplitude
     and conjugate amplitude because we will consider only
     $p_\perp$-integrated rates.
     (See, for example, section 4.1 of
     ref.\ \cite{2brem} for a more explicit argument,
     although applied there to a more complicated diagram.)
     Nor need we follow them before
     the first emission because we approximate the initial particle
     as on-shell.
     Only one of the two time orderings that contribute to the
     rate is shown above.
  }
\end {center}
\end {figure}

Throughout this paper, we draw diagrams for contributions to
splitting rates using the conventions of ref.\ \cite{2brem}, which
are adapted from Zakharov's description of splitting rates
\cite{Zakharov1,Zakharov2,Zakharov3}.
Fig.\ \ref{fig:split}b gives an example for single-splitting
(e.g.\ $g \to gg$) in the medium.

\begin {figure}[t]
\begin {center}
  \includegraphics[scale=0.6]{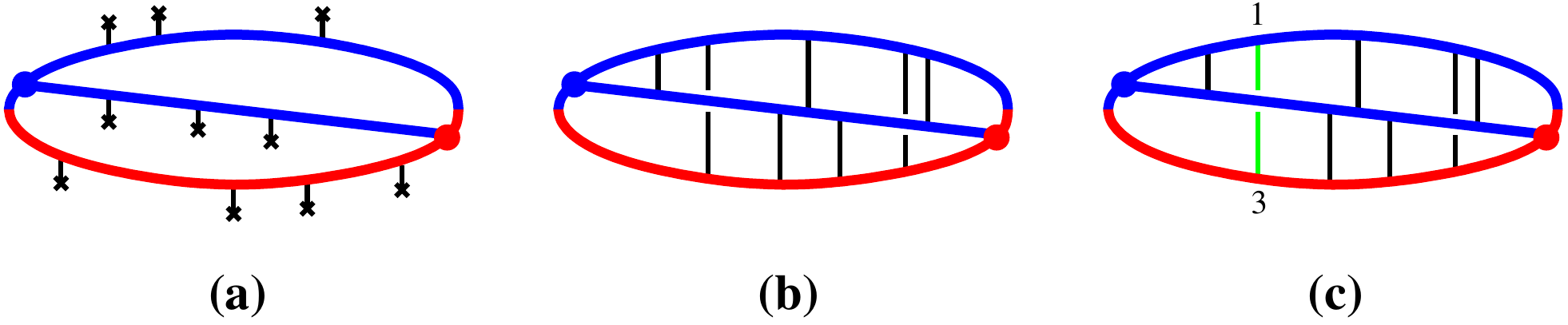}
  \caption{
     \label {fig:corr}
     (a) $g \to gg$ but now depicting interactions with the medium.
     Here, each black line ending in a cross represents an interaction
     of a high-energy particle with the gluon field in the medium.
     (b) The medium average of those interactions in the case of a
     weakly-coupled medium.  Here the black lines represent 2-point
     correlations of the medium interactions, which dominate for
     a weakly-coupled medium.  (The 2-point correlations can be
     written in terms of correlations $\langle A_\mu^a A_\nu^b \rangle$ of
     the background gluon fields present in the medium.)
     The correlations are drawn as vertical
     in this time-ordered diagram because, in the high-energy limit,
     the correlation lengths in the medium are parametrically
     small compared to the length (time) scale of the high-energy
     splitting process.  [Not shown but also present: short-time
     2-point correlations
     between two medium interactions of the {\it same} high-energy particle.]
     (c) One correlation between particles 1 and 3 is highlighted.
  }
\end {center}
\end {figure}

The high-energy particle lines shown in the figure are implicitly interacting
with, and scattering from, the gluon fields of the medium, as depicted
in fig.\ \ref{fig:corr}a, and the rate
is implicitly
averaged over the randomness of the medium.  Such interactions with
the medium change the color of each high-energy particle over time.
At first, it may seem like calculating the rate would require a complicated
analysis of the time dependence of the color of each such particle.
Fortunately, this is unnecessary for fig.\ \ref{fig:corr}a.%
\footnote{ Here and throughout, we will only be considering rates
  which are fully integrated over the transverse momenta of the
  daughter gluons.  Otherwise, the color dynamics is more complicated
  even for $g{\to}gg$.  See, for example,
  refs.\ \cite{NSZ6j,Zakharov6j}.
}
To get a flavor for the reason why, consider for a moment
the extreme case where the
medium itself is weakly-coupled.  Then (to leading order in the
coupling of the medium) the medium-averaged correlations of interactions
with the medium are 2-point correlations, as shown in fig.\ \ref{fig:corr}b.
Let's focus on one of these correlations,
such as the green line connecting particles 1 and 3
in fig.\ \ref{fig:corr}c.
Let $\Tgen_n^a$ represent color generators $\Tgen^a$ (in the appropriate
representation) that act on the color state of particle $n$.
The interaction of particle 1 with the gluonic field of the
medium comes with a factor of
$g \Tgen_1^a$.
The correlation of a pair of interactions of particles 1 and 3
with the medium
then comes with a factor of
$(g \Tgen_1^a)(g \Tgen_3^a) = g^2 \Tgen_1\cdot \Tgen_3$.
But this operator is quite trivial because, by color conservation
(after medium averaging),%
\footnote{
  Without medium averaging, the color neutrality of the 3-particle
  state would not be conserved over time.  That's because the interactions in
  fig.\ \ref{fig:corr}a (via gluon exchange with the medium)
  may randomly change the color of just
  one of the three high-energy particles at a given moment, and exchanging
  one gluon with the medium turns a 3-particle color singlet into a
  3-particle color octet.  After medium averaging, however, the interactions
  with the medium must be {\it correlated}, such as in fig.\ \ref{fig:corr}b,
  and so color cannot flow out of the 3-particle system since these correlations
  are instantaneous on the time scales relevant to splitting processes.
  (In perturbative language, the medium-averaged correlator
  $\langle A_\mu^a A_\nu^b \rangle$ of background gluon gauge fields
  vanishes unless $a=b$.)
  The situation is analogous to translation invariance of a gas in
  thermal equilibrium:
  any particular configuration of the molecules is not translation invariant,
  but translation invariance is recovered after thermal averaging.
}
the three high-energy particles in
fig.\ \ref{fig:split}b must form a color singlet, which means
$\Tgen_1{+}\Tgen_2{+}\Tgen_3=0$.  So, $\Tgen_1{+}\Tgen_3 = -\Tgen_2$ and thus%
\footnote{
  This argument is a simple generalization of an
  argument from ordinary, non-relativistic quantum mechanics.  Imagine three
  non-relativistic particles with spin angular momenta
  $\S_1$, $\S_2$, and $\S_3$.  If the three-particle system forms
  a spin singlet $|\chi\rangle$, then the operator $\S_1+\S_2+\S_3$ applied
  to $|\chi\rangle$ gives zero.  That means that
  $(\S_1+\S_3)|\chi\rangle = -\S_2|\chi\rangle$ and
  so (since the $\S_n$ for different particles commute with each other)
  $(\S_1+\S_3)^2|\chi\rangle = (\S_2)^2|\chi\rangle$.
  From this, one finds
  $\S_1\cdot\S_3|\chi\rangle
   = \frac12 [(\S_2)^2 - (\S_1)^2 - (\S_3)^2] |\chi\rangle$.
  So, on the subspace of spin-singlet states,
  $\S_1\cdot\S_3 = \frac12 \bigl[ s_2(s_2+1) - s_1(s_1+1) - s_3(s_3+1) \bigr]$.
  Eq.\ (\ref{eq:TTexample}) is just the generalization of this argument
  from the (covering) group SU(2) of rotations to other Lie groups
  such as SU(3).  The $s_n(s_n+1)$ in this footnote
  are just the quadratic Casimirs $C_n$ of SU(2).
  As is conventional in quantum mechanics, we are sloppy about explicitly
  writing identity operators.  In terms of single-particle operators,
  our $\S_1$ above is really $\S_1 \otimes \openone_2 \otimes \openone_3$,
  our $\S_2$ is really $\openone_1 \otimes \S_2 \otimes \openone_3$, etc.;
  our operator identity (\ref{eq:TTexample}) is only true when the operator
  $\Tgen_1\cdot \Tgen_3$ acts on the subspace of 3-particle
  color-singlet states; and
  the Casimirs on the right-hand side of (\ref{eq:TTexample}) are
  multiplied by the identity operator for that subspace.
  To make all color indices explicit, consider
  a color-singlet state
  $|\chi\rangle = c_{ijk} |ijk\rangle$ (implicit sum over indices),
  where $(i,j,k)$ are the appropriate (e.g.\ fundamental or adjoint) 
  color indices for particles (1,2,3)
  respectively,
  and $c_{ijk}$ are superposition coefficients that
  yield a color singlet.  Then eq.\ (\ref{eq:TTexample}) says that
  $(\Tgen^a_{R_1})_{ii'} (\Tgen^a_{R_3})_{kk'} c_{i'jk'}
  = \tfrac12 (C_2  - C_1 - C_3) c_{ijk}$, where the matrices
  $\Tgen_{R_n}^a$ are the
  generators associated with the color representation $R_n$
  (e.g.\ fundamental or adjoint) of particle $n$.
} 
\begin {equation}
  \Tgen_1\cdot \Tgen_3
  = \tfrac12 \bigl[ (\Tgen_1+\Tgen_3)^2 - \Tgen_1^2 - \Tgen_3^2 \bigr]
  = \tfrac12 (\Tgen_2^2  - \Tgen_1^2 - \Tgen_3^2)
  = \tfrac12 (C_2  - C_1 - C_3) ,
\label {eq:TTexample}
\end {equation}
where $C_i$ is the quadratic Casimir associated with the color representation
of particle $i$.  That means that $\Tgen_1\cdot \Tgen_3$ reduces to a simple
{\it fixed number} in this context.
(Specifically
$\Tgen_1\cdot \Tgen_3 = -\CA/2 = -\Nc/2$ in the case of $g \to gg$.)
Because of (\ref{eq:TTexample}),
we do not need to keep track of the dynamics of the individual colors of
the three high-energy particles
in order to calculate the rate for fig.\ \ref{fig:split}.

This conclusion can be generalized to strongly-coupled media as well
when one describes medium interactions using the $\hat q$
approximation.  See ref.\ \cite{Vqhat} for the argument.


\subsection {SU(3) color states for overlapping, double splitting}

\begin {figure}[t]
\begin {center}
  \includegraphics[scale=0.6]{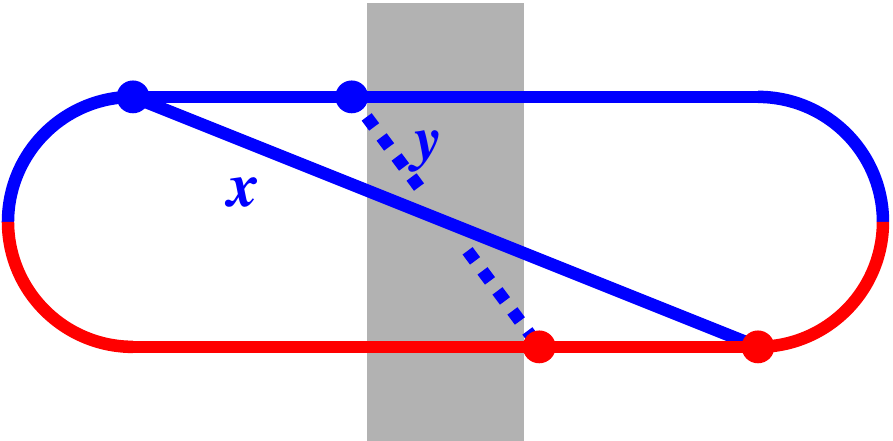}
  \caption{
     \label {fig:cross}
     One diagrammatic contribution \cite{2brem}
     to the rate for double splitting,
     such as $g \to ggg$.
  }
\end {center}
\end {figure}

Fig.\ \ref{fig:cross} shows an example of a contribution to the rate
for overlapping double splitting such as $g\to ggg$.
In the shaded region, the system has four high-energy particles
(three in the amplitude and one in the conjugate amplitude).
Again by color conservation, those four particles together must form a
color singlet. Unfortunately, unlike the 3-particle case,
color neutrality
$\Tgen_1{+}\Tgen_2{+}\Tgen_3{+}\Tgen_4=0$ is not enough to uniquely
determine combinations like $\Tgen_i\cdot\Tgen_j$ which appear
in correlations between high-energy particles' interactions with
the medium.  A similar uncertainty arises in more general arguments
\cite{Vqhat} in the context of the $\hat q$ approximation.

\begin {figure}[t]
\begin {center}
  \includegraphics[scale=0.4]{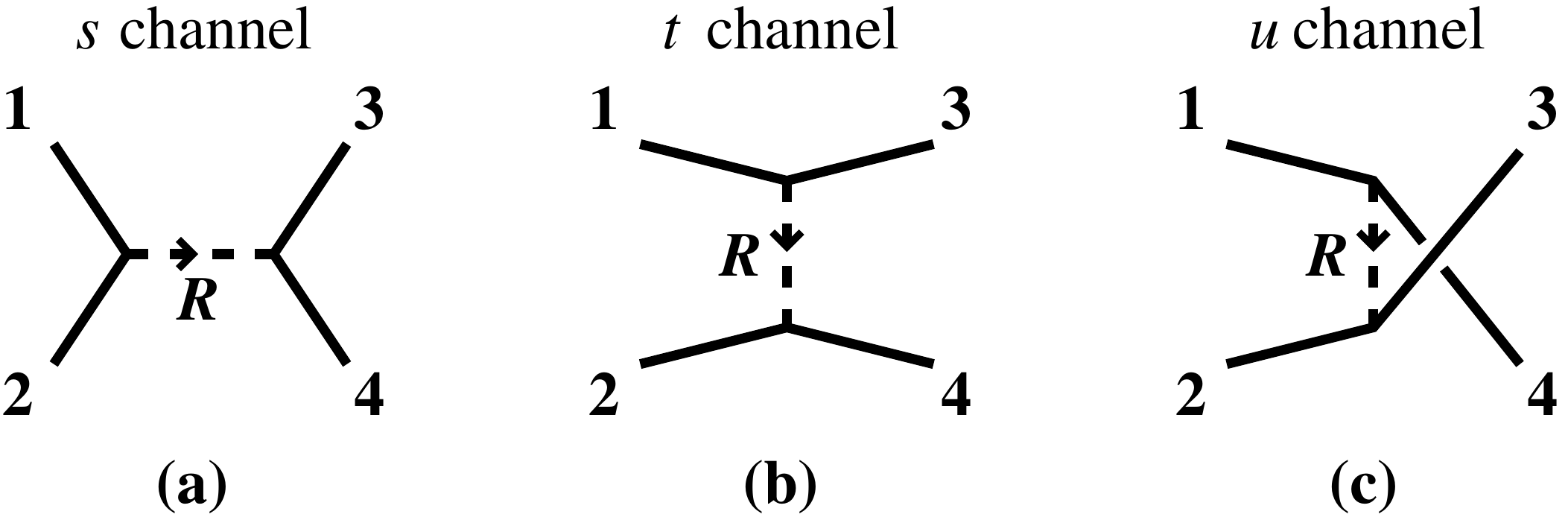}
  \caption{
     \label {fig:channels}
     Three ways to form bases for 4-gluon color singlet states.
     (The terms $s$-channel, etc.\ are merely evocative here;
     we are {\it not}\/ referring to $2{\leftrightarrow}2$ scattering.)
  }
\end {center}
\end {figure}

The source of this ambiguity is that there are many different ways one
can make a color singlet out of four gluons (similar to how
there are many ways to make a spin singlet out of four spin-1
particles in ordinary quantum mechanics).
In SU(3), the color representations of two gluons can
be combined as
\begin {equation}
   \bm{8}\otimes\bm{8} =
   \bm{1}_{\rm s} \oplus \bm{8}_{\rm a} \oplus \bm{8}_{\rm s} \oplus
     \bm{10}_{\rm a} \oplus \overline{\bm{10}}_{\rm a} \oplus \bm{27}_{\rm s} ,
\label {eq:8x8}
\end {equation}
where the subscripts ``$\rm s$'' and ``$\rm a$'' indicate
symmetric vs.\ anti-symmetric color combinations of the two gluons.
We could make a color singlet out of four gluons by combining the
first two gluons into any color representation $R$ appearing on the
right-hand side of (\ref{eq:8x8}), then combine the other two gluons
into its complex conjugate $\bar R$, and then combine the resulting
$R$ and $\bar R$ into a color singlet.  This process is depicted
schematically in fig.\ \ref{fig:channels}a, labeled
``$s$-channel.''  These $s$-channel color
states form a basis for all 4-gluon color
singlet states.  Alternatively, one may instead choose a ``$t$-channel''
or ``$u$-channel'' basis, as indicated in the figure.%
\footnote{
  For a variety of papers related to these constructions (and
  discussion of the color generalization of $6j$-symbols to relate
  different channels), see, for example,
  refs.\ \cite{color,NSZ6j,Zakharov6j,Sjodahl,Kaplan,Bickerstaff,CvitanovicUn,
  Cvitanovic}.
}

In this paper,
we find it convenient to work in the $u$-channel basis
because of particle numbering conventions in earlier papers on
overlapping formation times \cite{2brem,seq}.  We will label the
$u$-basis singlet states as $|R\rangle_u$.  Our initial basis for
discussing 4-gluon singlets is then%
\begin {equation}
  |{\bm 1}\rangle_u,
  |{\bm 8}_{\rm aa}\rangle_u,\,
  |{\bm 8}_{\rm as}\rangle_u,\,
  |{\bm 8}_{\rm sa}\rangle_u,\,
  |{\bm 8}_{\rm ss}\rangle_u,\,
  |{\bm{10}}\rangle_u,\,
  |\overline{\bm{10}}\rangle_u,\,
  |{\bm{27}}\rangle_u .
\label {eq:SU3states}
\end {equation}
For the case where $R={\bm 8}$, we have to label whether each pair of
the four particles formed the ${\bm 8}$ by a
symmetric ($\rm s$) or anti-symmetric  (${\rm a}$) combination, as
distinguished in (\ref{eq:8x8}).
As explained in the present context in ref.\ \cite{color},%
\footnote{
  This 5-dimensional subspace was also discussed earlier in a closely related
  context by refs.\ \cite{NSZ6j,Zakharov6j}.
}
only a 5-dimensional
subspace of (\ref{eq:SU3states}) appears in calculations of overlapping
formation times (e.g.\ fig.\ \ref{fig:cross}):%
\footnote{
  The fact that the states in (\ref{eq:5dim}) are designated as $u$-channel
  is irrelevant.  The analogous $s$-channel or $t$-channel states would
  span the same 5-dimensional subspace.
}
\begin {equation}
  |{\bm 1}\rangle_u,\,
  |{\bm 8}_{\rm aa}\rangle_u,\,
  |{\bm 8}_{\rm ss}\rangle_u,\,
  |{\bm{10}}{+}\overline{\bm{10}}\rangle_u,\,
  |{\bm{27}}\rangle_u ,
\label {eq:5dim}
\end {equation}
where
\begin {equation}
  |{\bm{10}}{+}\overline{\bm{10}}\rangle_u
  \equiv
  \tfrac{1}{\sqrt2}\bigl(
      |{\bm{10}}\rangle_u + |\overline{\bm{10}}\rangle_u
  \bigr).
\end {equation}

Soon, we will discuss how the color singlet state of the four gluons
evolves in the subspace (\ref{eq:5dim}) as the gluons travel through
the medium.  But first, we wish to discuss the generalization from
SU(3) to SU($N$).


\subsection {SU($N$) color dynamics for overlapping, double splitting}

For the sake of compactness, we will refer to the number of quark colors
as $N$ rather than $\Nc$ in the rest of this paper.
The generalization of the preceding discussion to $N > 3$ is
that the tensor product (\ref{eq:8x8}) of two gluon colors becomes%
\footnote{
  The SU($N$) Young tableaux
  corresponding to (\ref{eq:AxA}) and the actual
  dimensions of the representations may be found,
  for example, in eqs.\ (5.1) and (5.2) of ref.\ \cite{color}.
}
\begin {equation}
 {\rm A} \otimes {\rm A} =
 \bm{1}_{\rm s} \oplus {\rm A}_{\rm a} \oplus {\rm A}_{\rm s}
 \oplus \quotea{\bm{10}} \oplus \quotea{\overline{\bm{10}}}
 \oplus \quotes{\bm{27}}
 \oplus \quotes{\bm{0}} \,,
\label {eq:AxA}
\end {equation}
where ${\bm 1}$ is the singlet representation,
${\rm A}$ is the adjoint representation of SU($N$),
and, for example, $\quote{\bm{27}}$ means the SU($N$) representation
that generalizes the 27-dimensional representation of SU(3).
The scare quotes just mean that, though we quote the size
of the representation for $N{=}3$, we really mean the corresponding
representation of SU($N$).  Note that there is one more term in
(\ref{eq:AxA}) than in the original SU(3) product (\ref{eq:8x8}).
This representation $\quote{\bm{0}}$ of SU($N$) smoothly decouples and
disappears as one approaches $N \to 3$ from above.

For SU($N$) with $N>3$, there is a 6-dimensional (rather than 5-dimensional)
subspace of color singlet states relevant to calculations of
overlapping formation times, which is spanned by the basis
\cite{color,NSZ6j}
\begin {equation}
  |{\bm 1}\rangle_u,\,
  |{\rm A}_{\rm aa}\rangle_u,\,
  |{\rm A}_{\rm ss}\rangle_u,\,
  |\quote{\bm{10}{+}\overline{\bm{10}}}\rangle_u,\,
  |\quote{\bm{27}}\rangle_u ,\,
  |\quote{\bm 0}\rangle_u .
\label {eq:6dim}
\end {equation}
This generalizes (\ref{eq:5dim}).

In this paper we will quote some results about 4-particle color singlet
states from ref.\ \cite{color},
but we have found it convenient to use slightly
different overall sign conventions for the definitions of the
$u$-channel states (\ref{eq:6dim}).
The details of the relation between our sign conventions here
and those of ref.\ \cite{color} may be found in appendix \ref{app:signs}.

In Zakharov's version of the BDMPS-Z calculation of single splitting rates,
the problem is recast as two-dimensional quantum mechanics (in the transverse
plane) with an imaginary-valued ``potential energy'' $V$.
Ref.\ \cite{2brem} extended this picture, in the large-$N$ limit, to
calculations of overlap effects in double splitting, such as the
contribution to the rate represented by fig.\ \ref{fig:cross}.
The 4-gluon potential needed to treat the
shaded region of fig.\ \ref{fig:cross} for {\it finite} $N$
was worked out in ref.\ \cite{color}
for the $\hat q$ approximation.
The resulting 2-dimensional Hamiltonian for the 4-gluon evolution in the
shaded region of fig.\ \ref{fig:cross} was found to be%
\footnote{
  See appendix \ref{app:signs} for details of how the $s$-channel
  result of ref.\ \cite{color} was translated to the $u$-channel version
  in (\ref{eq:HVeffective}).
}
\begin {subequations}
\label {eq:HVeffective}
\begin {equation}
  \mtrx{H} =
  \frac{P_{41}^2}{2 x_4 x_1 (x_4{+}x_1) E}
  +
  \frac{P_{23}^2}{2 x_2 x_3 (x_2{+}x_3) E}
  +
  \mtrx{V}(\C_{41},\C_{23})
\label {eq:Heffective}
\end {equation}
with potential
\begin {multline}
  \mtrx{V}(\C_{41},\C_{23})
  =
  - \tfrac{i}{4} \hat q_{\rm A}
  \Bigl\{
     (x_4^2 + 2 x_4 x_1 \mtrx{S}_u + x_1^2) C_{41}^2
     +
     (x_2^2 + 2 x_2 x_3 \mtrx{S}_u + x_3^2) C_{23}^2
\\
     +
     2
     \bigl[
        \tfrac12 (x_4-x_1) (x_2-x_3) (\mtrx{S}_u-1)
        - (x_4+x_1)(x_2+x_3) \mtrx{T}_u
     \bigr] \C_{41}\cdot\C_{23}
   \Bigr\} .
\label {eq:Veffective}
\end {multline}
\end {subequations}
Above, symmetries have been used to reduce the 4-gluon quantum mechanics
problem with transverse positions $(\b_1,\b_2,\b_3,\b_4)$
to an effective 2-particle quantum mechanics problem
\cite{2brem,seq} written in terms of $(\C_{41},\C_{23})$ with
$\C_{ij} \equiv (\b_i{-}\b_j)/(x_i{+}x_j)$.
The $\P_{ij}$ are the canonical momenta conjugate to the $\C_{ij}$,
and $E$ is the energy of the initial particle in the double-splitting
process.
The $x_i$ represent the longitudinal momentum fractions of the four gluons.
The underlined quantities in (\ref{eq:HVeffective}) represent
$6\times6$ matrices (for $N > 3$)
that act on the 6-dimensional space of relevant 4-gluon color singlet states.
The matrices $\mtrx{S}_u$ and $\mtrx{T}_u$ encode results for the action
of $\Tgen_i\cdot\Tgen_j$ on this space in the $u$-channel basis
(\ref{eq:6dim}), encoded as%
\footnote{%
\label{foot:TdotT}%
   Because $\Tgen_1{+}\Tgen_2{+}\Tgen_4{+}\Tgen_4=0$ implies
   $(\Tgen_4{+}\Tgen_1)^2 = (\Tgen_2{+}\Tgen_3)^2$, and because
   all $\Tgen_i^2 = \CA^2$ (since all four particles are gluons), we
   have the additional relation that
   $\mtrx{\Tgen}_2\cdot\mtrx{\Tgen}_3 = \mtrx{\Tgen}_4\cdot\mtrx{\Tgen}_1$.
   Similarly,
   $\mtrx{\Tgen}_3\cdot\mtrx{\Tgen}_1 = \mtrx{\Tgen}_4\cdot\mtrx{\Tgen}_2$
   and
   $\mtrx{\Tgen}_1\cdot\mtrx{\Tgen}_2 = \mtrx{\Tgen}_4\cdot\mtrx{\Tgen}_3$.
}
\begin {equation}
   \mtrx{\Tgen}_4\cdot\mtrx{\Tgen}_1
     = - C_{\rm A} \mtrx{S}_u \,,
   \qquad
   \mtrx{\Tgen}_4\cdot\mtrx{\Tgen}_2
     = C_{\rm A}
     \bigl[ \tfrac12 (\mtrx{S}_u-\mtrx{\openone}) - \mtrx{T}_u \bigr] \,,
   \qquad
   \mtrx{\Tgen}_4\cdot\mtrx{\Tgen}_3
     = C_{\rm A}
     \bigl[ \tfrac12 (\mtrx{S}_u-\mtrx{\openone}) + \mtrx{T}_u \bigr]
\label {eq:TdotT}
\end {equation}
with
\begin {subequations}
\label {eq:ST}
\begin {equation}
   \mtrx{S}_u \equiv
     \left(
     \begin{array}{cccccc}
       ~1 &&&&& \\
       & ~\tfrac12 &&&& \\
       && ~\tfrac12 &&& \\
       &&& ~0 && \\
       &&&& -\tfrac{1}{N} & \\
       &&&&& \tfrac{1}{N}
     \end{array}
     \right)
\end {equation}\
and
\begin {equation}
   \mtrx{T}_u \equiv
     \left(
     \begin{array}{cccccc}
       0 & \tfrac{1}{\sqrt{N^2-1}} & 0 & 0 & 0 & 0\\
       \frac{1}{\sqrt{N^2-1}} & 0 & \tfrac14 & 0
           & \tfrac{1}{2N} \sqrt{\tfrac{N+3}{N+1}}
           & \tfrac{1}{2N} \sqrt{\tfrac{N-3}{N-1}} \\
       0 & \tfrac14 & 0 & \tfrac{1}{\sqrt{2(N^2-4)}} & 0 & 0\\
       0 & 0 & \tfrac{1}{\sqrt{2(N^2-4)}} & 0
           & \tau_+
           & \tau_- \\
       0 & \tfrac{1}{2N} \sqrt{\tfrac{N+3}{N+1}} & 0
           & \tau_+ & 0 & 0 \\
       0 & \tfrac{1}{2N} \sqrt{\tfrac{N-3}{N-1}} & 0
           & \tau_- & 0 & 0 \\
     \end{array}
     \right) ,
\end {equation}
\end {subequations}
where
\begin {equation}
   \tau_\pm \equiv \tfrac{1}{2N}\sqrt{\tfrac{(N\mp2)(N\pm1)(N\pm3)}{2(N\pm2)}} .
\end {equation}

In this paper, we will need to solve for the 4-gluon evolution of the
Hamiltonian (\ref{eq:HVeffective}) in perturbation theory in $1/N$ about the
$N{=}\infty$ limit.


\subsection {$N{=}\infty$ limit}

In the $N{\to}\infty$ limit, (\ref{eq:ST}) becomes
\begin {equation}
   \mtrx{S}_u \to
     \left(
     \begin{array}{cccccc}
       ~1 &&&&& \\
       & ~\tfrac12 &&&& \\
       && ~\tfrac12 &&& \\
       &&& ~0 && \\
       &&&& ~0 & \\
       &&&&& ~0
     \end{array}
     \right) ,
   \qquad
   \mtrx{T}_u \to
     \left(
     \begin{array}{cccccc}
       0 & & & & & \\
       & 0 & ~\tfrac14 & & & \\
       & ~\tfrac14 & 0 & & & \\
       & & & 0 & \tfrac1{2\sqrt2} & \tfrac1{2\sqrt2} \\
       & & & \tfrac1{2\sqrt2} & 0 & 0 \\
       & & & \tfrac1{2\sqrt2} & 0 & 0
     \end{array}
     \right)
  \quad\mbox{in basis}\quad
  \begin{array}{l}
    |{\bm 1}\rangle_u \\
    |{\rm A}_{\rm aa}\rangle_u \\
    |{\rm A}_{\rm ss}\rangle_u \\
    |\quote{{\bm{10}}{+}\overline{\bm{10}}}\rangle_u \\
    |\quote{\bm {27}}\rangle_u \\
    |\quote{\bm 0}\rangle_u  \,.
  \end{array}
\label {eq:STinfOriginal}
\end {equation}
Unlike the case of finite $N$,
the matrices $\underline{S}_u$ and $\underline{T}_u$ commute for $N = \infty$.
It is therefore possible to find a new basis that simultaneously diagonalizes
both matrices:
\begin {align}
  \sone &\equiv |{\bm 1}\rangle_u \,,
\nonumber\\
  \sAp &\equiv 
    \tfrac{1}{\sqrt2} |{\rm A}_{\rm aa}\rangle_u
    + \tfrac{1}{\sqrt2} |{\rm A}_{\rm ss}\rangle_u \,,
\nonumber\\
  \sAm &\equiv 
    \tfrac{1}{\sqrt2} |{\rm A}_{\rm aa}\rangle_u
    - \tfrac{1}{\sqrt2} |{\rm A}_{\rm ss}\rangle_u \,,
\nonumber\\
  \sAx &\equiv 
    \tfrac{1}{\sqrt2} |\quote{\bm{27}}\rangle_u
    - \tfrac{1}{\sqrt2} |\quote{\bm 0}\rangle_u \,,
\nonumber\\
  \sonexp &\equiv 
    \tfrac12 |\quote{\bm{27}}\rangle_u
    + \tfrac12 |\quote{\bm 0}\rangle_u
    + \tfrac{1}{\sqrt2} |\quote{{\bm{10}}{+}\overline{\bm{10}}}\rangle_u \,,
\nonumber\\
  \sonexm &\equiv 
    \tfrac12 |\quote{\bm{27}}\rangle_u
    + \tfrac12 |\quote{\bm 0}\rangle_u
    - \tfrac{1}{\sqrt2} |\quote{{\bm{10}}{+}\overline{\bm{10}}}\rangle_u \,,
\label {eq:basis}
\end {align}
in terms of which the $N{=}\infty$ limits (\ref{eq:STinfOriginal}) become
\begin {equation}
   \mtrx{S}^\Ninf \equiv
     \left(
     \begin{array}{cccccc}
       ~1 &&&&& \\
       & ~\tfrac12 &&&& \\
       && ~\tfrac12 &&& \\
       &&& ~0 && \\
       &&&& ~0 & \\
       &&&&& ~0
     \end{array}
     \right) ,
   \qquad
   \mtrx{T}^\Ninf \equiv
     \left(
     \begin{array}{cccccc}
       ~0 & & & & & \\
       & ~\tfrac14 & & & & \\
       & & -\tfrac14 & & & \\
       & & & ~0 & & \\
       & & & & ~\tfrac12 &  \\
       & & & & & -\tfrac12
     \end{array}
     \right)
  \quad\mbox{in basis}\quad
  \begin{array}{l}
    \sone \\
    \sAp \\
    \sAm \\
    \sAx \\
    \sonexp \\
    \sonexm \,.
  \end{array}
\label {eq:ST0}
\end {equation}
We will explain our naming convention for the basis states (\ref{eq:basis})
shortly.  We have dropped the subscript $u$ on $\mtrx{S}^\Ninf$
and $\mtrx{T}^\Ninf$ just to keep our notation from becoming too cluttered.

Because $\mtrx{S}^\Ninf$ and $\mtrx{T}^\Ninf$ are both diagonal, the potential
(\ref{eq:Veffective}), and so the Hamiltonian,
does not mix the states (\ref{eq:basis}) for $N{=}\infty$.
Each of these states propagates independently for $N{=}\infty$,
with non-matrix
potentials given by using the corresponding eigenvalues
from (\ref{eq:ST0}) in place
of the matrices $\mtrx{S}_u$ and $\mtrx{T}_u$ in (\ref{eq:Veffective}).
We will only encounter transitions between these color singlet
states when we later
investigate the $O(1/N)$ perturbations to $\mtrx{S}^\Ninf$ and $\mtrx{T}^\Ninf$.

The motivation for the names $\sone$ and $\sApm$ in
(\ref{eq:basis}) should be clear enough.
One may use the conversion matrices between
bases given in appendix \ref{app:signs} to see that
the state $\sAx$ defined in terms of $u$-channel color singlet states
is equivalent, in the limit $N\to\infty$, to the combination
$\bigl(|{\rm A}_{\rm aa}\rangle_s + |{\rm A}_{\rm ss}\rangle_s\bigr)/\sqrt2$
of $s$-channel basis states.  Similarly, the state $\sonexm$ is equivalent to
the $s$-channel basis state $|1\rangle_s$, and $\sonexp$ is equivalent to the
$t$-channel basis state $|1\rangle_t$.
So we may think of the cross ``$\times$'' in the notation
$\Ax$ or $\onexpm$ as meaning
that, for $N = \infty$, the state involves
the representation $R = {\rm A}$ or $R = 1$ in
a {\it cross}-channel different from our usual $u$-channel representation.

Later we will also use the definitions (\ref{eq:basis}) of basis
states when analyzing large but finite $N$.
In that case the equivalences just discussed (and so the motivation
for the notation) are not exactly correct.  So, for $N < \infty$,
one may also interpret the cross $\times$ in the colloquial sense
of ``crossed out'': a warning that the motivation for
the notation is no longer precise for those states.


\subsection{An aside:
  Diagrammatic interpretation of basis states for $N = \infty$}

We make a brief detour to present another way to characterize the basis
(\ref{eq:basis}) for $N=\infty$.  This alternative characterization
can offer insight and will be used for some detailed arguments in
section \ref{sec:perms},
but is not strictly necessary for most of our calculation.

\begin {figure}[t]
\begin {center}
  \includegraphics[scale=0.6]{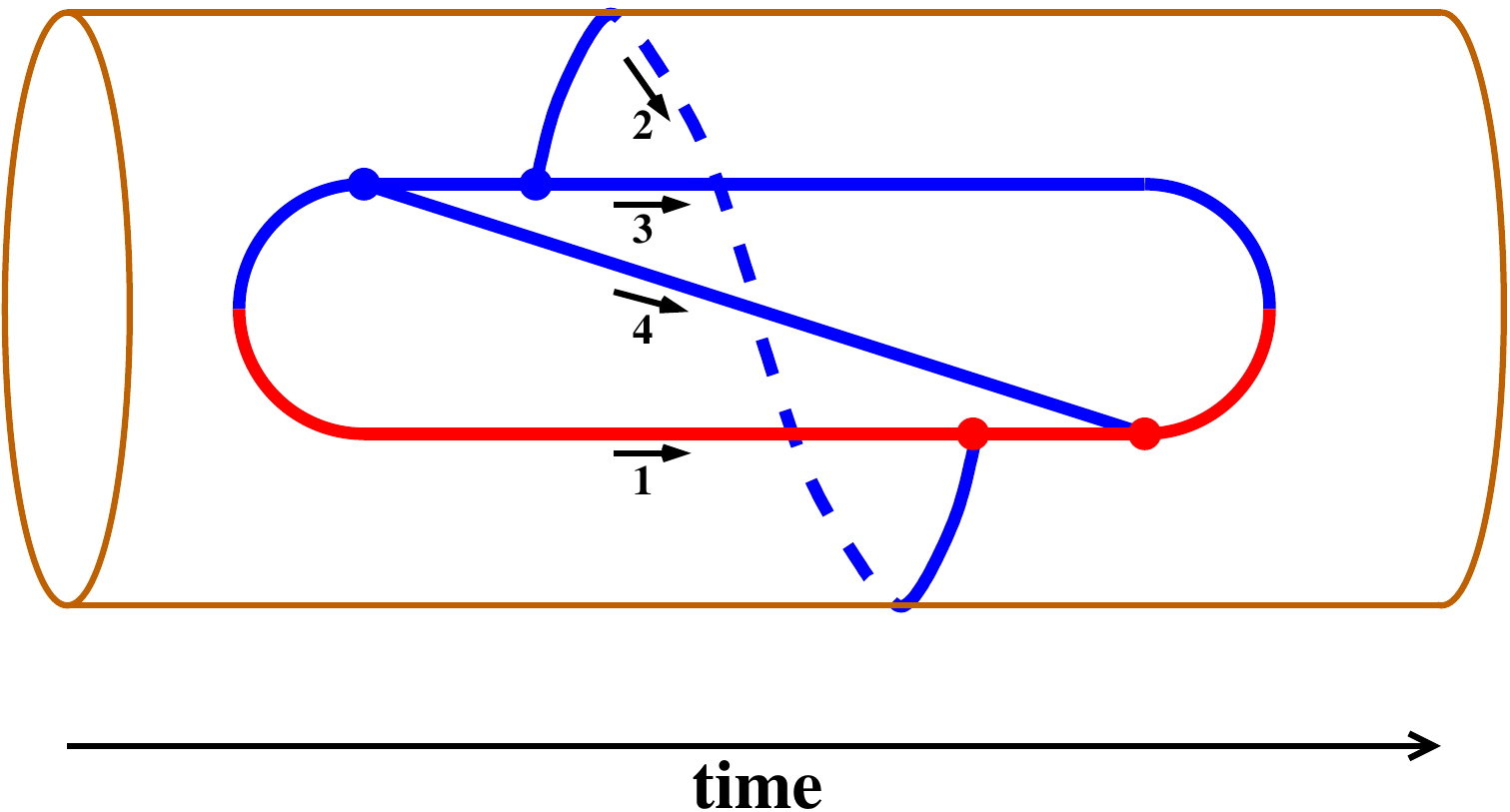}
  \caption{
     \label {fig:crosscyl}
     Fig.\ \ref{fig:cross} drawn on a cylinder.
     Here, solid lines indicate lines drawn on the front of the cylinder,
     and dashed lines indicate lines wrapping around the back.
  }
\end {center}
\end {figure}

Refs.\ \cite{2brem,seq} discuss
drawing time-ordered diagrams, such as fig.\ \ref{fig:cross} and others
on the surface of a cylinder, where time runs along the length of the cylinder.
The large-$N$ requirement that $N{=}\infty$ diagrams be ``planar''
\cite{tHooft}
can be translated to say that no lines should cross on the surface of
the cylinder.
So, for instance, fig.\ \ref{fig:cross} can be drawn on the cylinder
as in fig.\ \ref{fig:crosscyl}, where we have numbered the lines during
the 4-particle part of the evolution according to the convention of
ref.\ \cite{2brem}, which for this diagram corresponds
to identifying the longitudinal momentum fractions
of the gluons as $(x_1,x_2,x_3,x_4)=(-1,y,1{-}x{-}y,x)$.
Correlation lines, such as the black lines drawn
in fig.\ \ref{fig:corr}b (and also higher-point correlations), must also be
part of the ``planar'' diagram and so must lie along the surface of
the cylinder without crossing any other lines.  As a result, for
$N{=}\infty$, there can only be correlations between
high-energy particles that
are neighbors of each other as one goes around the circumference of
the cylinder.  So, during the 4-gluon phase of the time evolution
in fig.\ \ref{fig:crosscyl}, the medium interactions of particle 1
can be correlated with those of particles 2 and 4 but not with
particle 3.  We will indicate this particular sequence as
$(1234)$.
Any cyclic permutation, such as $(2341)$, would be an equivalent designation,
and so would the reverse order $(4321)$ or its cyclic permutations.
All that matters for discussing the interactions among the particles
in large $N$ is which of the four high-energy gluons are neighbors.

With this notation
the color singlet states (\ref{eq:basis}) may be identified
as (see appendix \ref{app:1234})
\begin {align}
  \sAp \to (1324),~~ \qquad \sAm &\to (1234), \qquad ~~\sAx \to (1243),
\nonumber\\
  \sone \to (41)(23), \qquad \sonexp &\to (13)(24), \qquad \sonexm \to (12)(34)
\label {eq:1234}
\end {align}
when $N = \infty$.
Above, the notation $(ij)(kl)$ means that particles $i$ and $j$ are contracted
into a color singlet and that particles $k$ and $l$ are also contracted into
a color singlet.

\begin {figure}[t]
\begin {center}
  \includegraphics[scale=0.4]{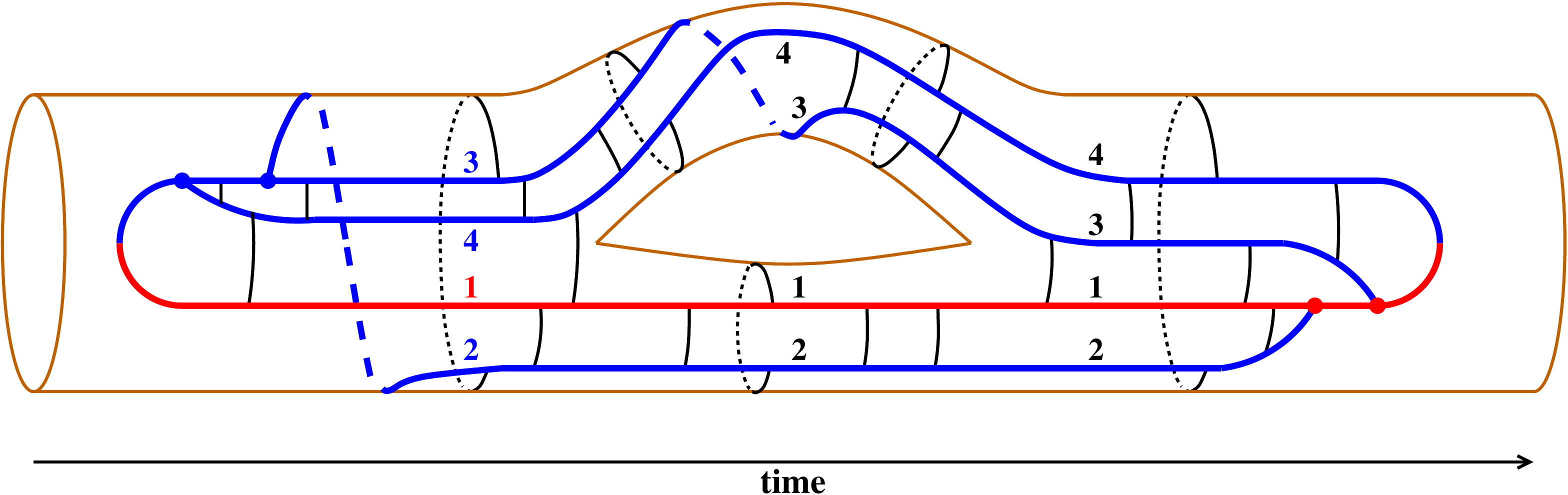}
  \caption{
     \label {fig:handle}
     A topological depiction of the $O(1/N^2)$ transition
     $(1234) \to (12)(34) \to (1243)$
     between $N{=}\infty$ color singlet states during the
     4-gluon phase of evolution of fig.\ \ref{fig:cross}.
     The black lines indicate (2-point) examples of correlations
     of interactions with the medium, which for $N{=}\infty$ are
     allowed only between neighbors.
  }
\end {center}
\end {figure}

In terms of the cylinder picture of fig.\ \ref{fig:crosscyl}, representing
states like $(ij)(kl)$ requires two separate cylinders: one for each
singlet pair.  This is a useful convention because it corresponds
naturally to the large-$N$ topological principle that diagrams
requiring handles are suppressed.
Specifically,
as a preview of what we will see later, fig.\ \ref{fig:handle} shows
one type of $1/N^2$ correction to fig.\ \ref{fig:crosscyl}.
As time progresses during the 4-gluon part of the evolution,
there is a $1/N$ suppressed transition from the
$(1234)$ color singlet state to the $(12)(34)$ color singlet state,
and then later another such transition to the $(1243)$ color singlet state.
In our notation (\ref{eq:basis}), that's
$\sAm \to \sonexm \to \sAx$, where each transition will be
due to $1/N$ corrections to the Hamiltonian.
Some examples of
(2-point%
\footnote{
  There is no reason to only include 2-point correlations
  here: They are simply easier to draw.  All that matters is that
  no lines cross when the diagram and correlations are drawn on the surface.
}%
)
correlations of medium interactions are shown by the black lines.
In the language of large $N$ diagrammatics, the resulting diagram
(interpreted here to {\it include} the medium correlations shown)
cannot be drawn as
a planar diagram, which is why it is $1/N^2$ suppressed.
In general, there is a suppression by $1/N^2$ for
every handle needed to draw a diagram on
a surface without crossing lines \cite{tHooft}.%
\footnote{
  See also Coleman's excellent ``$1/N$'' summer school lecture in
  ref.\ \cite{Coleman}.
}

\subsection {$1/N$ and $1/N^2$ corrections to the potential}

We can now work out corrections to the $N{=}\infty$ limit by expanding
the original Hamiltonian (\ref{eq:HVeffective}) in powers of $1/N$.
The dependence on $N$ appears only in the $\mtrx{S}_u$ and $\mtrx{T}_u$
matrices (\ref{eq:ST}), which can be expanded in powers of $1/N$.
But we will want to express the result in the basis (\ref{eq:basis})
of states that decouple in the $N{=}\infty$ limit, not the original
basis (\ref{eq:6dim}) used for presenting $\mtrx{S}_u$ and
$\mtrx{T}_u$.  After that change of basis,
\begin {equation}
   \mtrx{S}_u = \mtrx{S}^\Ninf + \delta\mtrx{S} ,
   \qquad
   \mtrx{T}_u = \mtrx{T}^\Ninf + \delta\mtrx{T} + \delta^2\mtrx{T} + O(N^{-3})
\end {equation}
with $\mtrx{S}^\Ninf$ and $\mtrx{T}^\Ninf$ as in (\ref{eq:ST0}) and
\begin {multline}
   \delta \mtrx{S} =
   \frac{1}{\sqrt2 \, N}
   \begin {pmatrix}
      ~0 & & & & & \\
      & ~0 & & & & \\
      & & ~0 & & & \\
      & & &  0 & -1 & -1 \\
      & & & -1 & 0 & 0 \\
      & & & -1 & 0 & 0
   \end{pmatrix} ,
   \qquad
   \delta \mtrx{T} =
   \frac{1}{\sqrt2 \, N}
   \begin {pmatrix}
      ~0 & ~1 & ~1 & ~0 & ~0 & ~0 \\
      ~1 & ~0 & ~0 & ~0 & ~1 & ~0 \\
      ~1 & ~0 & ~0 & ~0 & ~0 & ~1 \\
      ~0 & ~0 & ~0 & ~0 & ~0 & ~0 \\
      ~0 & ~1 & ~0 & ~0 & ~0 & ~0 \\
      ~0 & ~0 & ~1 & ~0 & ~0 & ~0
   \end{pmatrix} ,
\\
   \mbox{and}\quad
   \delta^2 \mtrx{T} =
   \frac{1}{N^2}
   \begin {pmatrix}
      ~0 & ~0 & ~0 & ~0 & 0 & 0 \\
      ~0 & ~0 & ~0 & ~\tfrac12 & 0 & 0 \\
      ~0 & ~0 & ~0 & ~\tfrac12 & 0 & 0 \\
      ~0 & ~\tfrac12 & ~\tfrac12 & ~0 & 0 & 0 \\
      ~0 & ~0 & ~0 & ~0 & -\tfrac54 & 0 \\
      ~0 & ~0 & ~0 & ~0 & 0 & \tfrac54 \\
   \end{pmatrix}
  \quad\mbox{in basis}\quad
  \begin{array}{l}
    \sone \\
    \sAp \\
    \sAm \\
    \sAx \\
    \sonexp \\
    \sonexm \,.
  \end{array}
\label {eq:dST}
\end {multline}


\section {Sequential diagrams}
\label {sec:seq}

\begin {figure}[t]
\begin {center}
  \includegraphics[scale=0.6]{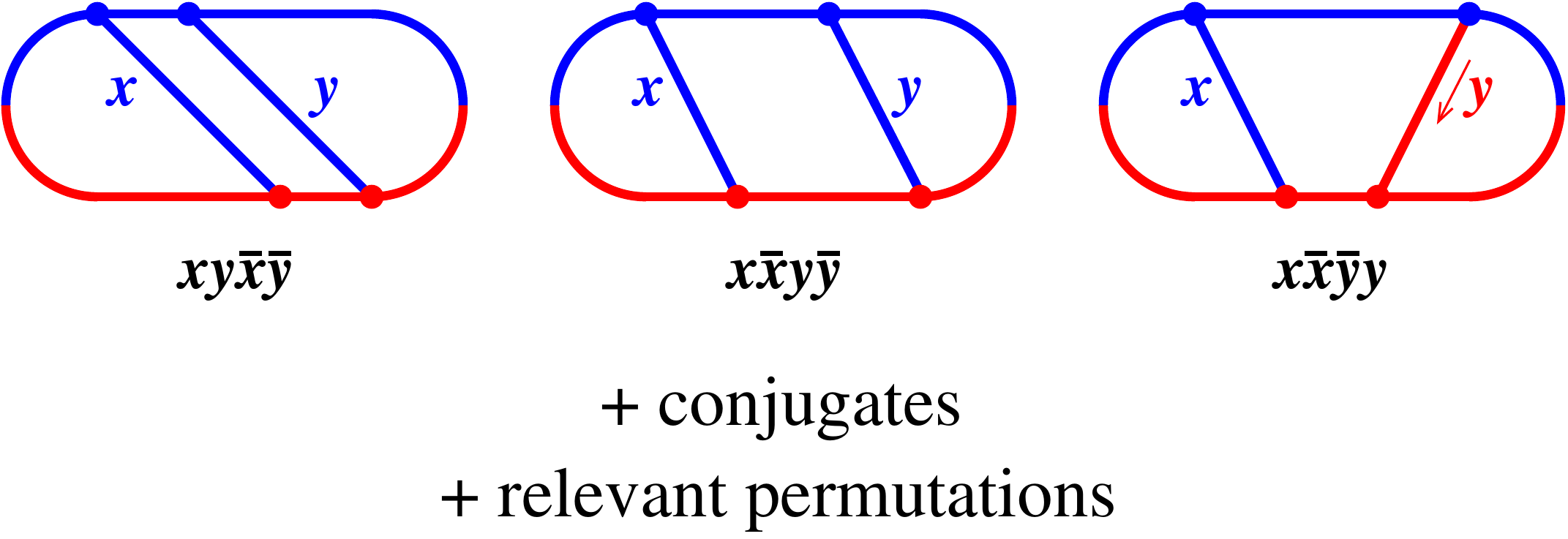}
  \caption{
     \label {fig:seq}
     The above diagrams contributing to double splitting $g{\to}ggg$
     are called the ``sequential diagrams'' in
     ref.\ \cite{seq}.  As in refs.\ \cite{2brem,seq}, the
     diagrams are individually named ($xy\bar x\bar y$, etc.)\
     by the time order of the vertices.
     The relevant permutations referenced above are those permutations of
     the daughters $x$, $y$, and $z \equiv 1{-}x{-}y$ that create
     distinct diagrams.
  }
\end {center}
\end {figure}

The sample diagram we have been
showing so far is called a {\it crossed}\/ diagram \cite{2brem}
because two lines
cross when it is drawn as in fig.\ \ref{fig:cross} (as opposed to the
drawing in fig.\ \ref{fig:crosscyl} of the same diagram on the cylinder).
To study $1/N^2$ corrections, it will be simpler to
start with a different class of diagrams called {\it sequential}\/ diagrams
\cite{seq}, shown in fig.\ \ref{fig:seq}.  However, only the first
diagram $xy\bar x\bar y$
(and its complex conjugate and permutations) will generate
$1/N^2$ corrections.
That's because, as discussed earlier,
there is no interesting color dynamics for 3-particle
propagation, which means that there are no finite-$N$ corrections
needed for those propagators provided one uses the value of $\hat q$
appropriate for the desired value of $N$.
(The same is true of 2-particle propagators.)
Only the $xy\bar x\bar y$ diagram in fig.\ \ref{fig:seq}
has a region of 4-particle evolution
and so non-trivial color dynamics, denoted by the shaded region
in fig.\ \ref{fig:xyxy}.

\subsection {Set-up and allowed color singlet transitions}

\begin {figure}[t]
\begin {center}
  \includegraphics[scale=0.6]{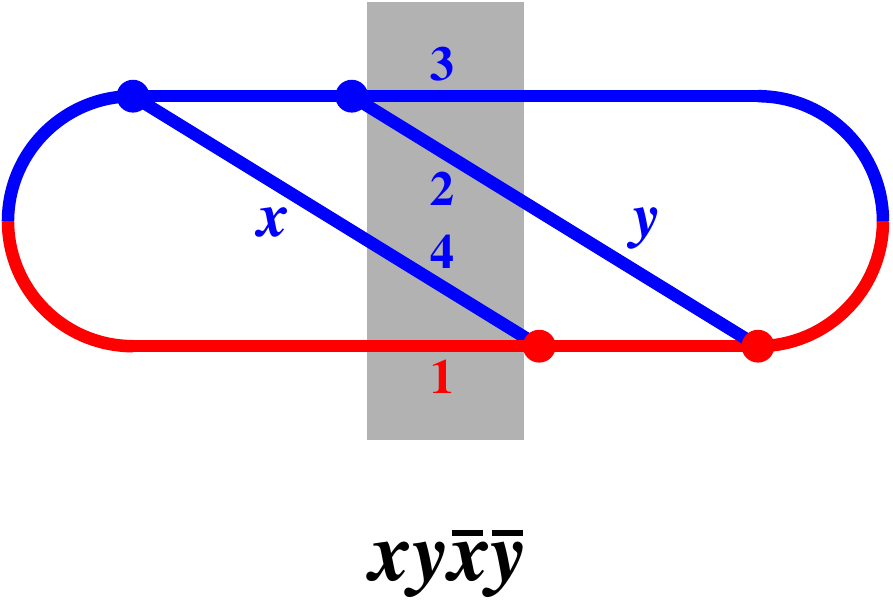}
  \caption{
     \label {fig:xyxy}
     The canonical ``sequential'' diagram for which finite-$N$
     corrections must be calculated.
  }
\end {center}
\end {figure}

We now focus exclusively on the $xy\bar x\bar y$ diagram.
In fig.\ \ref{fig:xyxy},
our numbering of particles in the region of 4-particle evolution follows
the same convention as refs.\ \cite{2brem,seq}.
This diagram gives a contribution to the rate for overlapping double
splitting $g{\to}ggg$ that is proportional to%
\footnote{
   Eq.\ (\ref{eq:xyxyfacs}) isolates the factors we want to discuss here
   from the $N{=}\infty$ expression in eq.\ (E.1) of ref.\ \cite{seq}.
   Technically, integrating over {\it all}\/ of the times
   $(t_\xx < t_\yx < t_\xbx < t_\ybx)$ gives probability, not rate.
   We should
   integrate only over time differences, but that detail
   is unimportant for the present discussion.
}
\begin {align}
   \int_{t_\xx < t_\yx < t_\xbx < t_\ybx}
   \int_{\B^\yx,\B^\xbx}
   &
   \nabla^{\bar n}_{\B^\ybx}
   \langle\B^\ybx,t_\ybx|\B^\xbx,t_\xbx\rangle
   \Bigr|_{\B^\ybx=0}
\nonumber\\ &\times
   \nabla^{\bar m}_{\C_{41}^\xbx}
   \nabla^n_{\C_{23}^\yx}
   \langle\C_{41}^\xbx,\C_{23}^\xbx,t_\xbx|\C_{41}^\yx,\C_{23}^\yx,t_\yx\rangle
   \Bigr|_{\C_{41}^\xbx=0=\C_{23}^\yx; ~ \C_{23}^\xbx=\B^\xbx; ~ \C_{41}^\yx=\B^\yx}
\nonumber\\ &\times
   \nabla^m_{\B^\xx}
   \langle\B^\yx,t_\yx|\B^\xx,t_\xx\rangle
   \Bigr|_{\B^\xx=0} .
\label {eq:xyxyfacs}
\end {align}
Above $(t_\xx,t_\yx,t_\xbx,t_\ybx)$ are the times of the four vertices
in fig.\ \ref{fig:xyxy} from left (earliest) to right (latest).
The factors $\langle\B^\yx,t_\yx|\B^\xx,t_\xx\rangle$ and
$\langle\B^\ybx,t_\ybx|\B^\xbx,t_\xbx\rangle$ represent the propagators
for the 3-particle evolution respectively before and after
the shaded region of the figure.
The factor
$\langle\C_{41}^\xbx,\C_{23}^\xbx,t_\xbx|\C_{41}^\yx,\C_{23}^\yx,t_\yx\rangle$
represents the propagator for the 4-particle evolution inside of the
shaded region.  There is a gradient $\grad$ (corresponding to a factor
of transverse momentum) associated with each
splitting vertex.
We have not shown here other overall factors, including
how those gradients are contracted
together by helicity-dependent DGLAP splitting functions.

The only non-trivial corrections to $N{=}\infty$ come from the
color dynamics of the 4-particle propagator, which we now write as
\begin {equation}
   G(\C_{41}^\xbx,\C_{23}^\xbx,t_\xbx; \C_{41}^\yx,\C_{23}^\yx,t_\yx)
   \equiv
   \langle\C_{41}^\xbx,\C_{23}^\xbx,t_\xbx|\C_{41}^\yx,\C_{23}^\yx,t_\yx\rangle .
\label {eq:Gdef}
\end {equation}
Our specification of the initial and final 4-particle states in the
propagator (\ref{eq:Gdef}) is incomplete: We will also need to
specify what 4-particle {\it color singlet states} we start and end in.
We find it convenient to rewrite (\ref{eq:Gdef}) as
\begin {equation}
  G(\vbxi^{\,\xbx},\Delta t,\lambda^\xbx; \vbxi^{\,\yx},0,\lambda^\yx)
  ,
\label {eq:G1}
\end {equation}
where
\begin {equation}
   \vbxi \equiv
   \begin {pmatrix} {\bm C}_{41} \\ {\bm C}_{23} \end{pmatrix}
\label {eq:xidef}
\end {equation}
is a 2-dimensional vector (with elements that are in turn 2-dimensional
vectors in the transverse plane) encoding the transverse position state
of the system at a given time;
\begin {equation}
  \Delta t \equiv t_\xbx - t_\yx
\end {equation}
is the total duration of the 4-particle evolution;
and $\lambda^\yx$ and $\lambda^\xbx$ label
the initial and final 4-particle color singlet states
for that evolution.

Ref.\ \cite{color} explains that those initial and final singlet states
are each $|{\rm A}_{\rm aa}\rangle_u$ for the diagram of
fig.\ \ref{fig:xyxy}.%
\footnote{
   See section 2.3 of \cite{color}.  Because of different labeling
   of the four particles there (our 1234 here is DBAC in fig.\ 6 of
   ref.\ \cite{color}), what we call $u$-channel here is what
   is called $s$-channel there.
}
A quick, graphical way to understand why is that
(i) as far as color representations are concerned,
everything to the left of the shaded region of fig.\ \ref{fig:xyxy}
looks like the $u$-channel diagram of fig.\ \ref{fig:channels}c
with a gluon ($R{=}{\rm A}$) for the internal line, corresponding to
$|{\rm A}\rangle_u$;
(ii) 3-gluon vertices combine gluons anti-symmetrically via the
group structure constants $f^{abc}$, therefore specializing to
$|{\rm A}_{\rm aa}\rangle_u$; and (iii) there is no color
dynamics for 3-particle evolution, which means that the interactions
with the medium in the actual diagram of fig.\ \ref{fig:xyxy} will
not affect the correspondence with the color-contraction diagram of
fig.\ \ref{fig:channels}c.  A similar argument applies
to everything to the right of the shaded region of fig.\ \ref{fig:xyxy}.

In terms of our $N{=}\infty$ eigenstates (\ref{eq:basis}), the
initial and final color-singlet states of the 4-particle evolution
are then
\begin {equation}
  |{\rm A}_{\rm aa}\rangle_u
  =
  \tfrac{1}{\sqrt2} \sAp + \tfrac{1}{\sqrt2} \sAm .
\label {eq:Aaa}
\end {equation}
So, we will be interested in 4-particle Green functions
(\ref{eq:G1}) where the initial state can be
$\lambda^\yx = \Ap$ or $\Am$ and the final state can be
$\lambda^\xbx = \Ap$ or $\Am$.

From the texture of the finite-$N$ corrections (\ref{eq:dST}) to the
$\mtrx{S}_u$ and $\mtrx{T}_u$ matrices that appear in the Hamiltonian
(\ref{eq:HVeffective}), we can now identify what 4-particle
color-singlet transitions contribute to the
$1/N^2$ correction to the $xy\bar x\bar y$ sequential diagram of
fig.\ \ref{fig:xyxy}.  As just discussed, the sequence of transitions
must start and end with $\Apm$.  The transition sequences allowed
by (\ref{eq:dST}) are then
\def\ssone{\kern0.31em\sone\kern0.31em}  
\begin {subequations}
\label {eq:seqtransitions}
\begin {align}
   \sAm \xrightarrow{\delta T}{} & \ssone   \xrightarrow{\delta T} \sAm ,
\\
   \sAm \xrightarrow{\delta T}{} & \sonexm  \xrightarrow{\delta T} \sAm ,
\\
   \sAm \xrightarrow{\delta T}{} & \ssone   \xrightarrow{\delta T} \sAp ,
\\
   \sAp \xrightarrow{\delta T}{} & \ssone   \xrightarrow{\delta T} \sAp ,
\\
   \sAp \xrightarrow{\delta T}{} & \sonexp  \xrightarrow{\delta T} \sAp ,
\\
   \sAp \xrightarrow{\delta T}{} & \ssone   \xrightarrow{\delta T} \sAm .
\end {align}
\end {subequations}
Note that neither $\delta\mtrx{S}$ nor $\delta^2\mtrx{T}$
contribute to any
allowed $O(N^{-2})$ corrections for this diagram.

There are no $O(N^{-1})$ corrections to the diagram:
neither $\delta\mtrx{S}$ nor $\delta\mtrx{T}$ produce a
direct $\sAp \rightarrow \sAm$ or $\sAm \rightarrow \sAp$ transition.
This is consistent with the fact that, for purely gluonic processes,
corrections in a large-$N$ analysis should appear in powers of $1/N^2$
\cite{tHooft}.

In passing, we note that the allowed transitions (\ref{eq:seqtransitions})
can be written in the alternative language of (\ref{eq:1234}) as
\begin {subequations}
\label {eq:seq1234}
\begin {align}
   (1234) \xrightarrow{\delta T}{} & (41)(23) \xrightarrow{\delta T} (1234) ,
\\
   (1234) \xrightarrow{\delta T}{} & (12)(34) \xrightarrow{\delta T} (1234) ,
\\
   (1234) \xrightarrow{\delta T}{} & (41)(23) \xrightarrow{\delta T} (1324) ,
\\[8pt]
   (1324) \xrightarrow{\delta T}{} & (41)(23) \xrightarrow{\delta T} (1324) ,
\\
   (1324) \xrightarrow{\delta T}{} & (13)(24) \xrightarrow{\delta T} (1324) ,
\\
   (1324) \xrightarrow{\delta T}{} & (41)(23) \xrightarrow{\delta T} (1234) .
\end {align}
\end {subequations}
Note that the last three sequences may be obtained
from the first three sequences by exchanging ($2{\leftrightarrow}3$)
particles 2 and 3.
But the only thing differentiating particles 2 and 3 in the
$xy\bar x\bar y$ diagram of fig.\ \ref{fig:xyxy} is their longitudinal
momentum fractions $y$ and $z\equiv 1{-}x{-}y$.
This means that instead
of calculating the contributions of all six sequences
(\ref{eq:seq1234}), one could, if desired, use only the first three
sequences but then add (i) that result to (ii) the same calculation with
the value of $y$ changed to $1{-}x{-}y$.


\subsection{$1/N$ perturbation theory for 4-particle propagator}

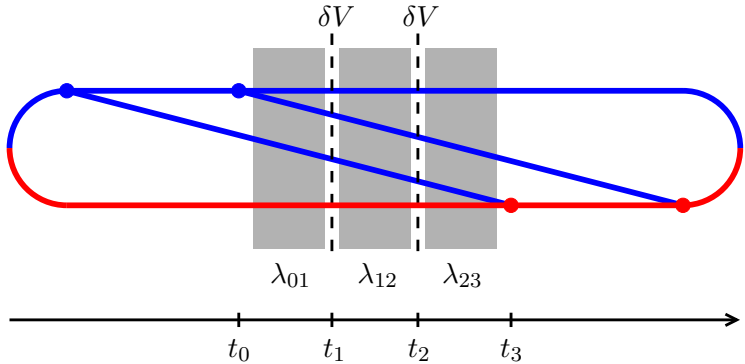
\begin {figure}[t]
\begin {center}
  \begin{picture}(277,140)(0,0)
    \put(0,15){\includegraphics[scale=0.6]{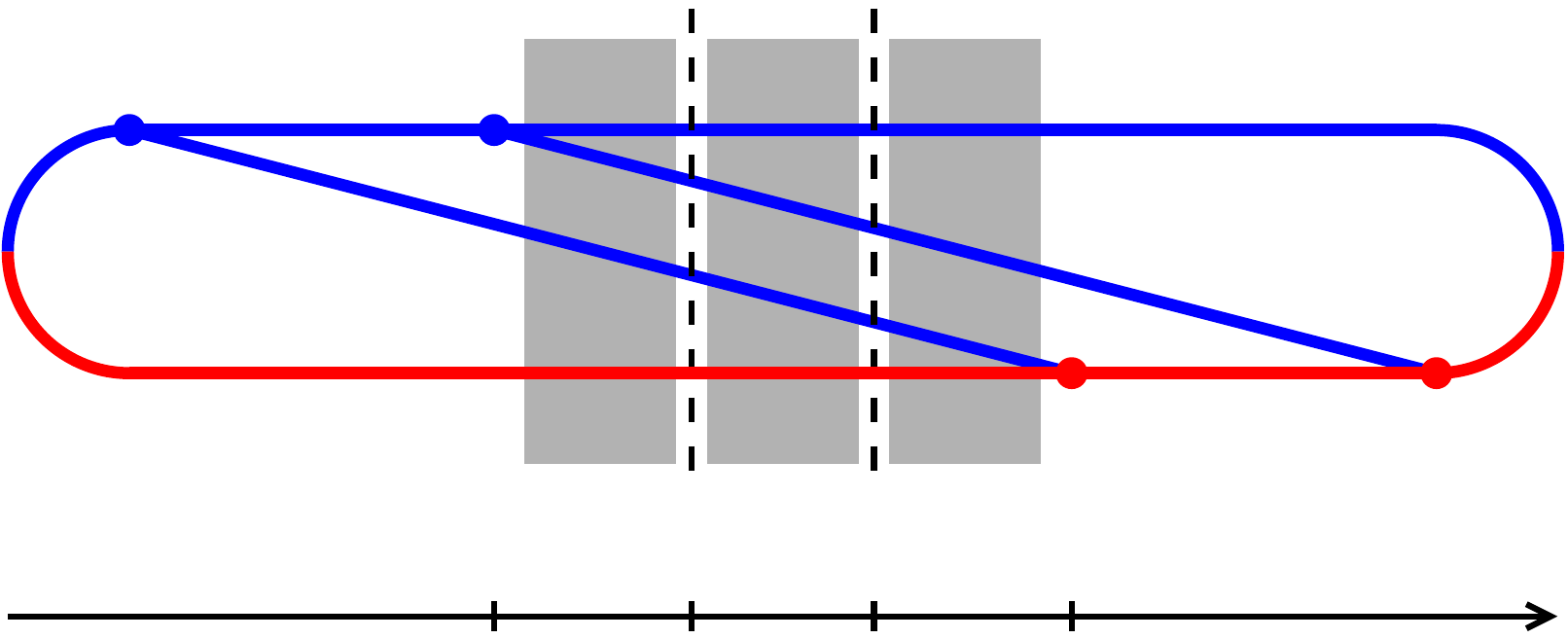}}
    \put(84,5){$t_0$}
    \put(120,5){$t_1$}
    \put(152,5){$t_2$}
    \put(187,5){$t_3$}
    \put(100,33){$\lambda_{01}$}
    \put(133,33){$\lambda_{12}$}
    \put(165,33){$\lambda_{23}$}
    \put(118,130){$\delta V$}
    \put(150,130){$\delta V$}
  \end{picture}
  \caption{
     \label {fig:dGseq}
     A depiction of the 2nd-order perturbative correction (\ref{eq:d2G1})
     in $1/N$ to 4-particle evolution.
     The shading shows regions
     of 4-particle propagation where $N{=}\infty$ propagators are used.
     The dashed lines represent insertions of the $1/N$ correction $\delta V$
     to the potential at intermediate
     times $t_1$ and $t_2$, which are integrated over.
  }
\end {center}
\end {figure}

Let $G^\Ninf_\lambda$ represent the $N{=}\infty$ 4-particle propagator
for any of the $N{=}\infty$ color singlet eigenstates of
(\ref{eq:basis}), indexed by $\lambda$.
In perturbation theory in $1/N$, then the transitions
(\ref{eq:seqtransitions}) correspond to $O(N^{-2})$ corrections to
the propagator of the form
\begin {align}
  \delta^2 & G_{\lambda_{23}\leftarrow\lambda_{01}}
             (\vbxi_3,\Delta t; \vbxi_0,0)
\nonumber\\ & \qquad
  =
  (-i)^2
  \sum_{\lambda_{12}}
  \int_{0<t_1<t_2<\Delta t} dt_1 \> dt_2 \>
  \int_{\vbxi_1,\vbxi_2}
  G^\Ninf_{\lambda_{23}}(\vbxi_3,\Delta t; \vbxi_2,t_2) \,
  \delta V_{\lambda_{23}\leftarrow\lambda_{12}}^{(\delta T)}\!(\vbxi_2) \,
\nonumber\\ & \hspace{10em} \times
  G^\Ninf_{\lambda_{12}}(\vbxi_2,t_2; \vbxi_1,t_1) \,
  \delta V_{\lambda_{12}\leftarrow\lambda_{01}}^{(\delta T)}\!(\vbxi_1) \,
  G^\Ninf_{\lambda_{01}}(\vbxi_1,t_1; \vbxi_0,t_0) .
\label {eq:d2G1}
\end {align}
Above, $t_0 = t^\yx$ and $t_3 = t^\xbx$
are the initial and final times of the 4-particle evolution (the
shaded region) in fig.\ \ref{fig:xyxy}.
The two $O(N^{-1})$
perturbations to $N{=}\infty$ evolution (caused by $\delta\mtrx{T}$)
occur at intermediate times
$t_1$ and $t_2$, as depicted in fig.\ \ref{fig:dGseq}.
Each $\lambda_{ij}$ designates an $N{=}\infty$ eigenstate from (\ref{eq:basis}).
As discussed previously, the initial color singlet state $\lambda_{01}$
and the final color singlet state $\lambda_{23}$ must be
$\sAp$ or $\sAm$ as in (\ref{eq:Aaa}) and (\ref{eq:seqtransitions}).
$\delta V^{(\delta T)}_{\lambda\leftarrow\lambda'}$ represents the
$(\lambda,\lambda')$ matrix element of the $\delta\mtrx{T}$ contribution
to the potential (\ref{eq:Veffective}).  The non-zero matrix elements
are all the same because the non-zero matrix elements of
$\delta\mtrx{T}$ in (\ref{eq:dST}) are all the same:
\begin {equation}
  \delta V^{(\delta T)}\!(\vbxi) =
  \frac{i\hat q_{\rm A}}{2\sqrt2\,N} \, (x_4+x_1)(x_2+x_3) \,
  \C_{41}\cdot\C_{23} \,.
\label {eq:dVdT0}
\end {equation}
In order to focus on structure over details,
and also to allow for later generalizations,
we will find it useful to introduce some short-hand notation
for (\ref{eq:dVdT0}) and also to distinguish the earlier-time and later-time
insertions of $\delta V$ in (\ref{eq:d2G1}):
\begin {equation}
  \delta V^{(\delta T)}_{\lambda_{12}\leftarrow\lambda_{01}}\!(\vbxi_1) =
  \tfrac12 \vbxi_1^{\,\top} R_1 \, \vbxi_1 ,
  \qquad
  \delta V^{(\delta T)}_{\lambda_{23}\leftarrow\lambda_{12}}\!(\vbxi_2) =
  \tfrac12 \vbxi_2^{\,\top} R_2 \, \vbxi_2
  \qquad
  \mbox{(for allowed transitions)}
\label {eq:dVdT}
\end {equation}
with
\begin {equation}
  R_1 = R_2 = R^{(\delta T)} \equiv
    -\frac{i\hat q_{\rm A}}{2\sqrt2\,N} \, (x_1{+}x_4)^2
    \begin{pmatrix} 0 & 1 \\ 1 & 0 \end {pmatrix}
\label {eq:R}
\end {equation}
(where we have used the fact that $x_1{+}x_2{+}x_3{+}x_4=0$).
Here, $R^{(\delta T)}$ is a $2{\times}2$ matrix that mixes the two components
$(\C_{41},\C_{23})$ of the vector $\vbxi$ defined
by (\ref{eq:xidef}).  It does not do anything to the transverse position
space in which each $\C$ lives except to contract the transverse indices,
as in (\ref{eq:dVdT0}).  If one wants to be explicit, one could think of
the matrices $R_i$ shown in (\ref{eq:dVdT}) as really being
$R_i \otimes \openone$,
where the $2{\times}2$ identity matrix
acts on transverse position space.  However, in our discussion,
we will not speak explicitly about the transverse space.  So, for example,
we will refer to $\vbxi$ throughout this paper
as a ``2-dimensional'' (rather than 4-dimensional) vector, and we will
correspondingly refer to the matrices in (\ref{eq:dVdT}) as the
$2{\times}2$ matrices (\ref{eq:R}).

We will see that the integration over all intermediate transverse positions
in our diagram can be performed analytically in $\hat q$ approximation.
That will leave three time integrals ($t_1$, $t_2$, and
$\Delta t \equiv t_3-t_0$) to later be performed numerically.

To continue, we need the structure of the $N{=}\infty$ 4-particle
propagators.  In the $\hat q$ approximation, these are 2-dimensional
harmonic oscillator propagators for a coupled set of two oscillators
$(C_{41},C_{23})$.  Adapting the notation of ref.\ \cite{2brem,seq}, we will
refer to the two complex normal-mode
frequencies of this system as $\Omega_\pm^{(\lambda)}$ and define the
$2{\times}2$ diagonal matrix
\begin {equation}
  \Omega_{(\lambda)}
  \equiv
  \begin {pmatrix}
     \Omega_+^{(\lambda)} & \\ & ~\Omega_-^{(\lambda)}
  \end {pmatrix} .
\label {eq:OmMatrixDef}
\end {equation}
For our context here, we have introduced the subscript or superscript
$\lambda$ to
indicate which color singlet-state (\ref{eq:basis}) we are finding
the $N{=}\infty$ propagators for.
Again adapting the notation of refs.\ \cite{2brem,seq}, we will
make a matrix $a$ whose columns are the corresponding normal mode
vectors:
\begin {equation}
   a_{(\lambda)} =
   \begin{pmatrix}
      C^+_{41} & C^-_{41} \\
      C^+_{23} & C^-_{23}
   \end{pmatrix}_{\!(\lambda)} .
\label {eq:adef}
\end {equation}
We will leave for later the details of exactly what $\Omega_\pm$ and $a$ are
for each $N{=}\infty$ color singlet state $\lambda$.  For now, we
have enough to write out the structure of the
harmonic-oscillator propagator, which is%
\footnote{
  It is because we are working in the same basis $(C_{41},C_{23})$ throughout
  the 4-particle evolution that the first and last terms in the exponent
  of (\ref{eq:Ginf}) have the same matrix $A_\lambda$.
  This is unlike the original $N{=}\infty$
  analysis of diagrams in ref.\ \cite{2brem,seq},
  where it was found more convenient to use a different basis at the two ends
  of the propagator.
}
\begin {equation}
  G_\lambda^\Ninf
    (\vbxi,t; \vbxi^{\,\prime},0)
  =
  f_\lambda(t)
  \exp\Bigl[
     - \tfrac12 \vbxi^{\,\,\top} \! A_\lambda(t) \,\vbxi
     + \vbxi^{\,\,\top} B_\lambda(t) \,\vbxi^{\,\prime}
     - \tfrac12 \vbxi^{\,\prime\top} \! A_\lambda(t) \,\vbxi^{\,\prime}
  \Bigr] ,
\label {eq:Ginf}
\end {equation}
where
\begin {align}
   A_\lambda(t) &\equiv
   -i \bigl[ (a^{\top})^{-1} \Omega\cot(\Omega t) \, a^{-1} \bigr]_{(\lambda)} \,,
\\
   B_\lambda(t) &\equiv
   -i \bigl[ (a^{\top})^{-1} \Omega\csc(\Omega t) \, a^{-1} \bigr]_{(\lambda)} \,,
\label {eq:Bdef}
\end {align}
and the prefactor%
\footnote{
  \label{foot:UV}
  For $N{=}\infty$, calculations of individual time-ordered diagrams
  were ultraviolet (UV) divergent (even for tree-level processes),
  which was treated
  with dimensional regularization in ref.\ \cite{dimreg}.  Those divergences,
  however, were associated with 4-particle evolution times $\Delta t \to 0$
  and so with the vacuum limit of the 4-particle propagators $G$.
  For vacuum evolution, there is no interesting color dynamics, and it
  is color dynamics that our $1/N$ corrections describe.  As a
  result, there will be no UV divergences in our calculations of corrections
  in this paper, which means that we do not need to use dimensional
  regularization and so may use the 2-transverse dimensional formula
  (\ref{eq:fdef}) for $f_\lambda$.
}
\begin {equation}
  f_\lambda(t) \equiv
  \det\left( \frac{B_\lambda(t)}{2\pi} \right) .
\label {eq:fdef}
\end {equation}


\subsection{Integrating over $\vbxi_1$ and $\vbxi_2$}

The integrals over $\vbxi_1$ and $\vbxi_2$ in the expression
(\ref{eq:d2G1}) for $\delta^2G$ are related to Gaussian integrals and
so may be done analytically.  We find the results are more compact if
we first combine the two integrals into a single Gaussian integral by
defining a 4-dimensional vector
\begin {equation}
   \vec{\bm \Xi} \equiv
   \begin{pmatrix} \vbxi_1 \\[3pt] \vbxi_2 \end{pmatrix}
\label {eq:Xidef}
\end {equation}
from the two intermediate position vectors $\vbxi_1$ and $\vbxi_2$.
Similarly, define
\begin {equation}
   \vec{\bm z} \equiv
   \begin{pmatrix} \vbxi_0 \\[3pt] \vbxi_3 \end{pmatrix}
\label {eq:zdef}
\end {equation}
to be a 4-dimensional vector composed of the initial and final position
vectors $\vbxi_0$ and $\vbxi_3$ for the 4-particle evolution.
Then the expression (\ref{eq:d2G1}) for $\delta^2G$,
together with (\ref{eq:dVdT}) for $\delta V$ and (\ref{eq:Ginf}) for $G^\Ninf$,
can be rewritten in the form
\begin {multline}
  \delta^2 G_{\lambda_{23}\leftarrow\lambda_{01}}
  (\vbxi_3,\Delta t; \vbxi_0,0)
  =
  (-i)^2 \sum_{\substack{{\rm allowed}\\\lambda_{12}}}
  f_{(01)} f_{(12)} f_{(23)}
  \int_{0<t_1<t_2<\Delta t} dt_1\> dt_2 \>
  e^{ -\frac12 \vec{\bm z}^\top {\cal A} \, \vec{\bm z} }
\\ \times
  \frac{\partial}{\partial j_1} \, \frac{\partial}{\partial j_2}
  \int d^4\Xi \>
     e^{ - \frac12 \vec{\bm \Xi}^\top {\cal U} \vec{\bm \Xi} }
     e^{ \vec{\bm z}^\top {\cal B} \, \vec{\bm \Xi} }
  \>
  \biggr|_{j_1=j_2=0} ,
\label {eq:d2GXi}
\end {multline}
where we define the $4{\times}4$ matrices
\begin {subequations}
\label {eq:calUABdef}
\begin {equation}
  {\cal U} \equiv
  \begin {pmatrix}
     A_{(01)} + A_{(12)} - j_1 R_1 & -B_{(12)} \\[2pt]
     -B_{(12)} & A_{(12)} + A_{(23)} - j_2 R_2
  \end {pmatrix} ,
\label {eq:calUdef}
\end {equation}
\begin {equation}
  {\cal A} \equiv
  \begin {pmatrix} A_{(01)} & \\ & A_{(23)} \end {pmatrix} ,
  \qquad
  {\cal B} \equiv
  \begin {pmatrix} B_{(01)} & \\ & B_{(23)} \end {pmatrix} .
\label {eq:calABdef}
\end {equation}
\end {subequations}
Above, we use the shorthand notation
\begin {equation}
   A_{(ij)} \equiv A_{\lambda_{ij}}^{}\!(t_j{-}t_i) ,
   \qquad
   B_{(ij)} \equiv B_{\lambda_{ij}}^{}\!(t_j{-}t_i) ,
   \qquad
   f_{(ij)} \equiv f_{\lambda_{ij}}^{}\!(t_j{-}t_i) .
\end {equation}
The parameters $j_1$ and $j_2$ are dummy source term coefficients
used to generate the two factors (\ref{eq:dVdT}) of $\delta V$
in (\ref{eq:d2G1})
from the Gaussian integral appearing in (\ref{eq:d2GXi}).
Doing that Gaussian integral gives%
\footnote{
   Even though we have written the Gaussian integral as a 4-dimensional
   integral $\int d^4\Xi\>\cdots$, it is secretly an 8-dimensional
   integral because each of the four components of
   $\Xi$ is itself a 2-dimensional position vector $\C$ in the transverse plane.
   For this reason, the Gaussian integral produces an exponential prefactor
   $\det(2\pi {\cal U}^{-1}) = (2\pi)^4\det({\cal U}^{-1})$
   [where $\det$ is the 4-dimensional determinant]
   instead of $\sqrt{\det(2\pi{\cal U}^{-1})}$.
} 
\begin {multline}
  \delta^2 G_{\lambda_{23}\leftarrow\lambda_{01}}
  (\vbxi_3,\Delta t; \vbxi_0,0)
  =
  (-i)^2 (2\pi)^4 \sum_{\substack{{\rm allowed}\\\lambda_{12}}}
  f_{(01)} f_{(12)} f_{(23)}
  \int_{0<t_1<t_2<\Delta t} dt_1\> dt_2 \>
\\ \times
  \frac{\partial}{\partial j_1} \, \frac{\partial}{\partial j_2}
  \biggl[
    \det({\cal U}^{-1})
    e^{
       -\frac12 \vec{\bm z}^\top
       ( {\cal A} - {\cal B}\,{\cal U}^{-1} {\cal B} ) 
       \vec{\bm z}
    }
 \biggr]_{j_1=j_2=0} .
\label {eq:d2Gdet}
\end {multline}


\subsection{Evaluating the $xy\bar x\bar y$ diagram}

We could now go through all the additional steps of (\ref{eq:xyxyfacs})
for evaluating the $xy\bar x\bar y$ diagram, which involve taking
gradients of the 4-particle propagator,
including the initial and final 3-particle propagators, integrating
analytically over the intermediate position $\B^\yx$ and $\B^\xbx$,
integrating analytically over the first and last vertex times
$t_\xx$ and $t_\ybx$, and correctly keeping track of all the prefactors
not shown explicitly in (\ref{eq:xyxyfacs}).
Instead, we are going to use a trick to bypass all of that by realizing
that we can adapt the final result of the same steps that were
applied in the original
$N{=}\infty$ calculations of refs.\ \cite{2brem,seq}.
The trick will be to cast the 4-particle propagator (\ref{eq:d2Gdet})
for our $1/N^2$ correction into the same schematic form as the 4-particle
propagator originally used in $N{=}\infty$ calculations.
Let's first discuss the latter to introduce notation
$(X,Y,Z)$ that was used in refs.\ \cite{2brem,seq,dimreg}.

In the original $N{=}\infty$ analysis of the $xy\bar x\bar y$ diagram
in ref.\ \cite{seq}, there were two color routings that had to be
considered, which in the language of our paper here correspond to
taking the full 4-particle propagator
$\langle\C_{41}^\xbx,\C_{23}^\xbx,t_\xbx|\C_{41}^\yx,\C_{23}^\yx,t_\yx\rangle$
for this diagram in (\ref{eq:xyxyfacs}) to be either
$G^\Ninf_{\Ap}$ or $G^\Ninf_{\Am}$, corresponding to
the two $N{=}\infty$ eigenstates that appear in (\ref{eq:Aaa}).
The calculations in ref.\ \cite{seq} focused on the color routing
called here $\Am = (1234)$, to which the result for the other color routing
could be related by swapping the daughters $y$ and $z \equiv 1{-}x{-}y$.
In evaluating the $\Am$ color routing, ref.\ \cite{seq} organized the
calculation (following the method of ref.\ \cite{2brem})
by writing the exponential piece of the
corresponding harmonic oscillator propagator in the form%
\footnote{
  Our (\ref{eq:XYZexp}) is not shown explicitly in ref.\ \cite{seq}.
  There the argument, in appendix E.2,
  proceeds by analogy with section 5.3 of ref.\ \cite{2brem} and
  skips over this explicit formula.  The analogous formula is
  eq.\ (5.41) of ref.\ \cite{2brem}.
}
\begin {multline}
  \langle\C_{41}^\xbx,\C_{23}^\xbx,t_\xbx|\C_{41}^\yx,\C_{23}^\yx,t_\yx\rangle
  =
\\
   f
   \exp\Biggl[
     - \frac12
     \begin{pmatrix} \C_{41}^\yx \\ \C_{23}^\yx \end{pmatrix}^{\!\!\top} \!
       \begin{pmatrix}
          \calX_\yx^\seq & Y_\yx^\seq \\
          Y_\yx^\seq & Z_\yx^\seq 
       \end{pmatrix}
       \begin{pmatrix} \C_{41}^\yx \\ \C_{23}^\yx \end{pmatrix}
     -
     \frac12
     \begin{pmatrix} \C_{23}^\xbx \\ \C_{41}^\xbx \end{pmatrix}^{\!\!\top} \!
       \begin{pmatrix}
          \calX_\xbx^\seq & Y_\xbx^\seq \\
          Y_\xbx^\seq & Z_\xbx^\seq
       \end{pmatrix}
       \begin{pmatrix} \C_{23}^\xbx \\ \C_{41}^\xbx \end{pmatrix}
\\
     +
     \begin{pmatrix} \C_{41}^\yx \\ \C_{23}^\yx \end{pmatrix}^{\!\!\top} \!
       \begin{pmatrix}
          X_{\yx\xbx}^\seq & Y_{\yx\xbx}^\seq \\
          \Ybar_{\yx\xbx}^\seq & Z_{\yx\xbx}^\seq
       \end{pmatrix}
       \begin{pmatrix} \C_{23}^\xbx \\ \C_{41}^\xbx \end{pmatrix}
   \Biggr] ,
\label {eq:XYZexp}
\end {multline}
where the $\C_{ij}$-independent prefactor $f$ is unimportant at the moment.
The above equation just gives particular
names to the entries of the matrices that
in this paper we would call $A_{\Am}$ and $B_{\Am}$: namely%
\footnote{
  The relationship between $(\calX,Y,Z)_\yx^\seq$ and $(\calX,Y,Z)_\xbx^\seq$
  follows from eqs.\ (E.11-12) of ref.\ \cite{seq} and from our
  (\ref{eq:XvscalXseq}), which shows the relationship
  between our $\calX$ here and
  the $X$ in ref.\ \cite{seq}.
}
\begin {equation}
   A_{\Am} =
       \begin{pmatrix}
          \calX_\yx^\seq & Y_\yx^\seq \\
          Y_\yx^\seq & Z_\yx^\seq
       \end{pmatrix}
   =
       {\cal S}
       \begin{pmatrix}
          \calX_\xbx^\seq & Y_\xbx^\seq \\
          Y_\xbx^\seq & Z_\xbx^\seq
       \end{pmatrix}
       {\cal S} ,
   \qquad
   B_{\Am} =
       \begin{pmatrix}
          X_{\yx\xbx}^\seq & Y_{\yx\xbx}^\seq \\
          \Ybar_{\yx\xbx}^\seq & Z_{\yx\xbx}^\seq
       \end{pmatrix}
       {\cal S} ,
\label {eq:ABXYZ}
\end {equation}
where
\begin {equation}
  {\cal S} \equiv \begin{pmatrix} 0 & 1 \\ 1 & 0 \end{pmatrix}
  \label {eq:calS}
\end {equation}
is a matrix that flips the vectors $(C_{23},C_{41})$ appearing in parts
of (\ref{eq:XYZexp}) to the basis $(C_{41},C_{23})$ that we have
used exclusively in this paper.
For $N{=}\infty$, particular formulas for the $(X,Y,Z)$'s were
given in ref.\ \cite{seq}, which also figured out how to write
the final answer for the diagram in terms of the $(X,Y,Z)$'s.

Now compare the old $N{=}\infty$ formula above
to the contribution of a {\it particular} color singlet transition
sequence $\lambda_{01} \to \lambda_{12} \to \lambda_{23}$ in
(\ref{eq:d2Gdet}) if we {\it leave out}\/ the operation
$\partial_{j_1} \partial_{j_2} [ \cdots ]_{j_1=j_2=0}$.
The dependence on the $\C_{ij}$'s is then completely contained in
the 4-vector $\vec{\bm z}$ of (\ref{eq:zdef})
and so in the exponential factor
\begin {equation}
    e^{
       -\frac12 \vec{\bm z}^\top
       ( {\cal A} - {\cal B}\,{\cal U}^{-1} {\cal B} ) 
       \vec{\bm z}
    }
\label {eq:expzz}
\end {equation}
of (\ref{eq:d2Gdet}).
Comparing this exponential factor with the one in
(\ref{eq:XYZexp}), we see that it has the same form, except
that the $(X,Y,Z)$'s for the $N{=}\infty$ calculation are replaced by
alternate versions, which we'll call $(\tilde X,\tilde Y,\tilde Z)$,
given by
\begin {equation}
  \begin{pmatrix}
     \tilde\calX_\yx^\seq & \tilde Y_\yx^\seq
       & -\tilde Y_{\yx\xbx}^\seq & -\tilde X_{\yx\xbx}^\seq
       \\[4pt]
     \tilde Y_\yx^\seq & \tilde Z_\yx^\seq
       & -\tilde Z_{\yx\xbx}^\seq & -\tilde \Ybar_{\yx\xbx}^\seq
       \\[4pt]
     -\tilde Y_{\yx\xbx}^\seq & -\tilde Z_{\yx\xbx}^\seq
       & \tilde Z_\xbx^\seq & \tilde Y_\xbx^\seq
       \\[4pt]
     -\tilde X_{\yx\xbx}^\seq & -\tilde \Ybar_{\yx\xbx}^\seq
       & \tilde Y_\xbx^\seq & \tilde\calX_\xbx^\seq
  \end{pmatrix}
  =
  {\cal A} - {\cal B}\,{\cal U}^{-1} {\cal B}
  .
\label {eq:XYZreadout}
\end {equation}
If we calculate ${\cal A} - {\cal B}\,{\cal U}^{-1} {\cal B}$
from the formulas (\ref{eq:calUABdef}), we can then use
(\ref{eq:XYZreadout}) to read off the
corresponding values of the $(\tilde X,\tilde Y,\tilde Z)$'s.
We may then use {\it those} values in place of the $(X,Y,Z)$'s in the
final $N{=}\infty$ result, except we will also need to replace the
prefactor $f$ in (\ref{eq:XYZexp}) by the prefactors in
(\ref{eq:d2Gdet}), and sum over the allowed color singlet transition
sequences.  At the very end, we will also then need to restore the overall
operation $\partial_{j_1} \partial_{j_2} [ \cdots ]_{j_1=j_2=0}$ that
we strategically
ignored in order to relate the different calculations!

Our starting point, the result of ref.\ \cite{seq} for the color routing
$\Am = (1234)$, is%
\footnote{
  Specifically, see eq.\ (2.36) of ref.\ \cite{seq},
  where the $\Am$ color routing
  of $xy\bar x\bar y$ is called $xy\bar x\bar y_2$.
}
\begin {align}
  \left[\frac{d\Gamma}{dx\,dy}\right]_{\substack{xy\bar x\bar y\\(\Am)}}^\Ninf =
   -
   \int_0^\infty & d(\Delta t) \>
   \frac{\CA^2 \alphas^2 M_\ix M_\fx^\seq}{8\pi^2(x_1{+}x_4)^2E^4} \,
   f_{\Am}
\nonumber\\ &\times
   \Bigl\{
     (\bar\beta Y_\yx^\seq Y_\xbx^\seq
        + \bar\alpha \Ybar_{\yx\xbx}^{\,\seq} Y_{\yx\xbx}^\seq) I_0^\seq
     + (\bar\alpha+\bar\beta+2\bar\gamma) Z_{\yx\xbx}^\seq I_1^\seq
\nonumber\\ &\quad
     + \bigl[
         (\bar\alpha+\bar\gamma) Y_\yx^\seq Y_\xbx^\seq
         + (\bar\beta+\bar\gamma) \Ybar_{\yx\xbx}^{\,\seq} Y_{\yx\xbx}^\seq
        \bigr] I_2^\seq
\nonumber\\ &\quad
     - (\bar\alpha+\bar\beta+\bar\gamma)
       (\Ybar_{\yx\xbx}^{\,\seq} Y_\xbx^\seq I_3^\seq
        + Y_\yx^\seq Y_{\yx\xbx}^\seq I_4^\seq)
   \Bigl\}
\label {eq:xyxyNinf}
\end {align}
where
\begin {subequations}
\label {eq:I}
\begin {equation}
    I_0^\seq =
    \left[ \frac{4\pi^2}{X_\yx X_\xbx - X_{\yx\xbx}^2} \right]^\seq \,,
  \qquad
    I_1^\seq =
    -
    \left[
       \frac{2\pi^2}{X_{\yx\xbx}}
       \ln\left( \frac{X_\yx X_\xbx - X_{\yx\xbx}^2}{X_\yx X_\xbx} \right)
    \right]^{\seq} ,
\end {equation}
\begin {equation}
    I_2^\seq = \left[ I_0 - \frac{I_1}{X_{\yx\xbx}} \right]^\seq ,
  \qquad
    I_3^\seq = \left[ \frac{X_{\yx\xbx} I_0}{X_\xbx} \right]^\seq ,
  \qquad
    I_4^\seq = \left[ \frac{X_{\yx\xbx} I_0}{X_\yx} \right]^\seq
\end {equation}
\end {subequations}
and%
\footnote{
  Eq.\ (\ref{eq:flam}) is defined using our conventions in this paper.
  To obtain it,
  start by permuting eqs.\ (5.35--5.36) of ref.\ \cite{2brem}
  to the basis $(C_{41},C_{23})$ we use,
  giving
  $|\det a_{(\lambda)}|^{-1} = |x_1 x_2 x_3 x_4|^{1/2} |x_1{+}x_4| E$
  in our conventions here.
  Then our (\ref{eq:fdef}) and (\ref{eq:Bdef}) give (\ref{eq:flam}).
}
\begin {equation}
  f_\lambda =
   (2\pi i)^{-2} (-x_1 x_2 x_3 x_4) (x_1{+}x_4)^2 E^2
   \Omega_+^{(\lambda)}\csc\bigl(\Omega_+^{(\lambda)} t\bigr) \,
   \Omega_-^{(\lambda)}\csc\bigl(\Omega_-^{(\lambda)} t\bigr)
  .
\label {eq:flam}
\end {equation}
Formulas for $(\bar\alpha,\bar\beta,\bar\gamma)$, which represent
various combinations of helicity-dependent DGLAP splitting functions,
may be found in ref.\ \cite{seq}.
The variables $X_\yx^\seq$ and $X_\xbx^\seq$ are related to the variables
$\calX_\yx^\seq$ and $\calX_\xbx^\seq$ we introduced earlier in
(\ref{eq:XYZexp}) by
\begin {subequations}
\label{eq:XvscalXseq}
\begin {align}
   X_\yx^\seq &= |M_\ix|\Omega_\ix + \calX_\yx^\seq ,
\\
   X_\xbx^\seq &= |M_\fx^\seq| \Omega_\fx^\seq + \calX_\xbx^\seq ,
\end {align}
\end {subequations}
where the additional $|M|\Omega$ terms arise from the integration of the
3-particle propagators, as described in ref.\ \cite{2brem}.
Finally, the formulas for $M_\ix$, $\Omega_\ix$, $M_\fx^\seq$, and
$\Omega_\fx^\seq$ may be found in ref.\ \cite{seq}.  These have to do
with the 3-particle evolution (which has no interesting color
dynamics), and they remain the same in our problem.

We now obtain the desired $1/N^2$ correction to (\ref{eq:xyxyNinf})
by swapping the $(X,Y,Z)$'s to $(\tilde X,\tilde Y,\tilde Z)$'s
and replacing the prefactor $f_{(\Am)}$ in (\ref{eq:xyxyNinf}) by
the analogous non-exponential factors (and operations)
in (\ref{eq:d2Gdet}):
\begin {align}
   \delta^2 \left[\frac{d\Gamma}{dx\,dy}\right]_{xy\bar x\bar y} &= {}
   \frac{\CA^2 \alphas^2 M_\ix M_\fx^\seq}{8\pi^2 (x_1{+}x_4)^2 E^4}
   \sum_{\substack{{\rm allowed}\\ \lambda_{01},\lambda_{12},\lambda_{23} }}
   \int_{0<t_1<t_2<\Delta t} dt_1 \> dt_2 \> d(\Delta t) \>
    (2\pi)^4 f_{(01)} f_{(12)} f_{(23)}
\nonumber\\ &\quad \times
   \frac{d}{dj_1} \, \frac{d}{dj_2}
   \biggl[
    \det({\cal U}^{-1})
    \Bigl\{
     (\bar\beta \tilde Y_\yx \tilde Y_\xbx
        + \bar\alpha \tilde\Ybar_{\yx\xbx} \tilde Y_{\yx\xbx})
          \tilde I_0
     + (\bar\alpha+\bar\beta+2\bar\gamma) \tilde Z_{\yx\xbx} \tilde I_1
\nonumber\\ &\hspace{6em}
     + \bigl[
         (\bar\alpha+\bar\gamma) \tilde Y_\yx \tilde Y_\xbx
         + (\bar\beta+\bar\gamma)
           \tilde \Ybar_{\yx\xbx} \tilde Y_{\yx\xbx}
        \bigr] \tilde I_2
\nonumber\\ &\hspace{6em}
     - (\bar\alpha+\bar\beta+\bar\gamma)
       (\tilde\Ybar_{\yx\xbx} \tilde Y_\xbx \tilde I_3
        + \tilde Y_\yx \tilde Y_{\yx\xbx} \tilde I_4)
    \Bigl\}^\seq
   \biggr]_{j_1=j_2=0} .
\label{eq:xyxyresult}
\end {align}
We have summed over all color transition sequences in
(\ref{eq:seqtransitions}).

To forestall possible confusion, we should mention that the result
(\ref{eq:xyxyresult}) automatically includes the product
\begin {equation}
  {}_u \langle {\rm A}_{\rm aa} | \lambda_{23} \rangle
       \langle \lambda_{01} | {\rm A}_{\rm aa} \rangle_u
  = \tfrac12
\label {eq:implicit12corr}
\end {equation}
of overlap factors of the initial and final 4-particle color singlet states
(\ref{eq:Aaa}) with $\lambda_{01}{=}\Apm$ and $\lambda_{23}{=}\Apm$ respectively.
That's because the same set of factors, in the form of
\begin {equation}
  {}_u \langle {\rm A}_{\rm aa} | \Am \rangle
       \langle \Am | {\rm A}_{\rm aa} \rangle_u
  = \tfrac12 ,
\label {eq:implicit12}
\end {equation}
were already implicitly
included in the $N{=}\infty$ result (\ref{eq:xyxyNinf}) for the
single color routing $\Am$.%
\footnote{
  The language of color singlet state overlap factors does not appear in
  the original $N{=}\infty$ calculation of ref.\ \cite{seq}.
  But (\ref{eq:implicit12}) is
  equivalent to the $\tfrac12$ in the factor $\tfrac12 \CA^2$ discussed
  immediately after eq.\ (E.1) of ref.\ \cite{seq}.
}


\subsection{Correction to total sequential diagram rate}
\label {sec:seqtotal}

\begin {figure}[t]
\begin {center}
  \includegraphics[scale=0.5]{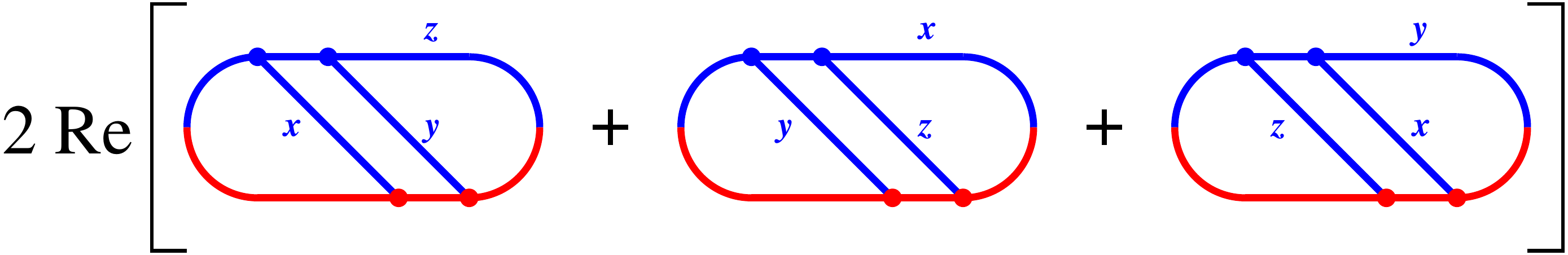}
  \caption{
     \label {fig:seqtotal}
     The sum of diagrams contributing to the total $1/N^2$ correction
     to the total sequential diagram rate.  Note that exchanging
     the daughters
     $y\leftrightarrow z$ in the first diagram does not generate
     a additional diagram if {\it all}\/ possible color transition
     possibilities
     have already been included in the evaluation of each diagram.
  }
\end {center}
\end {figure}

To get the $1/N^2$ correction to the total sequential diagram rate, we
need to (i) take $2\Re[\cdots]$ of (\ref{eq:xyxyresult}) in order
to include the correction to the conjugate diagram $\bar x\bar y x y$,
and (ii) add all permutations of the three final gluons $(x,y,z)$ which
generate distinct diagrams.
See fig.\ \ref{fig:seqtotal}.
Correspondingly, the total correction is
\begin {equation}
  \delta^2 \left[ \Delta \frac{d\Gamma}{dx\,dy} \right]_\seq
  =
  \delta^2 {\mathds A}_\seq(x,y)
  + \delta^2 {\mathds A}_\seq(y,z)
  + \delta^2 {\mathds A}_\seq(z,x)
\label {eq:seqtotal1}
\end {equation}
with
\begin {equation}
  \delta^2 {\mathds A}_\seq(x,y) \equiv
  2\Re\left\{
        \delta^2 \left[\frac{d\Gamma}{dx\,dy}\right]_{xy\bar x\bar y}
      \right\}
\label {eq:frakAdef}
\end {equation}
(The symbol ``$\Delta$'' on the left side of (\ref{eq:seqtotal1}) is
inessential to our present purpose and is included for the sake of
consistency with the $N{=}\infty$ discussion of ref.\ \cite{seq}.%
\footnote{
  See, in particular, section 1.1 of ref.\ \cite{seq}.  Because the
  $1/N^2$ corrections to sequential diagrams come {\it only} from the
  $xy\bar x\bar y$ diagram (and its conjugate and permutations), that
  distinction does not matter here.
}%
)

Alternatively, one may use the discussion about $y{\leftrightarrow}z$
after (\ref{eq:seq1234}) to write
\begin {align}
   \delta^2 \left[ \Delta \frac{d\Gamma}{dx\>dy} \right]_\seq
   = \quad
   & \delta^2{\cal A}_\seq(x,y) + \delta^2{\cal A}_\seq(y,z)
       + \delta^2{\cal A}_\seq(z,x)
\nonumber\\
   + ~ &
   \delta^2{\cal A}_\seq(y,x) + \delta^2{\cal A}_\seq(z,y)
       + \delta^2{\cal A}_\seq(x,z)
\label {eq:seqtotal2}
\end {align}
where $\delta^2{\cal A}_\seq(x,y)$ is also defined by (\ref{eq:frakAdef})
{\it except}\/ that the sum over allowed color sequences
$\lambda_{01}{\to}\lambda_{12}{\to}\lambda_{23}$ in (\ref{eq:xyxyresult})
is taken over only the {\it first three} sequences of (\ref{eq:seq1234}).
The appeal of the version (\ref{eq:seqtotal2}) is just that it has
a similar form to how $N{=}\infty$ results have been previously
presented \cite{seq}.%
\footnote{
  See eq.\ (3.1) of ref.\ \cite{seq}.
}


\section{Color-representation dependent formulas}
\label {sec:reps}

In order to use the preceding formulas, we need for each 4-particle
color singlet state the corresponding normal
mode frequencies and normal mode vectors for 4-particle evolution, with the
vectors written in the $(C_{41},C_{23})$ basis that we have been using
throughout.  That is, we need formulas for the $\Omega_\pm^{(\lambda)}$
and matrix $a_{(\lambda)}$ of eqs.\ (\ref{eq:OmMatrixDef}) and (\ref{eq:adef}).
In this section, we will present this information for all of our
$N{=}\infty$ eigenstates $(1,\Ap,\Am,\Ax,\onexp,\onexm)$, not just
the states that appeared in the $xy\bar x\bar y$ transitions
(\ref{eq:seqtransitions}), because the other states will be useful later
on in the evaluation of $1/N^2$ contributions to crossed diagrams.

We will start from the results for the $\sAp$ and $\sone$ color singlets.
The others may be related to these using permutation symmetries,
for which the alternate notation (\ref{eq:1234}) for $N{=}\infty$
color singlet states will be very useful.


\subsection {Basics}

\subsubsection{$\sAm = (1234)$}

This is the canonical color state considered in the earlier,
$N{=}\infty$ papers such as \cite{2brem,seq}.  A convenient summary
of the relevant formulas for $\Omega_\pm$ and $a$ can be found in
eqs.\ (A.21--22) and (A.27--30) of ref.\ \cite{qcd}, where our
matrix $a$ in the $(C_{41},C_{23})$ basis used here corresponds to the
matrix called $a_\yx$ there.
We note for later reference that these formulas all depend on
the momentum fractions $(x_1,x_2,x_3,x_4)=(-1,y,1{-}x{-}y,x)$ of the
four gluons.  So
\begin {equation}
   \Omega_{(\Am)} = \Omega_{(\Am)}(x_1,x_2,x_3,x_4)
   \qquad
   \mbox{and}
   \qquad
   a_{(\Am)} = a_{(\Am)}(x_1,x_2,x_3,x_4) ,
\end {equation}
where $\Omega_{(\lambda)}$ is the matrix defined in (\ref{eq:OmMatrixDef}).


\subsubsection{$\sone = (41)(23)$}

In this note, the $u$-channel color singlet state $\sone$ refers to
the case where the particle pairs $(41)$ and $(23)$ are each contracted
into a singlet.  This yields simple normal modes in the
$(C_{41},C_{23})$ basis.  The 4-particle potential (\ref{eq:Veffective})
for $N{=}\infty$ acts on the $\sone$ state as
\begin {equation}
  \mtrx{V}(\C_{41},\C_{23})
  =
  - \tfrac{i}{4} (x_4+x_1)^2 \hat q_{\rm A}
    (C_{41}^2 + C_{23}^2) .
\end {equation}
The normal mode frequencies $\Omega_\pm$ and vectors
$(C_{41},C_{23})^\pm$ are
\begin {equation}
   \uOmega_{(\one)}
   \equiv
   \begin{pmatrix} \Omega^{(\one)}_+ & \\ & \Omega^{(\one)}_- \end{pmatrix}
   =
   \sqrt{
     -\frac{i\qhatA}{2 E}
     \begin{pmatrix}
        \tfrac{1}{x_1}{+}\tfrac{1}{x_4} & \\
        & \tfrac{1}{x_2}{+}\tfrac{1}{x_3}
     \end{pmatrix}
   }
\label {eq:41s23freqs}
\end {equation}
and
\begin {equation}
  a_{(\one)}
   \equiv
   \begin{pmatrix}
      C_{41}^+ & C_{41}^- \\ C_{23}^+ & C_{23}^-
   \end{pmatrix}_{\!\!(\one)}
   =
   \frac{1}{E^{1/2}}
   \begin{pmatrix}
      [x_1 x_4(x_1{+}x_4)]^{-1/2} & \\
      & [x_2 x_3(x_2{+}x_3)]^{-1/2}
   \end{pmatrix} .
\label {eq:41s23modes}
\end {equation}
Following refs.\ \cite{2brem,seq}, the normal modes have been
normalized so that
\begin {equation}
   \begin{pmatrix} \C^i_{41} \\ \C^i_{23} \end{pmatrix}^{\!\top}
   \!{\mathfrak M}'
   \begin{pmatrix} \C^{j}_{41} \\ \C^{j}_{23} \end{pmatrix}
   = \delta^{ij} ,
\label {eq:NMnorm}
\end {equation}
where
\begin {equation}
   {\mathfrak M}'
   =
   \begin{pmatrix}
      x_4 x_1 (x_4{+}x_1) & \\ & x_2 x_3 (x_2{+}x_3)
   \end {pmatrix} E
\label {eq:frakM}
\end {equation}
is the mass matrix whose inverse appears in the kinetic term of the
Hamiltonian (\ref{eq:Heffective}) for the basis $(C_{41},C_{23})$ that
we use here.%
\footnote{
  See the discussion of eqs.\ (5.16--18) of ref.\ \cite{2brem}.
  Here we work in the basis
  $(C_{41},C_{23})$ instead of $(C_{34},C_{12})$, and so the indices
  $1234$ there are relabeled $2341$ here.
}


\subsection {Permutations}
\label{sec:perms}

\subsubsection{$\sonexm = (12)(34)$}

By permuting indices $1\leftrightarrow3$ in the result
(\ref{eq:41s23freqs}) for the $(41)(23)$ state, we obtain
the eigenfrequencies for the $(43)(21) = (12)(34) = \sonexm$
color singlet state:
\begin {equation}
   \uOmega_{(\onexm)} =
   \sqrt{
     -\frac{i\qhatA}{2 E}
     \begin{pmatrix}
        \tfrac{1}{x_3}{+}\tfrac{1}{x_4} & \\
        & \tfrac{1}{x_2}{+}\tfrac{1}{x_1}
     \end{pmatrix}
   }.
\end {equation}
The corresponding modes (\ref{eq:41s23modes}) for $(41)(23)$ were
expressed in the $(C_{41},C_{23})$ basis.  So, by making the
same permutation $1\leftrightarrow3$ to (\ref{eq:41s23modes}), we obtain
normal modes for $\sonexm$
in the $(C_{43},C_{21})$ basis:
\begin {equation}
   \begin{pmatrix}
      C_{43}^+ & C_{43}^- \\ C_{21}^+ & C_{21}^-
   \end{pmatrix}_{\!\!(\sonexm)}
    =
   \frac{1}{E^{1/2}}
   \begin{pmatrix}
      [x_4 x_3(x_4{+}x_3)]^{-1/2} & \\
      & [x_2 x_1(x_2{+}x_1)]^{-1/2}
   \end{pmatrix} .
\label {eq:first}
\end {equation}
Since $C_{ij} = -C_{ji}$, we can convert to the $(C_{34},C_{12})$ basis
(which we'll see is useful in just a moment) by negating
(\ref{eq:first}) to get
\begin {equation}
   \begin{pmatrix}
      C_{34}^+ & C_{34}^- \\ C_{12}^+ & C_{12}^-
   \end{pmatrix}_{\!\!(\sonexm)}
   =
   -
   \frac{1}{E^{1/2}}
   \begin{pmatrix}
      [x_4 x_3(x_4{+}x_3)]^{-1/2} & \\
      & [x_2 x_1(x_2{+}x_1)]^{-1/2} 
   \end{pmatrix} .
\label {eq:12s34modes0}
\end {equation}
To convert to the $(C_{41},C_{23})$ basis used throughout this
paper, now use the relation \cite{2brem}%
\footnote{
  This relation comes from eq.\ (5.31) on ref.\ \cite{2brem}.
}
\begin {equation}
   \begin{pmatrix} C_{41} \\ C_{23} \end{pmatrix}
   =
   \frac{1}{(x_1{+}x_4)}
   \begin{pmatrix}
       -x_3 & -x_2 \\
        \phantom{-}x_4 &  \phantom{-}x_1
   \end {pmatrix}
   \begin{pmatrix} C_{34} \\ C_{12} \end{pmatrix} .
\label {eq:transform}
\end {equation}
to get
\begin {equation}
   a_{(\onexm)} =
   -
   \frac{1}{(x_1{+}x_4) E^{1/2}}
   \begin{pmatrix}
        -x_3 & -x_2 \\
        \phantom{-}x_4 & \phantom{-}x_1
   \end {pmatrix}
   \begin{pmatrix}
      [x_3 x_4(x_3{+}x_4)]^{-1/2} & \\
      & [x_1 x_2(x_1{+}x_2)]^{-1/2}
   \end{pmatrix} .
\label {eq:12s34modes}
\end {equation}


\subsubsection{$\sonexp = (13)(24)$}

Similarly, permuting indices $3\leftrightarrow4$ in
(\ref{eq:41s23freqs}) for the $(41)(23)$ state, we obtain
\begin {equation}
   \uOmega_{(\onexp)} =
   \sqrt{
     -\frac{i\qhatA}{2 E}
     \begin{pmatrix}
        \tfrac{1}{x_1}{+}\tfrac{1}{x_3} & \\
        & \tfrac{1}{x_2}{+}\tfrac{1}{x_4}
     \end{pmatrix}
   }
\end {equation}
and
\begin {equation}
   \begin{pmatrix}
      C_{31}^+ & C_{31}^- \\ C_{24}^+ & C_{24}^-
   \end{pmatrix}_{\!\!(\sonexp)}
    =
   \frac{1}{E^{1/2}}
   \begin{pmatrix}
      [x_1 x_3(x_1{+}x_3)]^{-1/2} & \\
      & [x_2 x_4(x_2{+}x_4)]^{-1/2}
   \end{pmatrix} .
\label {eq:firstbasis}
\end {equation}
Now permute the conversion (\ref{eq:transform}) by $1{\leftrightarrow}4$
and then use $C_{ij} = -C_{ji}$ to get
\begin {equation}
   \begin{pmatrix} \C_{41} \\ \C_{23} \end{pmatrix}
   =
   \frac{1}{(x_1{+}x_4)}
   \begin{pmatrix}
        x_3 & -x_2 \\
        x_1 & -x_4
   \end {pmatrix}
   \begin{pmatrix} \C_{31} \\ \C_{24} \end{pmatrix} .
\label {eq:transform2}
\end {equation}
Applying this transformation to (\ref{eq:firstbasis}) then gives
the normal modes in the desired basis:
\begin {equation}
   a_{(\onexp)} =
   \frac{1}{(x_1{+}x_4) E^{1/2}}
   \begin{pmatrix}
        x_3 & -x_2 \\
        x_1 & -x_4
   \end {pmatrix}
   \begin{pmatrix}
      [x_1 x_3(x_1{+}x_3)]^{-1/2} & \\
      & [x_2 x_4(x_2{+}x_4)]^{-1/2}
   \end{pmatrix} .
\end {equation}


\subsubsection{$\sAp = (1324)$}

We can get this from the formulas for
$\sAm = (1234)$ by similar permutation arguments.
Swapping $2\leftrightarrow 3$,
\begin {equation}
   \uOmega_{(\Ap)}(x_1,x_2,x_3,x_4)
   = \uOmega_{(\Am)}(x_1,x_3,x_2,x_4)
\end {equation}
and
\begin {equation}
   \begin{pmatrix}
      C_{41}^+ & C_{41}^- \\ C_{32}^+ & C_{32}^-
   \end{pmatrix}_{\!\!(\Ap)}
   =
     a_{(\Am)}(x_1,x_3,x_2,x_4) .
\end {equation}
Since $C_{32} = -C_{23}$, we may rewrite that as
\begin {equation}
   a_{(\Ap)}(x_1,x_2,x_3,x_4)
   =
   \begin{pmatrix} 1 & \\ & -1 \end{pmatrix}
   a_{(\Am)}(x_1,x_3,x_2,x_4) .
\end {equation}


\subsubsection{$\sAx = (1243)$}

Similarly, swapping $3\leftrightarrow 4$ in formulas for
$\sAm = (1234)$ gives
\begin {equation}
   \uOmega_{(\Ax)}(x_1,x_2,x_3,x_4)
   = \uOmega_{(\Am)}(x_1,x_2,x_4,x_3)
\end {equation}
and
\begin {equation}
   \begin{pmatrix}
      C_{31}^+ & C_{31}^- \\ C_{24}^+ & C_{24}^-
   \end{pmatrix}_{\!\!(\Ax)}
   =
     a_{(\Am)}(x_1,x_2,x_4,x_3) .
\end {equation}
Then use (\ref{eq:transform2}) to get
\begin {equation}
   a_{(\Ax)}(x_1,x_2,x_3,x_4)
   =
   \frac{1}{(x_1{+}x_4)}
   \begin{pmatrix}
        x_3 & -x_2 \\
        x_1 & -x_4
   \end {pmatrix}
   a_{(\Am)}(x_1,x_2,x_4,x_3) .
\end {equation}


\section {Crossed diagrams}
\label {sec:cross}

\begin {figure}[t]
\begin {center}
  \includegraphics[scale=0.6]{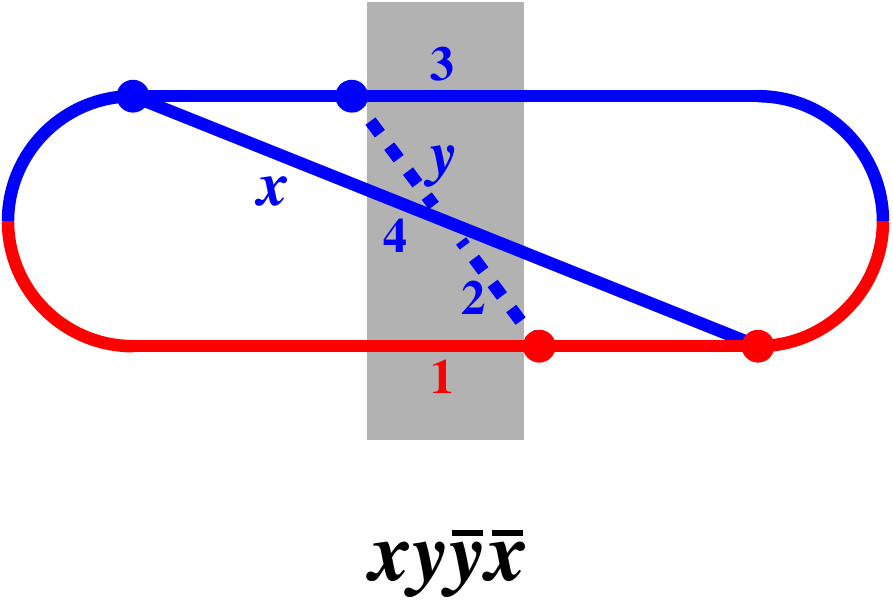}
  \caption{
     \label {fig:xyyx}
     The canonical ``crossed'' diagram.
     Particles in the (shaded) region of
     4-particle evolution are numbered according
     to the convention of ref.\ \cite{2brem}.
  }
\end {center}
\end {figure}

We now turn to crossed diagrams for $g \to ggg$.
The canonical crossed diagram, to which all others can be related \cite{2brem},
is the $xy\bar y\bar x$ diagram shown in fig.\ \ref{fig:xyyx}.

\subsection{Allowed Color Transitions}

At the start of the shaded region of 4-particle evolution, the particles
combine in the same way as for the sequential diagram of
fig.\ \ref{fig:xyxy}, and so the initial 4-particle color singlet state
is the same as before:
\begin {equation}
  |{\rm A}_{\rm aa}\rangle_u
  =
  \tfrac{1}{\sqrt2} \sAp + \tfrac{1}{\sqrt2} \sAm
  \qquad \mbox{(initial 4-particle state)}.
\label {eq:isinglet}
\end {equation}
However, the end of the shaded region is different: It is now gluons
1 and 2 that meet at a vertex.  So, the final state is the $s$-channel
version $|{\rm A}_{\rm aa}\rangle_s$
rather than $u$-channel version (\ref{eq:isinglet}).
In this paper, we find it convenient to always stick to the definition
(\ref{eq:basis}) of our basis states $(1,\Ap,\Am,\Ax,\onexp,\onexm)$, which
are defined in terms of $u$-channel singlet combinations.  We need
to figure out how to express our final ($s$-channel) color singlet state
$|{\rm A}_{\rm aa}\rangle_s$ in terms of this basis.
The matrix that converts (for any $N$) between the $s$-channel and
$u$-channel versions of the original basis states (\ref{eq:6dim}) is
given by \cite{color,NSZ6j,Sjodahl} (see appendix \ref{app:signs})
\begin {subequations}
\label {eq:suconvert}
\begin {equation}
  \begin{pmatrix}
    |{\bm 1}\rangle_s \\
    |{\rm A}_{\rm aa}\rangle_s \\
    |{\rm A}_{\rm ss}\rangle_s \\
    |\quote{\bm{10}{+}\overline{\bm{10}}}\rangle_s \\
    |\quote{\bm{27}}\rangle_s \\
    |\quote{\bm 0}\rangle_s
  \end{pmatrix}
  =
  \mathds{U}
  \begin{pmatrix}
    |{\bm 1}\rangle_u \\
    |{\rm A}_{\rm aa}\rangle_u \\
    |{\rm A}_{\rm ss}\rangle_u \\
    |\quote{\bm{10}{+}\overline{\bm{10}}}\rangle_u \\
    |\quote{\bm{27}}\rangle_u \\
    |\quote{\bm 0}\rangle_u
  \end{pmatrix}
\label {eq:su}
\end {equation}
with
\begin {equation}
  \mathds{U} =
  \begin{pmatrix}
    \tfrac{1}{N^2-1}
    & -\sqrt{\tfrac{1}{N^2-1}}
    & \sqrt{\tfrac{1}{N^2-1}}
    & -\sqrt{\tfrac{N^2-4}{2(N^2-1)}}
    & \tfrac{N}{2(N+1)} \sqrt{\tfrac{N+3}{N-1}}
    & \tfrac{N}{2(N-1)} \sqrt{\tfrac{N-3}{N+1}}
  \\[5pt]
  
    & \tfrac12
    & -\tfrac12
    & 0
    &  \tfrac12 \sqrt{\tfrac{N+3}{N+1}}
    & -\tfrac12 \sqrt{\tfrac{N-3}{N-1}}
  \\
  
    &
    & \tfrac{N^2-12}{2(N^2-4)}
    & \sqrt{\tfrac{2}{N^2-4}}
    & \tfrac{N}{2(N+2)} \sqrt{\tfrac{N+3}{N+1}}
    & -\tfrac{N}{2(N-2)} \sqrt{\tfrac{N-3}{N-1}}
  \\[5pt]

    &
    &
    & \tfrac12
    & \sqrt{\tfrac{(N-2)(N+3)}{8(N+1)(N+2)}}
    & \sqrt{\tfrac{(N+2)(N-3)}{8(N-1)(N-2)}}
  \\[5pt]
    \multicolumn{3}{c}{\rm (symmetric) }
    &
    & \tfrac{N^2+N+2}{4(N+1)(N+2)}
    & \tfrac{1}{4} \sqrt{\tfrac{N^2-9}{N^2-1}}
  \\[5pt]
  
    &
    &
    &
    &
    & \tfrac{N^2-N+2}{4(N-1)(N-2)}
  \\
  \end{pmatrix} .
\label {eq:Uconvert}
\end {equation}
\end {subequations}
For our present purpose, the only
piece of (\ref{eq:suconvert}) that we need is
\begin {equation}
  |{\rm A}_{\rm aa}\rangle_s =
  - \sqrt{\tfrac{1}{N^2-1}} \, |{\bm 1}\rangle_u
  + \tfrac12 \, |{\rm A}_{\rm aa}\rangle_u
  - \tfrac12 \, |{\rm A}_{\rm ss}\rangle_u
  + \tfrac12 \sqrt{\tfrac{N+3}{N+1}} \, |\quote{\bm{27}}\rangle_u
  - \tfrac12 \sqrt{\tfrac{N-3}{N-1}} \, |\quote{\bm 0}\rangle_u .
\end {equation}
Using (\ref{eq:basis}) to convert to the basis states
$(\one,\Ap,\Am,\Ax,\onexp,\onexm)$ that we
use for our analysis in this paper, and then expanding in $1/N$,
\begin {multline}
  |{\rm A}_{\rm aa}\rangle_s
  =
  \frac{ \sAm + \sAx }{\sqrt{2}}
  + \frac{\sonexp + \sonexm -2\sone}{2N}
  - \frac{3 \sAx}{2\sqrt2 \, N^2}
  + O(N^{-3})
\\
  \mbox{(final 4-particle state)}.
\label{eq:fsinglet}
\end {multline}

For future reference, note
that the overall sign of $|{\rm A}_{\rm aa}\rangle_s$ is
merely a phase convention choice for that state.
Different choices of this sign convention must lead to compensating
changes of sign in the rule for the diagrammatic vertex at the
end of the 4-particle evolution in fig.\ \ref{fig:xyyx}.
We will later discuss how to get the overall sign of our answer
right without having to drill down into such details.%
\footnote{
  We did not have to think about the phase convention in our
  discussion of sequential diagrams because the initial and final
  color singlet states were both the same: $|{\rm A}_{\rm aa}\rangle_u$.
  So changing sign convention
  $|{\rm A}_{\rm aa}\rangle_u \to -|{\rm A}_{\rm aa}\rangle_u$ would
  have no effect since the sign would appear twice in the calculation
  of the 4-particle evolution---once at the start and once at the end.
}

We may now using the initial and final singlet states (\ref{eq:isinglet})
and (\ref{eq:fsinglet}), together with the textures of the perturbations
$\delta\mtrx{S}$, $\delta\mtrx{T}$ and $\delta^2\mtrx{T}$ of
(\ref{eq:dST}), to list all possible 4-particle
color transition sequences that contribute
to $1/N^2$ corrections to the $xy\bar y\bar x$ diagram of
fig.\ \ref{fig:xyyx}.  They are listed in table \ref{tab:Xtransitions}.

\begin {table}[tp]
\begin {center}
\setlength{\tabcolsep}{7pt}
\renewcommand{\arraystretch}{1.4}
\begin{tabular}{ccccc}
\hline\hline
  transition & equivalent &
   \vbox{\hbox{$\delta S$, $\delta T$, $\delta^2 T$}\hbox{\kern1.1em factors}} &
   \vbox{\hbox{\kern0.4em color}\hbox{\phantom{.}}\hbox{overlap}} &
   $\phi$
\\ \hline
   $\sAm \xrightarrow{\delta T}{} \ssone \xrightarrow{\delta T} \sAm$
   & $(1234) \to (41)(23) \to (1234)$
   & $\tfrac{1}{2N^2}$
   & $\tfrac12$
   & $\frac{1}{2N^2}$
\\
   $\sAm \xrightarrow{\delta T}{} \sonexm \xrightarrow{\delta T} \sAm$
   & $(1234) \to (12)(34) \to (1234)$
   & $\tfrac{1}{2N^2}$
   & $\tfrac12$
   & $\frac{1}{2N^2}$
\\
   $\sAp \xrightarrow{\delta T}{} \ssone
               \xrightarrow{\delta T} \sAm$
   & $(1324) \to (41)(23) \to (1234)$
   & $\tfrac{1}{2N^2}$
   & $\tfrac12$
   & $\frac{1}{2N^2}$
\\ \hline
   $\sAm \xrightarrow{\delta T}{} \sonexm \xrightarrow{\delta S} \sAx$
   & $(1234) \to (12)(34) \to (1243)$
   & $-\tfrac{1}{2N^2}$
   & $\tfrac12$
   & $-\frac{1}{2N^2}$
\\
   $\sAp \xrightarrow{\delta T}{} \sonexp \xrightarrow{\delta S} \sAx$
   & $(1324) \to (13)(24) \to (1243)$
   & $-\tfrac{1}{2N^2}$
   & $\tfrac12$
   & $-\frac{1}{2N^2}$
\\ \hline
   $\sAm \xrightarrow{\delta^2 T}{} \sAx $
   & $(1234) \to (1243)$
   & $\tfrac{1}{2N^2}$
   & $\tfrac12$
   & $\frac{1}{2N^2}$
\\
   $\sAp \xrightarrow{\delta^2 T}{} \sAx $
   & $(1324) \to (1243)$
   & $\tfrac{1}{2N^2}$
   & $\tfrac12$
   & $\frac{1}{2N^2}$
\\
   $\sAm \xrightarrow{\delta T}{} \ssone $
   & $(1234) \to (41)(23)$
   & $\tfrac{1}{\sqrt2\,N}$
   & $-\tfrac1{\sqrt2\,N}$
   & $-\frac{1}{N^2}$
\\
   $\sAm \xrightarrow{\delta T}{} \sonexm $
   & $(1234) \to (12)(34)$
   & $\tfrac{1}{\sqrt2\,N}$
   & $\tfrac1{2\sqrt2\,N}$
   & $\frac{1}{2N^2}$
\\
   $\sAp \xrightarrow{\delta T}{} \ssone $
   & $(1324) \to (41)(23)$
   & $\tfrac{1}{\sqrt2\,N}$
   & $-\tfrac1{\sqrt2\,N}$
   & $-\frac{1}{N^2}$
\\
   $\sAp \xrightarrow{\delta T}{} \sonexp$
   & $(1324) \to (13)(24)$
   & $\tfrac{1}{\sqrt2\,N}$
   & $\tfrac1{2\sqrt2\,N}$
   & $\frac{1}{2N^2}$
\\[3pt]
\hline\hline
\end{tabular}
\end {center}
\caption{
   \label{tab:Xtransitions}
   Allowed 4-particle
   color transitions at order $1/N^2$ for the $xy\bar y\bar x$
   diagram, along with (i) the associated $\delta T$, $\delta S$ or
   $\delta^2 T$ factors, and (ii) the product
   of the initial and final
   color overlap factors $\langle\lambda_\ix | {\rm A}_{\rm aa}\rangle_u$
   and ${}_s\langle {\rm A}_{\rm aa} | \lambda_\fx\rangle$.
   Also shown is the product $\phi$ of (i) and (ii)
   {\it relative} to what it would be
   $
    \bigl[\,
      {}_s\langle {\rm A}_{\rm aa} | \Am \rangle
      \langle \Am | {\rm A}_{\rm aa} \rangle_u
      \,{=}\,\tfrac12
    \,\bigr]
   $
   in the $N{=}\infty$ calculation of the crossed diagram.
   The horizontal lines separate groups of processes that have to be
   handled differently:
   2nd order in $\delta V$ with two $\delta T$ transitions; 
   2nd order in $\delta V$ with a $\delta T$ and $\delta S$ transition; 
   1st order in $\delta V$ with a $T$-based perturbation.
   There are no non-zero $O(1/N^2)$ contributions at 0th order
   in $\delta\mtrx{V}$.  [Specifically, the $1/N^2$ term in
   (\ref{eq:fsinglet}) for the final state $|A_{\rm aa}\rangle_s$
   does not directly overlap the initial state
   $|A_{\rm aa}\rangle_u$ of (\ref{eq:Aaa}).]
}
\end{table}


\subsection {2nd order in $\delta V$}

\begin {figure}[t]
\begin {center}
  \begin{picture}(277,140)(0,0)
    \put(0,15){\includegraphics[scale=0.6]{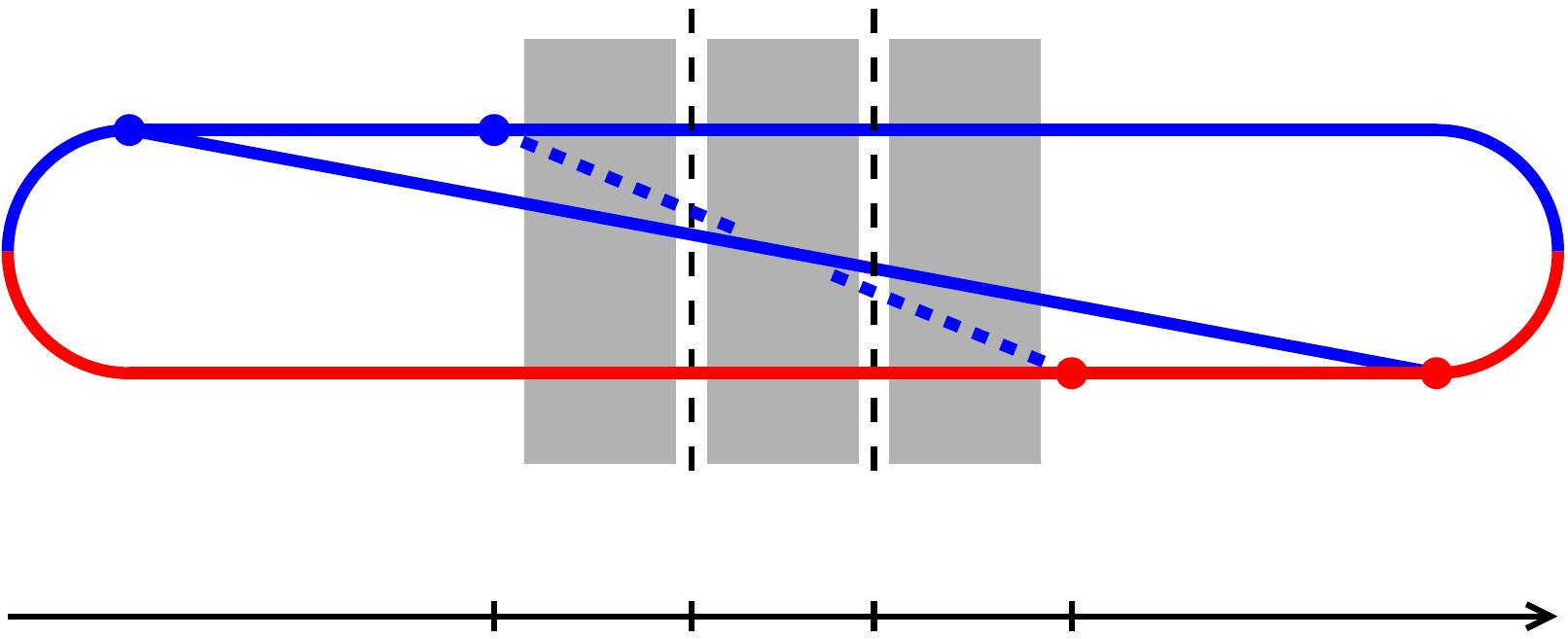}}
    \put(84,5){$t_0$}
    \put(120,5){$t_1$}
    \put(152,5){$t_2$}
    \put(187,5){$t_3$}
    \put(100,33){$\lambda_{01}$}
    \put(133,33){$\lambda_{12}$}
    \put(165,33){$\lambda_{23}$}
    \put(118,130){$\delta V$}
    \put(150,130){$\delta V$}
  \end{picture}
  \caption{
     \label {fig:dGcross}
     The analog of fig.\ \ref{fig:dGseq},
     now for the $xy\bar y\bar x$ crossed diagram.
  }
\end {center}
\end {figure}
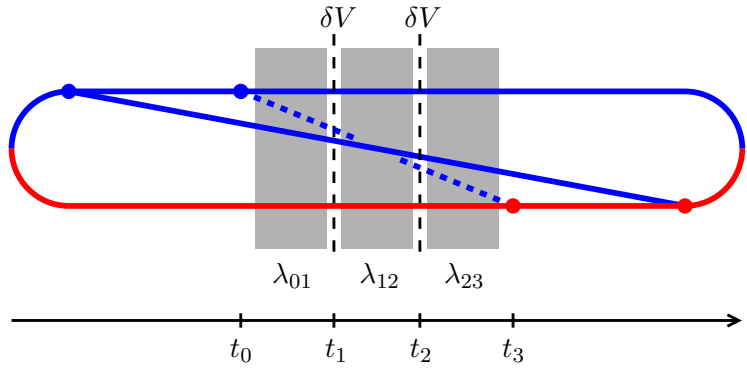

We start by examining the first five lines of table \ref{tab:Xtransitions},
which are the cases that involve {\it two} insertions of perturbations
$\delta\mtrx{V}$ in the 4-particle evolution.
Schematically, these cases correspond to
fig.\ \ref{fig:dGcross}, which is the crossed diagram analog of
fig.\ \ref{fig:dGseq}.  The formulas for these contributions
to the crossed diagram $xy\bar y\bar x$
are basically the same as the formulas we found in section \ref{sec:seq}
for the sequential diagram $xy\bar x\bar y$ except for
some minor modifications.  One modification is simply that we
should use the color transitions given by the first five lines of
table \ref{tab:Xtransitions} instead of the sequential diagram transitions
of (\ref{eq:seqtransitions}).  But there are other changes needed
as well.

\subsubsection{Modification: $(X,Y,Z)$}

The known $N{=}\infty$ rate for the $xy\bar y\bar x$ diagram has a form
similar to that quoted earlier for the $\Am{=}(1234)$ color routing of the
$xy\bar x\bar y$ diagram in
(\ref{eq:xyxyNinf}).  The $xy\bar y\bar x$ case is \cite{2brem}%
\footnote{
   Unlike $N{=}\infty$ sequential diagrams, $N{=}\infty$
   crossed diagrams have only a single
   color routing.
}
\begin {align}
  \left[\frac{d\Gamma}{dx\,dy}\right]_{xy\bar y\bar x} =
   -
   \int_0^\infty & d(\Delta t) \>
   \frac{\CA^2 \alphas^2 M_\ix M_\fx}{8\pi^2(x_1{+}x_4)^2E^4} \,
   f_{\Am}
\nonumber\\ &\times
   \Bigl\{
     (\beta Y_\yx Y_\ybx
        + \alpha \Ybar_{\yx\ybx} Y_{\yx\ybx}) I_0
     + (\alpha+\beta+2\gamma) Z_{\yx\ybx} I_1
\nonumber\\ &\quad
     + \bigl[
         (\alpha+\gamma) Y_\yx Y_\ybx
         + (\beta+\gamma) \Ybar_{\yx\ybx} Y_{\yx\ybx}
        \bigr] I_2
\nonumber\\ &\quad
     - (\alpha+\beta+\gamma)
       (\Ybar_{\yx\ybx} Y_\ybx I_3
        + Y_\yx Y_{\yx\ybx} I_4)
   \Bigl\} .
\label {eq:xyyxNinf}
\end {align}
The $(\alpha,\beta,\gamma)$ are different combinations of helicity-dependent
DGLAP splitting functions than those in the sequential case, and their
formulas may be found in ref.\ \cite{2brem}.
The $I_n$ here have the same form as the $I_n^\seq$ of (\ref{eq:I})
except that the superscript ``$\seq$'' should be removed from everything.
However, the $(X,Y,Z)$'s are somewhat different from the
$(X^\seq,Y^\seq,Z^\seq)$'s.
In the original $N{=}\infty$ calculation \cite{2brem}, they were defined
so that the exponential factor in the 4-particle propagator was%
\footnote{
  See eq.\ (5.41) of ref.\ \cite{2brem}, with the caveat that, similar to
  our previous discussion of the sequential case, our $\calX_\yx$
  and $\calX_\ybx$ here do not contain the effects of the initial and final
  3-particle evolution and are related to the $X_\yx$ and $X_\ybx$ of
  ref.\ \cite{2brem} by our eq.\ (\ref{eq:XvscalXcross}).
}
\begin {multline}
   \exp\Biggl[
     - \frac12
     \begin{pmatrix} \C_{41}^\yx \\ \C_{23}^\yx \end{pmatrix}^{\!\!\top} \!
       \begin{pmatrix}
          \calX_\yx & Y_\yx \\
          Y_\yx & Z_\yx 
       \end{pmatrix}
       \begin{pmatrix} \C_{41}^\yx \\ \C_{23}^\yx \end{pmatrix}
     -
     \frac12
     \begin{pmatrix} \C_{34}^\ybx \\ \C_{12}^\ybx \end{pmatrix}^{\!\!\top} \!
       \begin{pmatrix}
          \calX_\ybx & Y_\ybx \\
          Y_\ybx & Z_\ybx
       \end{pmatrix}
       \begin{pmatrix} \C_{34}^\ybx \\ \C_{12}^\ybx \end{pmatrix}
\\
     +
     \begin{pmatrix} \C_{41}^\yx \\ \C_{23}^\yx \end{pmatrix}^{\!\!\top} \!
       \begin{pmatrix}
          X_{\yx\ybx} & Y_{\yx\ybx} \\
          \Ybar_{\yx\ybx} & Z_{\yx\ybx}
       \end{pmatrix}
       \begin{pmatrix} \C_{34}^\ybx \\ \C_{12}^\ybx \end{pmatrix}
   \Biggr] ,
\label {eq:XYZexpCross}
\end {multline}
where
\begin {subequations}
\label{eq:XvscalXcross}
\begin {align}
   X_\yx &= |M_\ix|\Omega_\ix + \calX_\yx ,
\\
   X_\ybx &= |M_\fx| \Omega_\fx + \calX_\ybx
\end {align}
\end {subequations}
similar to (\ref{eq:XvscalXseq}).  Explicit formulas for
$M_\ix$, $\Omega_\ix$, $M_\fx$, $\Omega_\fx$ may be found in ref.\ \cite{2brem}.

The pattern common to the presentations
(\ref{eq:XYZexp}) and (\ref{eq:XYZexpCross})
of the sequential and crossed exponentials
is
that in each vector, the {\it bottom} $\C_{ij}^v$ is the one
for which lines $i$ and $j$ come together at the corresponding vertex $v$
of the diagram.  It was the use of this convention that made the
$N{=}\infty$ rate
formulas (\ref{eq:xyxyNinf}) and (\ref{eq:xyyxNinf}) for sequential
and crossed diagrams have similar structure.

Similar to what happened for the sequential diagram,
(\ref{eq:XYZexpCross}) for this crossed diagram just gives particular
names to the entries of the matrices that in this paper we would call
$A_{\Am}$ and $B_{\Am}$ --- the identically {\it same} matrices that
were relevant to the case of sequential diagrams in (\ref{eq:ABXYZ}).
Here the relations are
\begin {subequations}
\label {eq:ABXYZcross}
\begin {equation}
   A_{\Am} =
       \begin{pmatrix}
          \calX_\yx & Y_\yx \\
          Y_\yx & Z_\yx
       \end{pmatrix}
   =
       {\mathfrak S}^{-1\top}
       \begin{pmatrix}
          \calX_\ybx & Y_\ybx \\
          Y_\ybx & Z_\ybx
       \end{pmatrix}
       {\mathfrak S}^{-1} ,
   \qquad
   B_{\Am} =
       \begin{pmatrix}
          X_{\yx\ybx} & Y_{\yx\ybx} \\
          \Ybar_{\yx\ybx} & \tilde Z_{\yx\ybx}
       \end{pmatrix}
       {\mathfrak S}^{-1} ,
\end {equation}
\end {subequations}
where ${\mathfrak S}$ is the matrix from (\ref{eq:transform}) that
converts the $(C_{34},C_{12})$ basis into the $(C_{41},C_{23})$ basis:
\begin {subequations}
\label {eq:Sfrak}
\begin {equation}
   \begin{pmatrix} \C_{41} \\ \C_{23} \end{pmatrix}
   =
   {\mathfrak S}
   \begin{pmatrix} \C_{34} \\ \C_{12} \end{pmatrix}
\end {equation}
with
\begin {equation}
   {\mathfrak S} \equiv
   \frac{1}{(x_1{+}x_4)}
   \begin{pmatrix}
       -x_3 & -x_2 \\
        \phantom{-}x_4 &  \phantom{-}x_1
   \end {pmatrix} .
\end {equation}
\end {subequations}

Like we did for the sequential $xy\bar x\bar y$ diagram, we now want
to put the exponential factor
\begin {equation}
    e^{
       -\frac12 \vec{\bm z}^\top
       ( {\cal A} - {\cal B}\,{\cal U}^{-1} {\cal B} ) 
       \vec{\bm z}
    }
\label {eq:expzz2}
\end {equation}
for the $1/N^2$ correction into the same form as the exponential factor
(\ref{eq:XYZexpCross}) for the known $N{=}\infty$ result.
Since (\ref{eq:expzz2}) is exactly the same as before, the only difference
is in the identification of the $(X,Y,Z)$'s.  By comparing
(\ref{eq:ABXYZcross}) with the sequential version (\ref{eq:ABXYZ}),
we can read off the relation of the
$(X,Y,Z)$'s of the crossed diagram
with the previously identified $(X,Y,Z)^\seq$ of the sequential diagram:%
\footnote{
  If one removes all of the tildes, then the relations (\ref{eq:XYZcross})
  {\it also} relate the $N{=}\infty$ crossed and sequential
  formulas for $(X,Y,Z)$, which can be
  verified from the formulas for $(X,Y,Z)$ in refs.\ \cite{2brem,seq},
  once one uses (\ref{eq:XvscalXseq}) and (\ref{eq:XvscalXcross})
  to isolate what we call the $\calX$'s from the $X$'s.
}
\begin {subequations}
\label {eq:XYZcross}
\begin {equation}
  \begin{pmatrix}
    \tilde\calX_\yx & \tilde Y_\yx \\[2pt]
    \tilde Y_\yx & \tilde Z_\yx
  \end{pmatrix}
  =
  \begin{pmatrix}
    \tilde\calX_\yx^\seq & \tilde Y_\yx^\seq \\[2pt]
    \tilde Y_\yx^\seq & \tilde Z_\yx^\seq 
  \end{pmatrix}
  ,
  \qquad
  \begin{pmatrix}
    \tilde\calX_\ybx & \tilde Y_\ybx \\[2pt]
    \tilde Y_\ybx & \tilde Z_\ybx
  \end{pmatrix}
  =
  {\mathfrak S}^\top {\cal S}
  \begin{pmatrix}
    \tilde \calX_\xbx^\seq & \tilde Y_\xbx^\seq \\[2pt]
    \tilde Y_\xbx^\seq & \tilde Z_\xbx^\seq
  \end{pmatrix}
  {\cal S} {\mathfrak S}
  ,
\end {equation}
\begin {equation}
  \begin{pmatrix}
    \tilde X_{\yx\ybx} & \tilde Y_{\yx\ybx} \\[2pt]
    \tilde\Ybar_{\yx\ybx} & \tilde Z_{\yx\ybx}
  \end{pmatrix}
  =
  \begin{pmatrix}
    \tilde X_{\yx\xbx}^\seq & \tilde Y_{\yx\xbx}^\seq \\[2pt]
    \tilde\Ybar_{\yx\xbx}^\seq & \tilde Z_{\yx\xbx}^\seq
  \end{pmatrix}
  {\cal S} {\mathfrak S}
  ,
\end {equation}
\end {subequations}
where ${\cal S}$ is again defined by (\ref{eq:calS}).
So, to compute
$(\tilde X,\tilde Y,\tilde Z)$'s for the crossed diagrams, first compute
${\cal A} - {\cal B}\,{\cal U}^{-1} {\cal B}$ as in section \ref{sec:seq},
then read out the
$(\tilde X,\tilde Y,\tilde Z)^\seq$ values using (\ref{eq:XYZreadout}),
and finally convert those values using (\ref{eq:XYZcross}) above.


\subsubsection{Modification: the matrix $R_2$}

The identification of the $2{\times}2$ matrices
$R_1 = R_2 = R^{(\delta T)}$ back in (\ref{eq:R}) was
based on the fact that only $\delta T$ transitions (\ref{eq:seqtransitions})
were relevant
for the sequential diagram $xy\bar x\bar y$.
The same is true for the first three rows of table \ref{tab:Xtransitions},
which shows the allowed transition sequences for the crossed
diagram $xy\bar y\bar x$.  We will refer to those first three rows
as ``$\delta T\,\delta T$'' transition sequences.

In the next two rows of the table, however,
the second transition of each sequence is instead a $\delta S$ transition.
We will refer to these rows as ``$\delta T\,\delta S$'' transition
sequences.  $\mtrx{S}_u$ appears differently than $\mtrx{T}_u$ in the
potential (\ref{eq:Veffective}), and so its contribution to
$\delta\mtrx{V}$ matrix elements will be different than those of the
$\delta T$ contribution (\ref{eq:dVdT}).
The non-zero matrix elements associated with $\delta\mtrx{S}$ are
\begin {equation}
  \delta V^{(\delta S)}\!(\vbxi) =
  \frac{i\hat q_{\rm A}}{2\sqrt2\,N}
  \Bigl[
     x_4 x_1 C_{41}^2 + x_2 x_3 C_{23}^2 
     + \tfrac12 (x_4-x_1)(x_2-x_3) \C_{41}\cdot\C_{23}
  \Bigr] ,
\end {equation}
which can be written in the form of
$\tfrac12 \vbxi^{\,\top} R^{(\delta S)} \vbxi$ with
\begin {equation}
  R^{(\delta S)} = 
  + \frac{i \qhatA}{2\sqrt2\, N}
  \begin{pmatrix}
    2x_1 x_4 & \tfrac12(x_4{-}x_1)(x_2{-}x_3) \\
    \tfrac12(x_4{-}x_1)(x_2{-}x_3) & 2 x_2 x_3
  \end{pmatrix} .
\label {eq:RforS}
\end {equation}
So, the final rule is that we need to use
\begin {equation}
   (R_1,R_2) =
   \begin {cases}
      (R^{(\delta T)}, R^{(\delta T)})
         & \mbox{for $\delta T\,\delta T$ transitions};
      \\
      (R^{(\delta T)}, R^{(\delta S)})
         & \mbox{for $\delta T\,\delta S$ transitions}
   \end {cases}
\end {equation}
in the construction (\ref{eq:calUdef}) of the $4{\times}4$ matrix ${\cal U}$.


\subsubsection{Final result for 2nd order in $\delta V$}

With the preceding modifications, the final result for the first five
rows of table \ref{tab:Xtransitions} has the same relation to
(\ref{eq:xyyxNinf}) that the sequential result (\ref{eq:xyxyresult})
did to (\ref{eq:xyxyNinf}):
\begin {align}
   \delta^2 \left[\frac{d\Gamma}{dx\,dy}\right]_{xy\bar y\bar x}^{(\delta V)^2} &= {}
   \frac{\CA^2 \alphas^2 M_\ix M_\fx}{8\pi^2 (x_1{+}x_4)^2 E^4}
   \sum_{\substack{{\rm allowed}\\ \lambda_{01},\lambda_{12},\lambda_{23} }}
   \int_{0<t_1<t_2<\Delta t} dt_1 \> dt_2 \> d(\Delta t) \>
    (2\pi)^4 f_{(01)} f_{(12)} f_{(23)}
\nonumber\\ &\quad \times
   \frac{d}{dj_1} \, \frac{d}{dj_2}
   \biggl[
    \det({\cal U}^{-1})
    \Bigl\{
     (\beta \tilde Y_\yx \tilde Y_\ybx
        + \alpha \tilde\Ybar_{\yx\ybx} \tilde Y_{\yx\ybx})
          \tilde I_0
     + (\alpha+\beta+2\gamma) \tilde Z_{\yx\ybx} \tilde I_1
\nonumber\\ &\hspace{6em}
     + \bigl[
         (\alpha+\gamma) \tilde Y_\yx \tilde Y_\ybx
         + (\beta+\gamma)
           \tilde \Ybar_{\yx\ybx} \tilde Y_{\yx\ybx}
        \bigr] \tilde I_2
\nonumber\\ &\hspace{6em}
     - (\alpha+\beta+\gamma)
       (\tilde\Ybar_{\yx\ybx} \tilde Y_\ybx \tilde I_3
        + \tilde Y_\yx \tilde Y_{\yx\ybx} \tilde I_4)
    \Bigl\}
   \biggr]_{j_1=j_2=0} .
\label{eq:xyyxresult}
\end {align}

In the context of the sequential $xy\bar x\bar y$ diagram, we previously
discussed that equality of initial and final 4-particle color singlet
state overlap factors between (i) the calculation of $1/N^2$ corrections
in (\ref{eq:implicit12corr}) and (ii) the $N{=}\infty$ calculation
in (\ref{eq:implicit12}).  A similar match-up occurs for the
first five rows of table \ref{tab:Xtransitions}.
Specifically, the calculation of those $1/N^2$ corrections should
contain a factor of
\begin {equation}
  {}_s \langle {\rm A}_{\rm aa} | \lambda_{23} \rangle
       \langle \lambda_{01} | {\rm A}_{\rm aa} \rangle_u
  = \tfrac12 .
\label {eq:Ximplicit12corr}
\end {equation}
But we did not have to explicitly include that factor because
our starting point ---
the $N{=}\infty$ formula (\ref{eq:xyyxNinf}) for the crossed
diagram $xy\bar x\bar y$ ---
already implicitly
contained an equal factor of%
\footnote{
  Our (\ref{eq:Ximplicit12}) is equivalent to the
  $\tfrac12$ in the result $\tfrac12 \CA^2$ of eq.\ (4.17) of
  ref.\ \cite{2brem}.
}
\begin {equation}
  {}_s \langle {\rm A}_{\rm aa} | \Am \rangle
       \langle \Am | {\rm A}_{\rm aa} \rangle_u
  = \tfrac12 .
\label {eq:Ximplicit12}
\end {equation}


\subsection {A single $\delta T$ or $\delta^2 T$ perturbation}

\begin {figure}[t]
\begin {center}
  \begin{picture}(277,140)(0,0)
    \put(0,15){\includegraphics[scale=0.6]{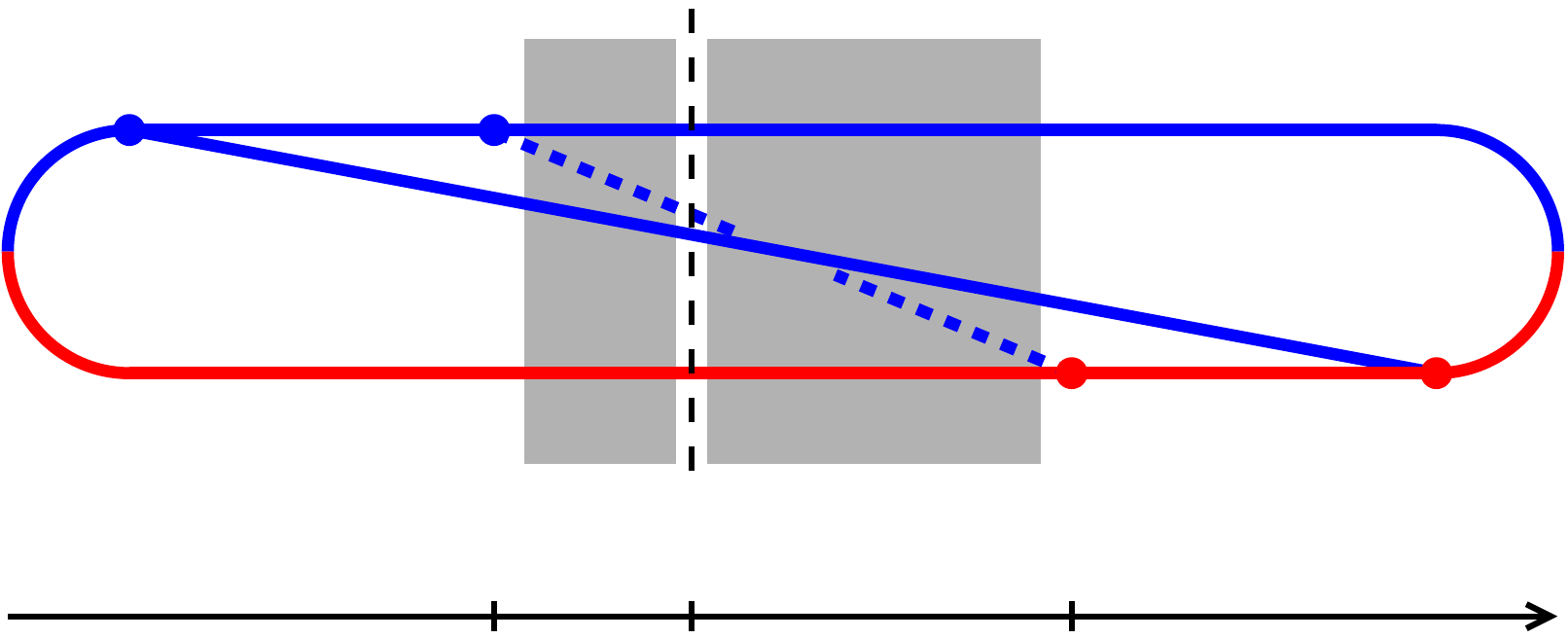}}
    \put(84,5){$t_0$}
    \put(120,5){$t_1$}
    \put(187,5){$t_2$}
    \put(100,33){$\lambda_{01}$}
    \put(149,33){$\lambda_{12}$}
    \put(118,130){$\delta V$}
  \end{picture}
  \caption{
     \label {fig:dG1}
     Like fig.\ \ref{fig:dGcross} except with only one insertion of
     $\delta V$ (or $\delta^2 V$) during the 4-particle time evolution.
  }
\end {center}
\end {figure}
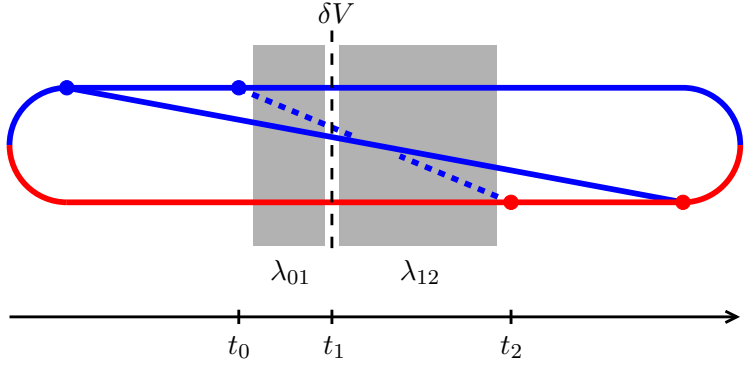

We now turn to the last group of transitions in table \ref{tab:Xtransitions},
where there is only one $\delta T$ or $\delta^2 T$ perturbation.
Schematically, this corresponds to fig.\ \ref{fig:dG1}.
The analog of (\ref{eq:d2G1}) is
\begin {align}
  \delta & G_{\lambda_{12}\leftarrow\lambda_{01}}
             (\vbxi_2,\Delta t; \vbxi_0,0)
\nonumber\\ & \qquad
  =
  -i
  \int_{0<t_1<\Delta t} dt_1 \>
  \int_{\vbxi_1}
  G^\Ninf_{\lambda_{12}}(\vbxi_2,t_2; \vbxi_1,t_1) \,
  \delta V_{\lambda_{12}\leftarrow\lambda_{01}}^{(\delta^n T)}\!(\vbxi_1) \,
  G^\Ninf_{\lambda_{01}}(\vbxi_1,t_1; \vbxi_0,t_0) ,
\end {align}
where $t_0 = t^\yx$ and $t_2 = t^\ybx$ are the initial and final times
of the 4-particle evolution.
The analog of (\ref{eq:d2GXi}) is then
\begin {multline}
  \delta G_{\lambda_{12}\leftarrow\lambda_{01}}
  (\vec{\bm \xi}_2,\Delta t; \vec{\bm \xi}_0,0)
  =
  -i
  f_{(01)} f_{(12)}
  \int_{0<t_1<\Delta t} dt_1\>
  e^{
    -\frac12 \vbxi_0^{\,\top} \! A_{(01)} \, \vbxi_0
    -\frac12 \vbxi_2^{\,\top} \! A_{(12)} \, \vbxi_2
  }
\\ \times
  \frac{d}{d j_1}
  \int d^2\xi_1 \>
     e^{ - \frac12 \vbxi_1^{\top} \! U \vbxi_1 }
     e^{ \, \vbxi_1^{\,\top} \left(B_{(01)}\vbxi_0 + B_{(12)}\vbxi_2\right) }
  \>
  \biggr|_{j_1=0} ,
\end {multline}
where
\begin {equation}
   U \equiv A_{(01)} + A_{(12)} - j_1 R_1 .
\label {eq:Udef}
\end {equation}
Doing the Gaussian integral over $\vec{\bm \xi}_1$ yields the
analog of (\ref{eq:d2Gdet}):
\begin {subequations}
\label {eq:dGdT}
\begin {align}
  &\delta G_{\lambda_{12}\leftarrow\lambda_{01}}
  (\vec{\bm \xi}_2,\Delta t; \vec{\bm \xi}_0,0)
  =
\nonumber\\ &\qquad
  -i (2\pi)^2
  f_{(01)} f_{(12)}
  \int_{0<t_1<\Delta t} dt_1\>
  \frac{d}{d j_1}
  \biggl[
   \det(U^{-1}) \,
    e^{
      -\frac12 \vbxi_0^{\,\top} \! A_{(01)} \, \vbxi_0
      -\frac12 \vbxi_2^{\,\top} \! A_{(12)} \, \vbxi_2
      +\frac12 \Jvec^{\,\top} \! U^{-1} \Jvec
    }
  \biggr]_{j_1=0}
\end {align}
with
\begin {equation}
   \Jvec \equiv B_{(01)}\vbxi_0 + B_{(12)}\vbxi_2 \,.
\end {equation}
\end {subequations}

By comparison of the exponential in (\ref{eq:dGdT}) to the $N{=}\infty$
exponential in (\ref{eq:XYZexpCross}), and accounting for the change
of basis (\ref{eq:Sfrak}), we find that for these processes
\begin {subequations}
\label {eq:XYZcrossdT}
\begin {equation}
  \begin{pmatrix}
    \tilde\calX_\yx & \tilde Y_\yx \\ \tilde Y_\yx & \tilde Z_\yx
  \end{pmatrix}
  =
  A_{(01)} - B_{(01)} U^{-1} B_{(01)}
  ,
  \qquad
  \begin{pmatrix}
    \tilde\calX_\ybx & \tilde Y_\ybx \\ \tilde Y_\ybx & \tilde Z_\ybx
  \end{pmatrix}
  =
  {\mathfrak S}^\top
  ( A_{(12)} - B_{(12)} U^{-1} B_{(12)} )
  {\mathfrak S}
  ,
\end {equation}
\begin {equation}
  \begin{pmatrix}
    \tilde X_{\yx\ybx} & \tilde Y_{\yx\ybx} \\
    \tilde\Ybar_{\yx\ybx} & \tilde Z_{\yx\ybx}
  \end{pmatrix}
  =
  B_{(01)} U^{-1} B_{(12)}
  {\mathfrak S}
  .
\end {equation}
\end {subequations}
Also, as before, the $\cal X$'s are related to the $X$'s by
(\ref{eq:XvscalXcross}).
The analog of (\ref{eq:xyyxresult}) is then
\begin {align}
   \delta^2 \left[\frac{d\Gamma}{dx\,dy}
            \right]_{xy\bar y\bar x}^{\delta T\,{\rm or}\,\delta^2 T} = {}
   &
   i \frac{\CA^2 \alphas^2 M_\ix M_\fx}{8\pi^2 (x_1{+}x_4)^2 E^4}
   \sum_{\substack{{\rm allowed}\\ \lambda_{01},\lambda_{12} }}
   \Phi_{\lambda_{01},\lambda_{12}}
   \int_{0<t_1<\Delta t} \!\!\!\! dt_1 \> d(\Delta t) \>
    (2\pi)^2 f_{(01)} f_{(12)}
\nonumber\\ &\quad \times
   \frac{d}{dj_1}
   \biggl[
    \det(U^{-1})
    \Bigl\{
     (\beta \tilde Y_\yx \tilde Y_\ybx
        + \alpha \tilde\Ybar_{\yx\ybx} \tilde Y_{\yx\ybx})
          \tilde I_0
     + (\alpha+\beta+2\gamma) \tilde Z_{\yx\ybx} \tilde I_1
\nonumber\\ &\hspace{6em}
     + \bigl[
         (\alpha+\gamma) \tilde Y_\yx \tilde Y_\ybx
         + (\beta+\gamma)
           \tilde \Ybar_{\yx\ybx} \tilde Y_{\yx\ybx}
        \bigr] \tilde I_2
\nonumber\\ &\hspace{6em}
     - (\alpha+\beta+\gamma)
       (\tilde\Ybar_{\yx\ybx} \tilde Y_\ybx \tilde I_3
        + \tilde Y_\yx \tilde Y_{\yx\ybx} \tilde I_4)
    \Bigl\}
   \biggr]_{j_1=0} ,
\label {eq:xyyxresultdT}
\end {align}
where the $(\tilde X,\tilde Y,\tilde Z)$'s are now those determined by
(\ref{eq:XYZcrossdT}) and the $\Phi_{\lambda_{01},\lambda_{12}}$ is a
normalization factor we discuss below.

For simplicity, we will take
\begin {equation}
  R_1 = R^{(\delta T)}
\label {eq:R1again}
\end {equation}
in the definition (\ref{eq:Udef}) of $U$ for all of the single
$\delta V$ processes summarized in table \ref{tab:PhiFactor}.
Because we are using (\ref{eq:R1again}) for $\delta^2 T$
as well as $\delta T$ perturbations, this leads to the first
of two normalization issues.

\begin {table}[tp]
\begin {center}
\renewcommand{\arraystretch}{1.4}
\setlength{\tabcolsep}{7pt}
\begin{tabular}{@{}ccccc@{}}
\hline\hline
  transition & equivalent & $\Phi$
\\ \hline
   $\sAm \xrightarrow{\delta^2 T}{} \sAx $
   & $(1234) \to (1243)$
   & $\frac{1}{\sqrt2\,N}$
\\
   $\sAp \xrightarrow{\delta^2 T}{} \sAx $
   & $(1324) \to (1243)$
   & $\frac{1}{\sqrt2\,N}$
\\
   $\sAm \xrightarrow{\delta T}{} \ssone $
   & $(1234) \to (14)(23)$
   & $\frac{-2}{\sqrt2\,N}$
\\
   $\sAm \xrightarrow{\delta T}{} \sonexm $
   & $(1234) \to (12)(34)$
   & $\frac{1}{\sqrt2 \, N}$
\\
   $\sAp \xrightarrow{\delta T}{} \ssone $
   & $(1324) \to (14)(23)$
   & $\frac{-2}{\sqrt2\,N}$
\\
   $\sAp \xrightarrow{\delta T}{} \sonexp$
   & $(1324) \to (13)(24)$
   & $\frac{1}{\sqrt2 \, N}$
\\[3pt]
\hline\hline
\end{tabular}
\end {center}
\caption{
   \label{tab:PhiFactor}
   The last group of transition sequences from table \ref{tab:Xtransitions},
   along with the corresponding factor $\Phi$ appearing in
   (\ref{eq:xyyxresultdT}).
 }
\end{table}

The operation ${\partial/\partial j_1} [ \cdots ]_{j_1=0}$ in
(\ref{eq:xyyxresultdT}) was constructed to introduce one
factor of
$\tfrac12 \vbxi_1^{\,\top} R_1 \vbxi_1$ (unexponentiated)
into the calculation of the overall result.
Based on (\ref{eq:dST}),
$\delta^2 T$ matrix elements relevant to
the transitions in table \ref{tab:PhiFactor} all have value
$1/2N^2$.  In contrast, the non-zero matrix elements of $\delta T$
in (\ref{eq:dST}) are all $1/\sqrt 2\,N$.  To correct for this
difference, our overall factor $\Phi$ in (\ref{eq:xyyxresultdT})
will need to contain (among other things) a factor of
\begin {align}
    \begin {cases}
       1 ,                   & \mbox{for $\,\delta T\,$ transition} ; \\
       \frac{1}{\sqrt2\,N} , & \mbox{for $\delta^2 T$ transition} .
    \end {cases}
\label {eq:phi1}
\end {align}

Earlier, we explained that our starting point --- the $N{=}\infty$ result ---
implicitly contains an initial and final color singlet overlap factor
(\ref{eq:Ximplicit12}), which equals the similar overlap factors
${}_s \langle {\rm A}_{\rm aa} | \lambda_{23} \rangle
       \langle \lambda_{01} | {\rm A}_{\rm aa} \rangle_u$
needed for both $\delta T\,\delta T$ and
$\delta T\,\delta S$ transition sequences.
From the ``color overlap'' column of table \ref{tab:Xtransitions},
however, we see that these overlap factors are different for
some of the other transition sequences.  Our overall normalization
in (\ref{eq:xyyxresultdT}) will need to account for this difference
as well.  Putting that correction together with (\ref{eq:phi1}), the
overall normalization correction we need in (\ref{eq:xyyxresultdT}) is
\begin {align}
 \Phi_{\lambda_{01},\lambda_{12}}
 &\equiv
 \frac{ 
    {}_s \langle {\rm A}_{\rm aa} | \lambda_{12} \rangle
    \langle \lambda_{01} | {\rm A}_{\rm aa} \rangle_u
  }{
    {}_s \langle {\rm A}_{\rm aa} | \Am \rangle
    \langle \Am | {\rm A}_{\rm aa} \rangle_u
  }
  \times
  \frac{
    \mbox{$\delta T$ or $\delta^2 T$ matrix element}
  }{
    \mbox{non-zero $\delta T$ matrix elements}
  }
\nonumber\\
  &=
    2 \,
    {}_s \langle {\rm A}_{\rm aa} | \lambda_{12} \rangle
    \langle \lambda_{01} | {\rm A}_{\rm aa} \rangle_u
    \times
    \begin {cases}
       1 ,                   & \mbox{for $\,\delta T\,$ transition} ; \\
       \frac{1}{\sqrt2\,N} , & \mbox{for $\delta^2 T$ transition} .
    \end {cases}
\end {align}
The values of $\Phi$
are shown explicitly in Table \ref{tab:PhiFactor}.
(They are the same as $\sqrt2\, N\phi$, where $\phi$ is the last column
of table \ref{tab:Xtransitions}.)

Note that the $1/N^2$ behavior of (\ref{eq:xyyxresultdT}) comes
from two places: One factor of $1/N$ comes from the factor
$R_1{=}R^{(\delta T)}$ (\ref{eq:R}) produced by
the operation ${\partial/\partial j_1} [ \cdots ]_{j_1=0}$, and the other comes
from the values of $\Phi$
in table \ref{tab:PhiFactor}.


\subsection{Correction to total crossed diagram rate}

The $1/N^2$ correction to the $xy\bar y\bar x$ diagram corresponds to
the sum of the results of (\ref{eq:xyyxresult}) and (\ref{eq:xyyxresultdT}),
each using the formulas for the $(\tilde X,\tilde Y,\tilde Z)$'s appropriate to
that particular process [(\ref{eq:XYZcross}) or (\ref{eq:XYZcrossdT})]
and each summed over the relevant entries of table \ref{tab:Xtransitions}.
In order to connect with previous work, it will be convenient to
give the name $\delta^2 C$ to the total $\Delta t$ integrand for the
$xy\bar y\bar x$ diagram:
\begin {equation}
   \int_0^\infty d(\Delta t) \>
   \delta^2 C(x_1,x_2,x_3,x_4,\alpha,\beta,\gamma,\Delta t)
   \equiv
   \delta^2 \left[\frac{d\Gamma}{dx\,dy}\right]_{xy\bar y\bar x}^{(\delta V)^2}
   +
   \delta^2 \left[\frac{d\Gamma}{dx\,dy}
            \right]_{xy\bar y\bar x}^{\delta T\,{\rm or}\,\delta^2 T} .
\label {eq:xyyxtot}
\end {equation}

\begin {figure}[t]
\begin {center}
  \includegraphics[scale=0.42]{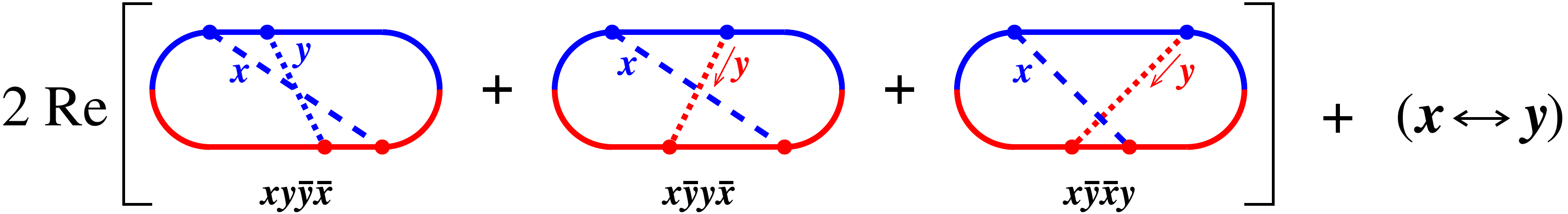}
  \caption{
     \label {fig:crossdiags}
     The sum of diagrams that define the quantity $A(x,y)$ in
     ref.\ \cite{2brem}.
  }
\end {center}
\end {figure}

To get the $1/N^2$ correction to the total crossed diagram rate from
eq.\ (\ref{eq:xyyxtot}) for the $xy\bar y\bar x$ diagram, we need to follow
the same steps as originally used for the $N{=}\infty$ calculation in
ref.\ \cite{2brem}.  There, the total rate was organized by
first summing over the diagrams represented by fig.\ \ref{fig:crossdiags}
to get
\begin {equation}
   \delta^2 A(x,y) \equiv
   \int_0^\infty d(\Delta t) \> 
     2 \Re \bigl[ \delta^2 B(x,y,\Delta t) + \delta^2 B(y,x,\Delta t) \bigr]
\end {equation}
with
\begin {align}
   \delta^2 B(x,y,\Delta t)
   &\equiv
       \delta^2 C({-}1,y,z,x,\alpha,\beta,\gamma,\Delta t)
       + \delta^2
            C\bigl({-}(1{-}y),{-}y,1{-}x,x,\beta,\alpha,\gamma,\Delta t\bigr)
\nonumber\\ &\qquad\qquad
       + \delta^2
            C\bigl({-}y,{-}(1{-}y),x,1{-}x,\gamma,\alpha,\beta,\Delta t\bigr) ,
\end {align}
where we've used the same notation ($A,B,C$) as ref.\ \cite{2brem}.%
\footnote{
  See eqs. (8.1--8.3) of ref.\ \cite{2brem}.
  But, for the $1/N^2$ term corrections being considered here, there are
  no additional ``pole'' terms, as previously discussed in footnote
  \ref{foot:UV}.  For the same reason, it is also unnecessary to make the
  vacuum subtraction of eq.\ (8.4) of ref.\ \cite{2brem}.
}
Finally, we need to sum over any remaining
permutations of the daughters $(x,y,z)$ that lead to distinct
diagrams.  Just as in the $N{=}\infty$ analysis of ref.\ \cite{2brem},
this gives%
\footnote{
  See eq.\ (8.1) of ref.\ \cite{2brem}.
}
\begin {equation}
   \delta^2 \left[ \frac{d\Gamma}{dx\>dy} \right]_{\rm crossed}
   =
   \delta^2 A(x,y) + \delta^2 A(z,y) + \delta^2 A(x,z) .
\label {eq:summary1}
\end {equation}

\section {Numerical results}
\label {sec:numerics}

\subsection {Main results}
\label {sec:nresults}

To get results for $1/N^2$ corrections, we numerically integrated
over $(t_1,t_2,\Delta t)$ or $(t_1,\Delta t)$.
A short discussion of our numerical method is given in appendix
\ref{app:nmethod}.
For comparison, the $N{=}\infty$ results from previous literature
only require numerical integration over $\Delta t$.%
\footnote{
   The $N{=}\infty$ results for crossed and sequential diagrams were
   derived in refs.\ \cite{2brem,seq,dimreg}, but a convenient summary
   of results may be found in appendix A.2 of ref.\ \cite{qcd}.
}
Dividing $1/N^2$ corrections by the corresponding $N{=}\infty$ result gives
the relative size of the corrections.

\begin {figure}[t]
\begin {center}
  \includegraphics[scale=0.6]{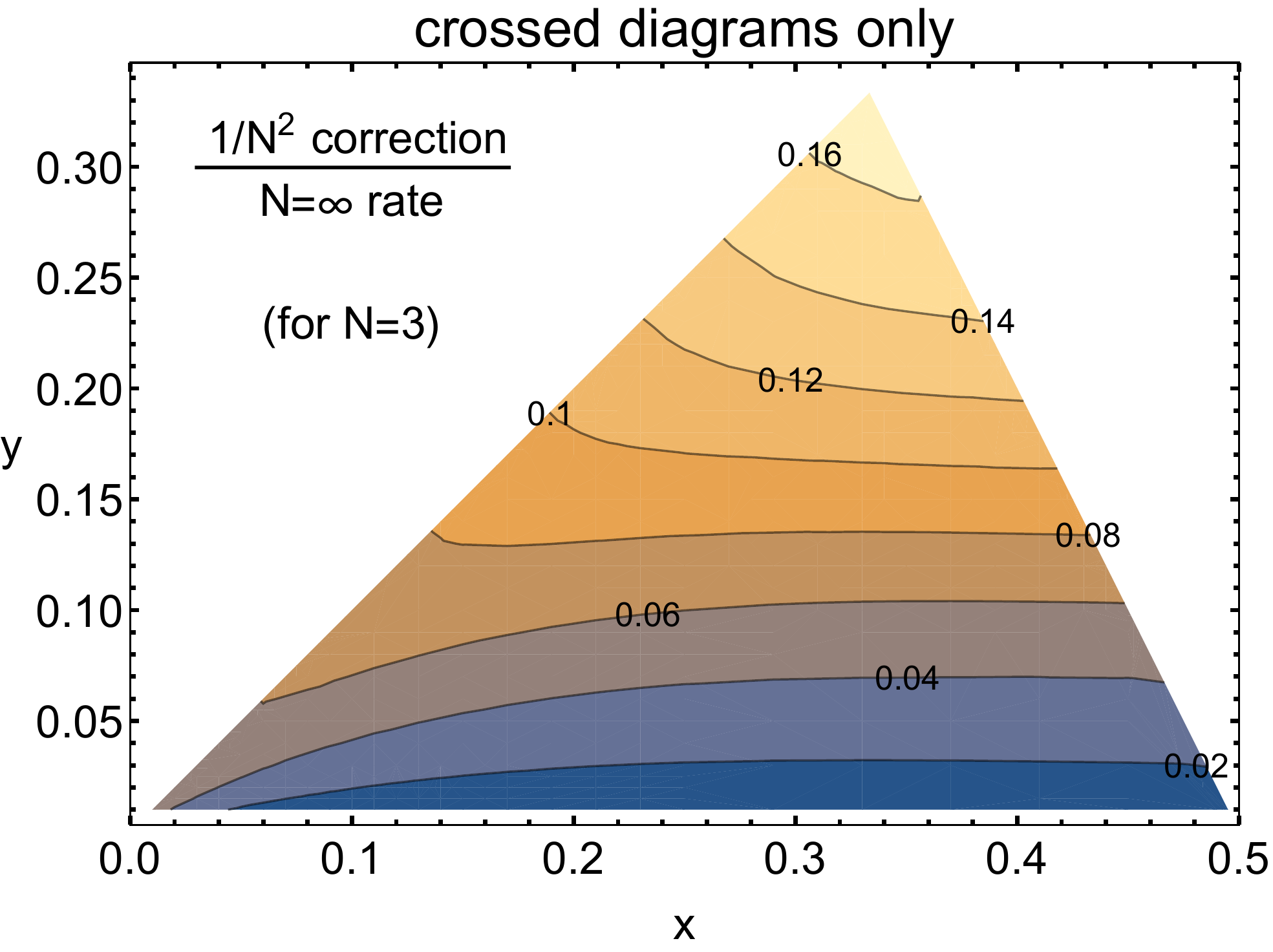}
  \caption{
     \label {fig:ratio_crossed}
     The ratio, for {\it crossed diagrams} only, of
     (i) the $1/N^2$ correction to (ii) the $N{=}\infty$ result
     for the differential rate
     $d\Gamma/dx\,dy$ for (the crossed diagram contribution to)
     overlapping double splitting $g \to ggg$.
     We have used $N{=}3$ in this plot, but one may multiply the
     results by $(3/N)^2$ to restore the $N$ dependence of the
     $1/N^2$ correction.
     Very tiny wiggles in the contour lines are an artifact
     of interpolation from a discrete set of numerical data points.
     We have left out $y < 0.01$ just to simplify the numerical effort
     that went into making this plot.
     The ratio goes to zero as $y {\to} 0$, as one may see from
     the later discussion of fig.\ \ref{fig:crossVseq} for a particular
     value of $x$.
  }
\end {center}
\end {figure}

\begin {figure}[t]
\begin {center}
  \includegraphics[scale=0.6]{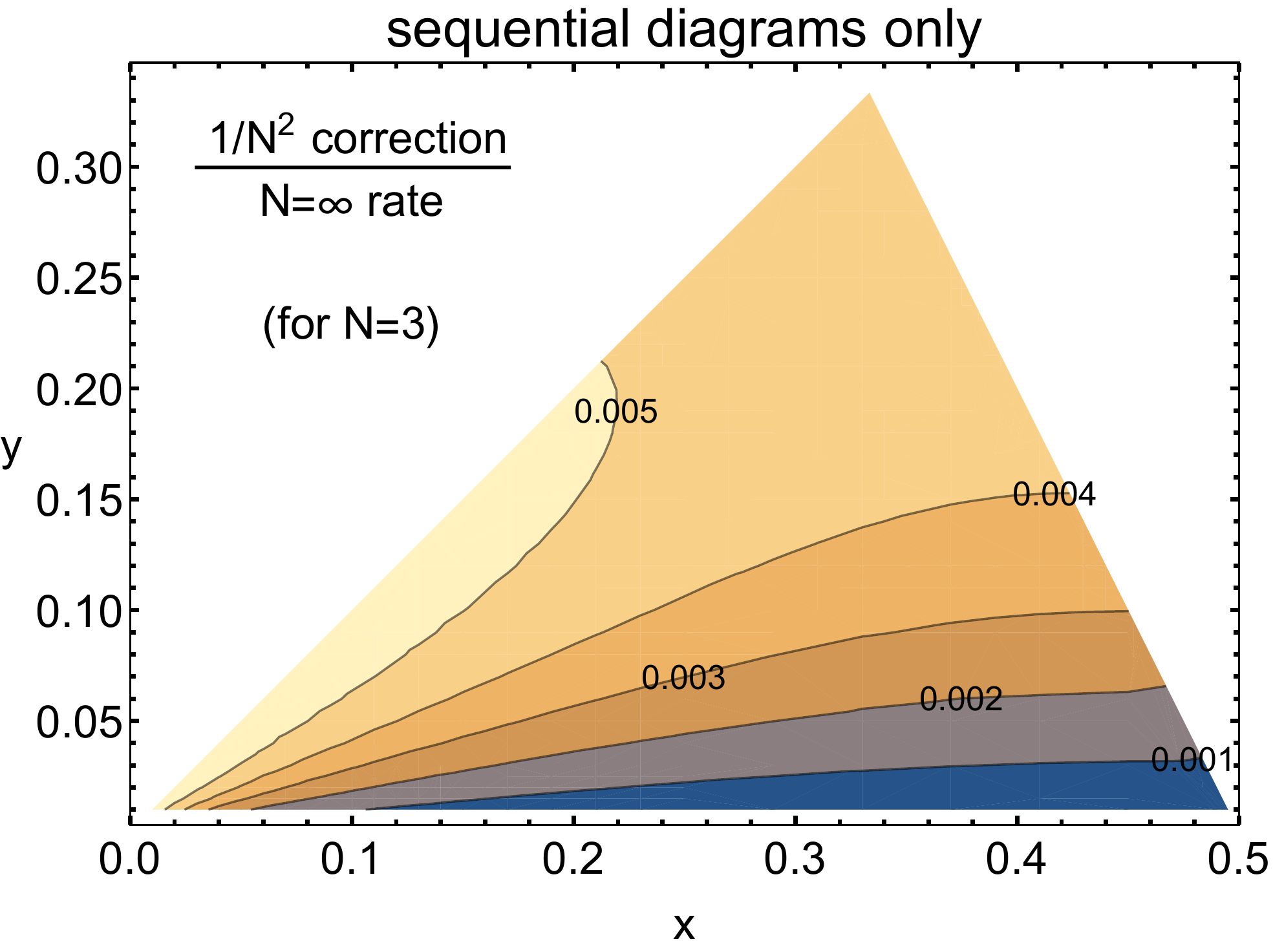}
  \caption{
     \label {fig:ratio_seq}
     Like fig.\ \ref{fig:ratio_crossed} except now for
     {\it sequential diagrams} instead of crossed diagrams.
  }
\end {center}
\end {figure}

The relative size of $1/N^2$ corrections to crossed and sequential
diagrams for overlapping double splitting $g{\to}ggg$ are shown,
respectively, in
figs.\ \ref{fig:ratio_crossed} and \ref{fig:ratio_seq}
for $N{=}3$ (QCD).
Our convention in these plots is to let $y$ represent the energy fraction
of the lowest-energy daughter, $x$ represent the next lowest, and then
$z=1{-}x{-}y$ represents the highest-energy daughter.
So the plot region has been restricted by these conventions to
$y < x < 1{-}x{-}y$.
The $1/N^2$ corrections to sequential diagrams are very small: less
than 1\%.  The corrections to crossed diagrams are substantially larger.
The largest relative correction occurs at the apex of the triangular region,
$(x,y,z)=(\frac13,\frac13,\frac13)$, where the correction
is roughly 17\%.

\begin {figure}[t]
\begin {center}
  \includegraphics[scale=0.6]{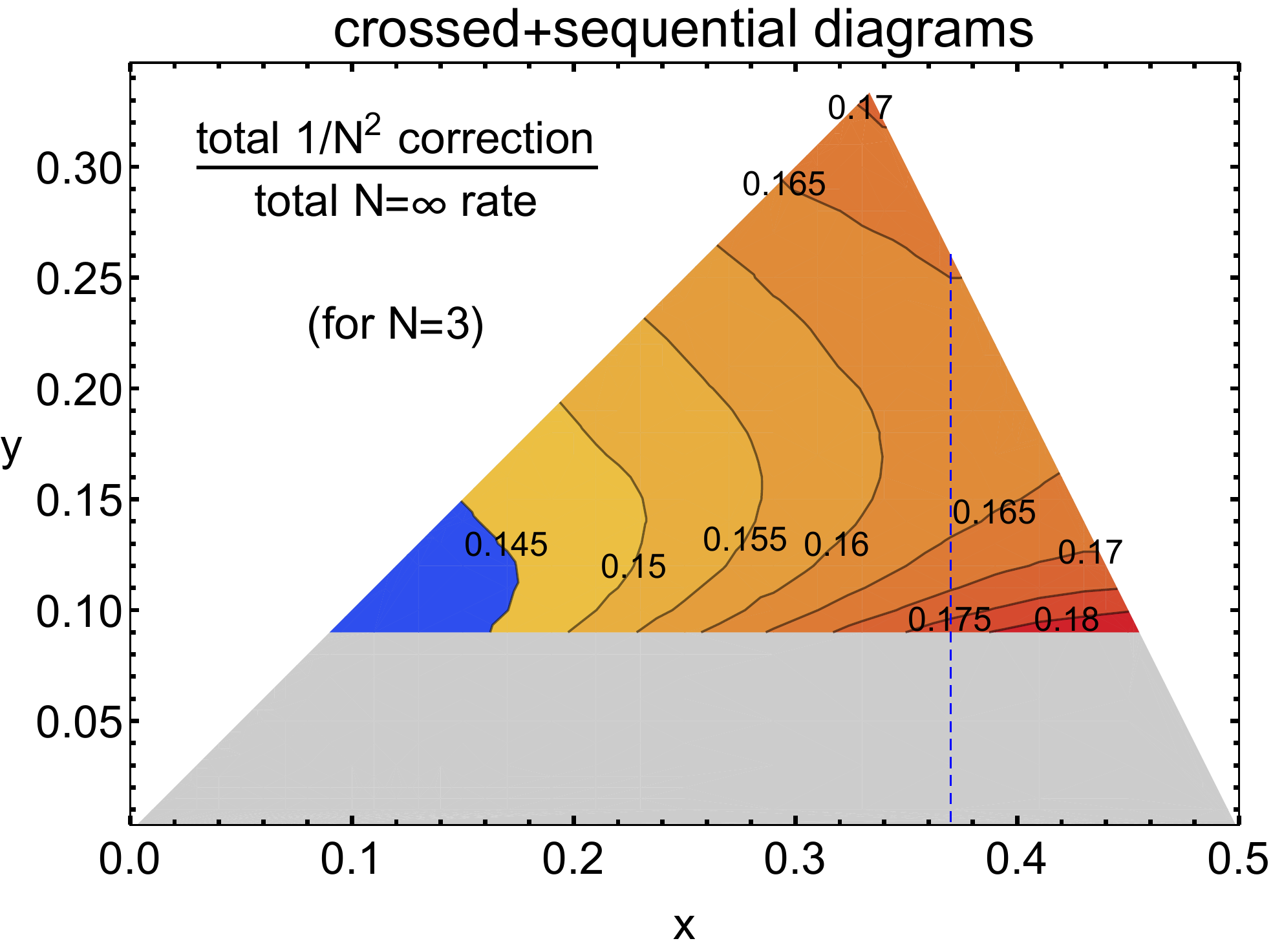}
  \caption{
     \label {fig:ratio_tot}
     Like figs.\ \ref{fig:ratio_crossed} and \ref{fig:ratio_seq}
     except now for the total $g\to ggg$ rate.
     The behavior for y<0.1 (the gray shaded region) is
     discussed in the main text.
  }
\end {center}
\end {figure}

We showed the separate crossed and sequential diagram ratios first because
there is a subtlety to discussing relative corrections to the
total rate (crossed plus sequential).%
\footnote{
   As discussed at the very end of section \ref{sec:intro},
   our ``total'' here, defined as the sum of crossed and sequential
   diagrams, does not quite contain every process
   that contributes to $g{\to}ggg$.
}
Fig.\ \ref{fig:ratio_tot}
shows a plot of the ratio
\begin {equation}
   \frac{
     \mbox{total $1/N^2$ correction}
   }{
     \mbox{total $N{=}\infty$ rate}
   }
\label {eq:ratio}
\end {equation}
for $g\to ggg$, but restricted to $y > 0.1$.  Similar to
fig.\ \ref{fig:ratio_crossed}, there is a (local) maximum at the
apex of the triangular region, where the $1/N^2$ correction is roughly
17\%.  Unlike fig.\ \ref{fig:ratio_crossed}, however,
around $y \sim 0.1$ the rate has started to grow with decreasing $y$.
As we will explain, this
small-$y$ growth is an artifact of how we have so far chosen to look
at the size of $1/N^2$ corrections.

\begin {figure}[t]
\begin {center}
  \includegraphics[scale=0.5]{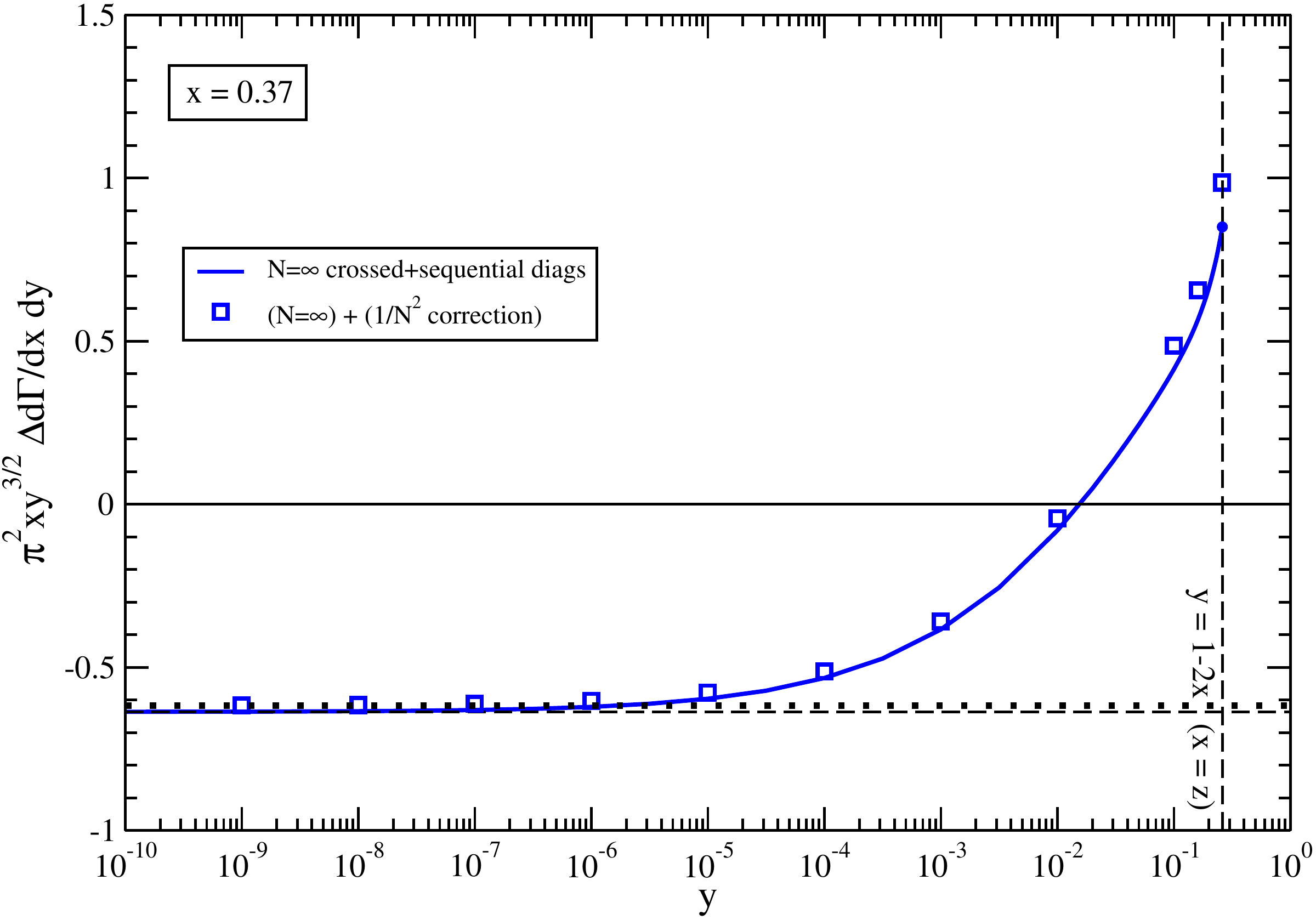}
  \caption{
     \label {fig:total}
     A plot for $x=0.37$ of the $y$ dependence of the
     total $\Delta\,d\Gamma/dx\,dy$, multiplied by $\pi^2 x y^{3/2}$
     [and in units of $N\alphas \sqrt{\qhatA/E}$, but remember that
     $N\alphas$ is held fixed as $N{\to}\infty$].
     The plot shows
     (solid curve) the $N{=}\infty$ value and (squares) the
     $N{=}\infty$ value plus the $1/N^2$ correction.
     The horizontal dashed line shows the limiting $y{\to}0$ behavior
     of the $N{=}\infty$ result, and the nearby horizontal dotted line shows
     the limiting behavior of the corrected result.
  }
\end {center}
\end {figure}

Instead of showing the ratio (\ref{eq:ratio}),
fig.\ \ref{fig:total} shows, for a particular value of $x$,
the small-$y$ behavior of
(i) the $N{=}\infty$ result for the total rate vs.\ (ii)
the sum of the $N{=}\infty$ result and the $1/N^2$ correction.
We've taken $x=0.37$, which corresponds
to the blue dashed line in fig.\ \ref{fig:ratio_tot}.%
\footnote{
  There is nothing special about the specific choice $x=0.37$.
}
At small $y$, both the $N{=}\infty$ results and the total
$1/N^2$ correction blow up%
\footnote{
  For a hand-waving qualitative explanation,
  see section 1.4 of ref.\ \cite{seq}.
}
as $1/y^{3/2}$,
and so we have followed the convention of ref.\ \cite{seq}
and instead plotted
\begin {equation}
   \pi^2 x y^{3/2} \left[ \Delta \frac{d\Gamma}{dx\,dy} \right]_{\rm total}
\label {eq:totalnorm}
\end {equation}
in fig.\ \ref{fig:total}.
The extremely tiny values of $y$ plotted in the figure are unlikely to ever
be relevant to any real-world physics because, at the very least, one needs
$yE \gg T$ for our high-energy approximations.%
\footnote{
  There's additionally the issue that, for small enough $y$,
  one would need to implement resummation of soft radiation.
}
Nonetheless, this figure is useful to understand the behavior of our
formulas.  The important feature is that the total $N{=}\infty$ result
crosses zero at $y \sim 0.01$ (for this value of $x$).  This is
possible because $\Delta \Gamma/dx\,dy$ does not represent a rate; it
represents the correction to a rate due to overlapping formation times
(see section 1.1 of ref.\ \cite{seq} for explanation),%
\footnote{
  Readers may wonder if one could instead divide the $1/N^2$
  corrections by a positive
  {\it complete}\/ $g{\to}ggg$ rate instead of
  dividing by just the (varying sign)
  {\it correction} $\Delta \Gamma/dx\,dy$ from overlap effects.
  Section 1.1 of ref.\ \cite{seq} explains why it is not meaningful
  to talk about such a ``complete'' rate of double splitting in an infinite
  medium.  It has to do with the fact that one way to achieve
  $g{\to}ggg$ is via two independent single emissions $g{\to}gg$
  that are abitrarily far separated in time.
}
and a
correction may be positive or negative.  Where the $N{=}\infty$ result
vanishes, the {\it relative} size (\ref{eq:ratio}) of the $1/N^2$
correction to that $N{=}\infty$ result will blow up to infinity, by
definition.  From fig.\ \ref{fig:total}, however, one sees that there
is very little difference between the $N{=}\infty$
curve and the corrected curve for $y < 0.1$.
In any case, in applications to energy loss and in-medium shower
development, the $1/y^{3/2}$
small-$y$ behavior of overlapping double splitting $g {\to} ggg$
is canceled \cite{qcd} by similar behavior of
virtual corrections to single splitting $g {\to} gg$, leaving behind
double-log divergences \cite{Blaizot,Iancu,Wu}
that are {\it independent}\/ of $N$.
So the small-$y$ behavior of fig.\ \ref{fig:total}, and in
particular the $1/N^2$ corrections to the small-$y$ behavior,
are not of much physical interest.%
\footnote{
  The ``double log'' behavior referred to above arises from
  $y^{-1} \ln y$ behavior in the combined real and virtual rates,
  producing a double
  logarithm when integrated over $y$.
  This is in contrast to the more
  divergent $y^{-3/2}$ infrared behavior shown in fig.\ \ref{fig:total}
  for the real rate by itself.
  The fact that the soft behavior is
  $y^{-1} \ln y$ when virtual corrections to single splitting
  are included has been known since the early work
  of refs.\ \cite{Blaizot, Iancu, Wu} on energy loss
  in the soft-$y$ approximation.
  The lack of $N$ dependence of those double-log
  results appears in their calculations as a special feature of
  the dynamics of the soft gluon emission limit.
  (An explicit calculation showing in detail the cancellation of
  $y^{-3/2}$ divergences between real and virtual diagrams
  for the case $N{=}\infty$
  may be found in ref.\ \cite{qcd}, which is focused on
  generic-$y$ results but also
  extracts their small-$y$ behavior.)
}
Furthermore, the original motivation for the large-$N$ approximation
in this problem was as a tool to be able to study
overlapping {\it hard}\/ splittings ($y \sim x \sim 1$).

In principle, the best way to
investigate the size of $1/N^2$ corrections would be to quote
the relative size of their effect on an (infrared-safe)
characteristic of in-medium shower development.
Since we do not have everything needed for that (such as
$1/N^2$ corrections for virtual diagrams), we interpret
fig.\ \ref{fig:total} to mean that a reasonable proxy is the
largest relative size of the $1/N^2$ corrections to $d\Gamma/dx\,dy$
for $y$ values that are {\it not}\/ small ---
namely, the roughly 17\% correction
at the apex of fig.\ \ref{fig:ratio_tot}.

\subsection{More detail on small-$y$ behavior of crossed vs.\ sequential}

\begin {figure}[t]
\begin {center}
  \includegraphics[scale=0.5]{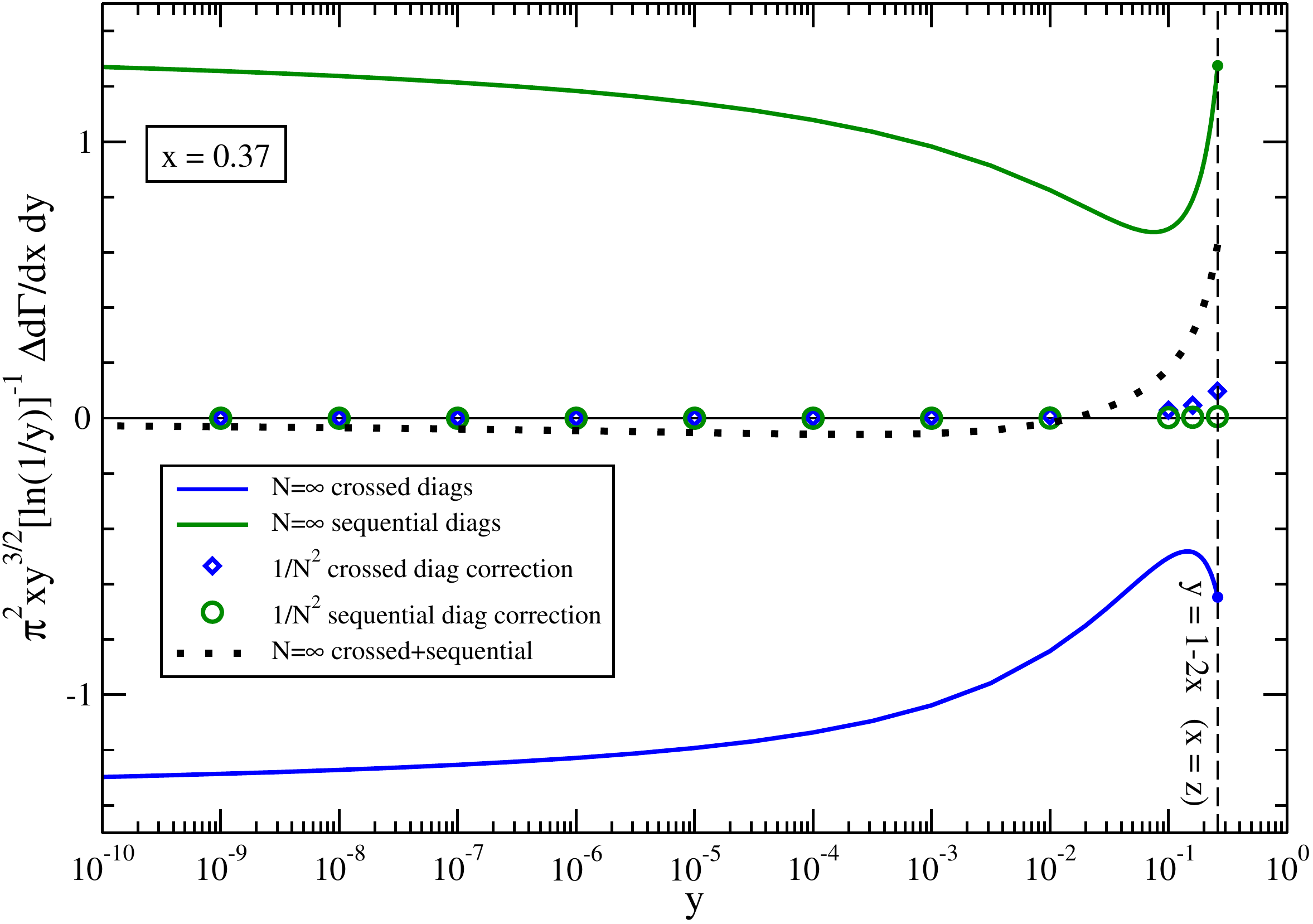}
  \caption{
     \label {fig:crossVseq}
     A plot for $x=0.37$ of the $y$ dependence of the
     different contributions to $\Delta\,d\Gamma/dx\,dy$,
     multiplied by $\pi^2 x y^{3/2}$ and then divided by $\ln(y^{-1})$.
     The dotted line shows the total crossed+sequential
     $N{=}\infty$ result, corresponding to the solid curve in
     fig.\ \ref{fig:total}.
     Note that the vertical axis is normalized differently here than in
     fig.\ \ref{fig:total}, but still presented in units of
     $N\alphas \sqrt{\qhatA/E}$.
     [The intended purpose of this plot is qualitative.
     See fig.\ \ref{fig:crossVseq3} if interested in the precise values
     corresponding to the $1/N^2$ data points.]
  }
\end {center}
\end {figure}

There are some interesting qualitative features about the small-$y$
behavior of crossed vs.\ sequential diagrams.
Fig.\ \ref{fig:crossVseq} shows various different elements that went
into the previous small-$y$ numerics of fig.\ \ref{fig:total}.

First, as noted originally in ref.\ \cite{seq},
the $N{=}\infty$ crossed and sequential results individually behave
like $\ln(y^{-1})/y^{3/2}$ even though their sum just behaves like
$1/y^{3/2}$.  Because of the log dependence of the individual
contributions, we have chosen to plot the contributions to
\begin {equation}
   \frac{\pi^2 x y^{3/2}}{\ln(y^{-1})} \,
   \left[ \Delta \frac{d\Gamma}{dx\,dy} \right]
\label {eq:cVsnorm}
\end {equation}
here instead of the normalization (\ref{eq:totalnorm}) used
for fig.\ \ref{fig:total}.

Notice that the absolute size of the $1/N^2$ correction to sequential
diagrams (green circles in fig.\ \ref{fig:crossVseq}) is quite
small compared to that for crossed diagrams (blue diamonds).
This means that, in absolute terms, the total $1/N^2$ correction
is overwhelmingly dominated by crossed diagrams.
[Sequential diagrams play a role in the size of relative corrections
shown in fig.\ \ref{fig:ratio_tot} because they affect the
$N{=}\infty$ denominator of (\ref{eq:ratio}) even though they
do not noticeably affect the $1/N^2$ numerator.]

\begin {figure}[t]
\begin {center}
  \includegraphics[scale=0.5]{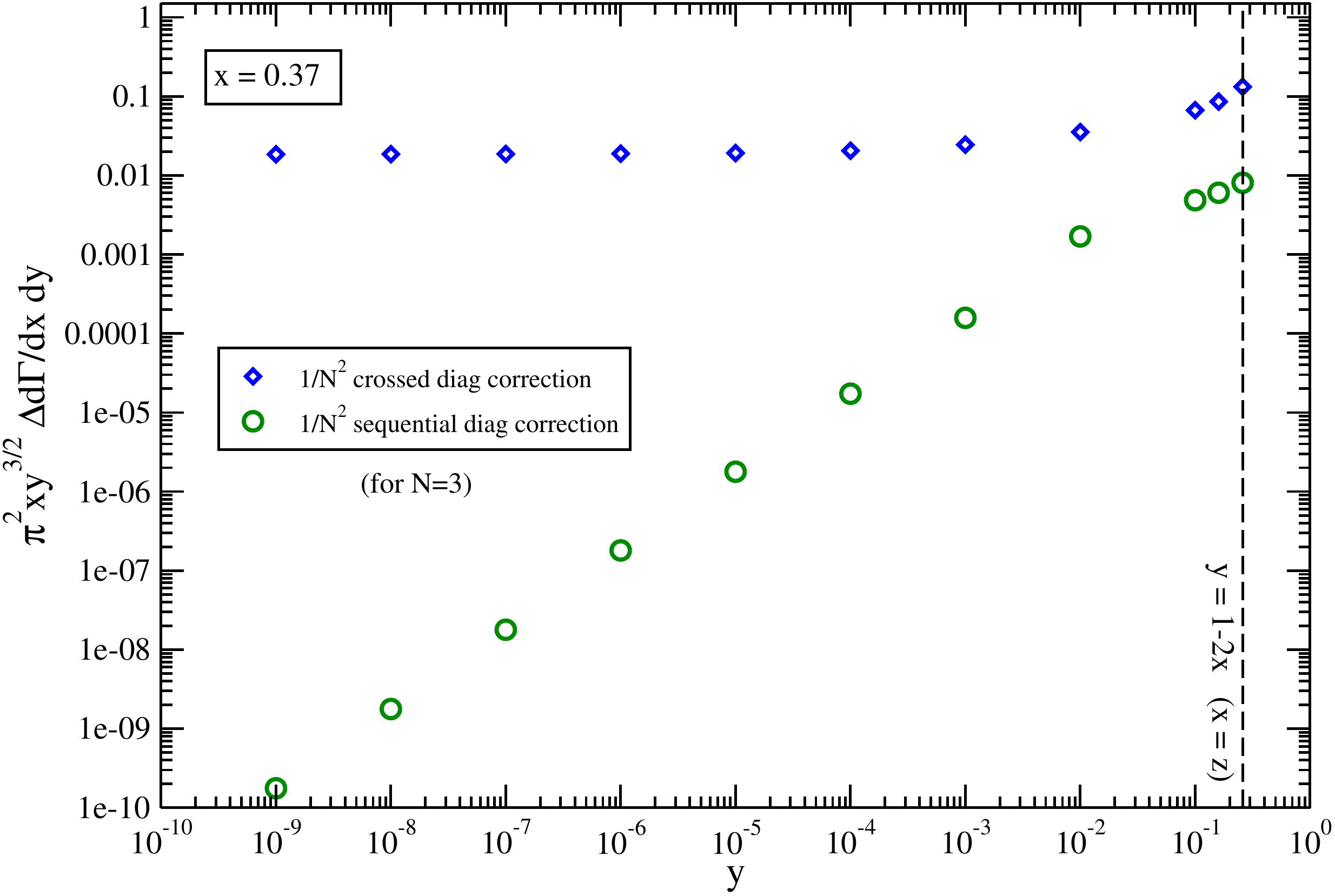}
  \caption{
     \label {fig:crossVseq3}
     A plot for $x=0.37$ of the $y$ dependence of the
     different $1/N^2$ contributions to $\Delta\,d\Gamma/dx\,dy$,
     multiplied by $\pi^2 x y^{3/2}$ as in fig.\ \ref{fig:total}.
     Note that, in contrast to fig.\ \ref{fig:crossVseq}, we have
     not divided by $\ln(1/y)$.
  }
\end {center}
\end {figure}

Fig.\ \ref{fig:crossVseq3}, where the normalization of the verical axis
is returned to
(\ref{eq:totalnorm}), shows that
the crossed and sequential $1/N^2$ corrections are different in another way
as well.
From this plot, we find numerically that
the crossed diagram correction behaves like $1/y^{3/2}$ for
small $y$, with no $\ln(y^{-1})$ enhancement.  This is why the
corresponding blue diamonds in fig.\ \ref{fig:crossVseq} approach
zero, due to the additional normalization factor $1/\ln(y^{-1})$
in that plot.
In contrast, we find that the sequential diagram correction has an
even milder dependence on small $y$, behaving like $1/y^{1/2}$.
When integrated over $y$ in applications,
this means that the $1/N^2$ correction from
crossed diagrams will contribute to infrared (IR) divergences (similar to
the IR divergences of the $N{=}\infty$ results discussed in
ref.\ \cite{qcd}), but the $1/N^2$ correction from sequential diagrams
will be IR finite.

\subsection{Comparison of size of $1/N^2$ corrections to related work}

In this paper, we have been focused on the problem of overlapping
formation times for the double splitting process $g{\to}ggg$.  For
simplicity, we have followed previous $N{=}\infty$ work on this
problem \cite{2brem,seq,qcd} and only considered rates
$\Delta\,d\Gamma/dx\,dy$ that have been integrated over the (small)
transverse momenta $\p_\perp$ of all three daughters.
For technical reasons,%
\footnote{
  See, for example, the argument in section 4.1 of ref.\ \cite{2brem}.
}
studying $\p_\perp$-integrated rates allows one
to ignore what happens to any daughter after it has been emitted in both
the amplitude and conjugate amplitude.  It's the reason, for example,
why the dynamics of the $y$ gluon is no longer relevant after the
first conjugate-amplitude (red) vertex in fig.\ \ref{fig:cross}
for $xy\bar y\bar x$ interference diagram.

\begin {figure}[t]
\begin {center}
  \includegraphics[scale=0.6]{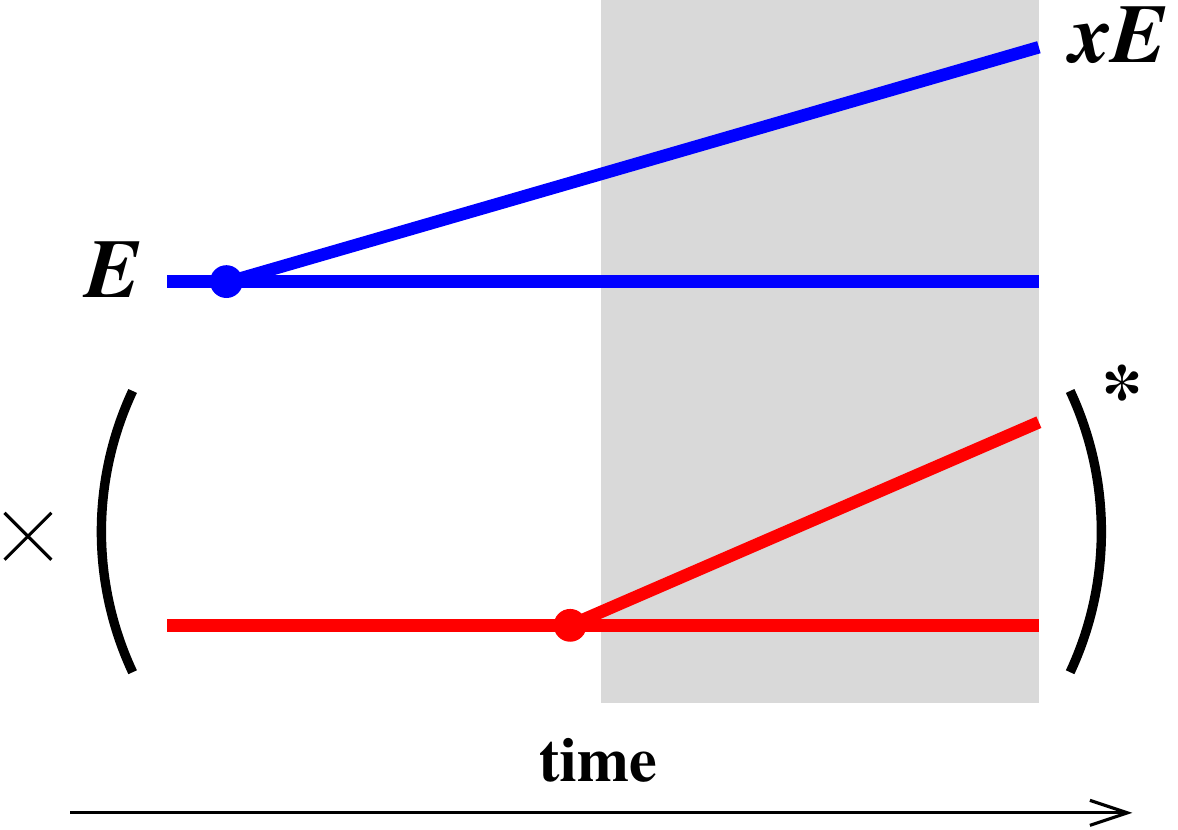}
  \caption{
     Similar to fig.\ \ref{fig:split}a for the rate of single splitting
     $g{\to}gg$, but here including later time-evolution of the daughters
     (shaded region) that must be included in order to study the
     $\p_\perp$ distribution of the daughters.
     In the shaded region, the above interference term contains four
     gluon lines which, in the language we have used in this
     paper, requires treating 4-gluon color singlet dynamics in the
     medium.
     \label{fig:splitpt}
  }
\end {center}
\end {figure}

However, there is a different type of problem (not about overlapping
formation times) where similar
issues of 4-gluon color-singlet dynamics also arise:
the {\it un}-integrated $\p_\perp$ distribution
$d\Gamma/dx\,d^2p_\perp$ for {\it single} splitting $g{\to}gg$ in
the medium.  Unlike fig.\ \ref{fig:split} for the $\p_\perp$-integrated
$g {\to} gg$ rate, one must instead
follow the color dynamics for a time
{\it after} the splitting has taken place in both amplitude and
conjugate amplitude, corresponding to the shaded region of
fig.\ \ref{fig:splitpt}.
During this time, one must treat the color dynamics of the four gluons
shown in the shaded region.%
\footnote{
   The color dynamics of the two daughters decouple after a time of
   order the formation time, often referred to in this context as
   the color decoherence time.
}
Refs.\ \cite{NSZ6j,Zakharov6j,Konrad} have investigated how to treat
this problem beyond the $N{=}\infty$ limit.
Unlike our work (on our different problem), their calculations
approximate the
trajectories of the high-energy particles as perfectly straight lines.
So they only include color dynamics and not the
dynamics of particle trajectories.
In this rigid geometry approximation (also known as the ``antenna''
approximation), they are able to more easily treat finite or expanding
media.

For our specific purposes here,
ref.\ \cite{Konrad} is interesting because the authors
explicitly calculate the $1/N^2$ correction to the $N{=}\infty$ limit
(using a different approach to calculate $1/N^2$ corrections than we have).
In their numerics, they study a medium of length $L$ with constant
$\hat q$.  Their results will depend on $L$, and
they study the case where the dimensionless
ratio $L \sqrt{\hat q/E}$ (which parametrically is the ratio of $L$
to what the formation length $L_{\rm f}$ would be in an infinite medium)
is $\simeq 0.55$.
In this particular case, they find that the $1/N^2$ correction
to the $N{=}\infty$ distribution for $g{\to}gg$ can be as large
as 16\%.
Direct
comparison to our roughly 17\% correction should not be taken too seriously,
however, since (i) the process we study is very different, and
(ii) their numerics hold quark $\hat q_{\rm F}^{}$ fixed as they vary $N$,
whereas we hold gluon $\qhatA$ fixed.


\section {Conclusion}
\label {sec:conclusion}

With two caveats, we have found that $1/N^2$ corrections to
$N{=}\infty$ results for overlapping double gluon splitting ($g\to
ggg$) can be as large as approximately 17\% for $N{=}3$ (QCD).
One caveat is the
technical one explained in section \ref{sec:nresults} that
measurements of {\it relative} corrections become meaningless when
the leading ($N{=}\infty$) answer goes through zero at small $y \sim 0.01$,
and so we have focused on the size of corrections for not-small $y$.
Also, small-$y$ emission is not the case where large-$N$ techniques
were necessary to simplify the problem, since previous work on
overlapping formation times with a {\it soft} emission \cite{Blaizot,Iancu,Wu}
(which included the effects of virtual emissions) was done without
resorting to the large-$N$ approximation.  Our interest here has been
to estimate the reliability of using $N{=}\infty$ results specifically
for the case where $y$ is not small.

The other caveat is
that, in this exploratory analysis, we have not included
diagrammatic contributions
to $g{\to}ggg$ that involve 4-gluon vertices nor, in Light-Cone
Perturbation Theory, instantaneous longitudinal gluon exchange.
Nonetheless, our provisional take-away is that the
$N{=}\infty$ limit taken in previous analysis is likely
a moderately good approximation.
Work on using the $N{=}\infty$ approximation to answer the ultimate
question about the effect of overlapping non-soft emission on
in-medium shower development is ongoing (using results for $N{=}\infty$
rates from ref.\ \cite{qcd} and the framework suggested by
ref.\ \cite{qedNfstop}).

Ultimately, a complete analysis of $1/N^2$ effects on energy loss
should also include calculation of virtual diagrams for $g \to gg$, as
discussed for $N{=}\infty$ in ref.\ \cite{qcd}.

Calculating virtual diagrams through order $1/N^2$ may also be
interesting for better understanding soft radiative corrections to
hard single splitting $g{\to gg}$.  Such radiative corrections give
rise to IR double logarithms \cite{Blaizot,Iancu,Wu}
and sub-leading IR single logarithms.
The single logarithms have been calculated for $N{=}\infty$ (for
infinite medium in the $\hat q$ approximation) in
refs.\ \cite{logs,logs2}.  We do not know for sure whether those
single logarithms have any non-trivial dependence on $N$.  It would be
interesting to be able to explicitly check at order $1/N^2$.

Finally, to answer a question proposed in the introduction, we note
that our roughly 17\% corrections for $N{=}3$ are roughly consistent
with (e.g.\ within a factor of 2 of)
the naive, {\it merely parametric} guess of
$O(1/N^2) \sim 10\%$.


\acknowledgments

This work was supported, in part, by the U.S. Department
of Energy under Grant Nos.~DE-SC0007974 and DE-SC0007984.


\appendix

\section{More on $(s,t,u)$ channel color singlet states}

\subsection{Sign conventions and conversions}
\label {app:signs}

The overall sign conventions for $u$-channel states in this paper
were set by how we translated the $s$-channel results for the potential
$V(\C_{12},\C_{34})$ in eqs.\ (4.3) and (3.12a) of ref.\ \cite{color} to the
$u$-channel version shown here in (\ref{eq:HVeffective}).
The translation is to simply relabel the particles $(1,2,3,4)$ there
as $(4,1,2,3)$ here.  We take our basis of color states (\ref{eq:6dim})
to be similarly permuted:
\begin {equation}
   (1,2,3,4)  \longrightarrow  (4,1,2,3) \qquad \mbox{[for $s \to u$]}
\label {eq:stou}
\end {equation}
to go from $s$-channel $|R\rangle_s$ to $u$-channel $|R\rangle_u$.
Because we have used the same permutation to define our $u$-channel
states, our formulas
(\ref{eq:ST}) here for the matrices $\underline{S}_u$ and $\underline{T}_u$ are
the {\it same} as the corresponding $s$-channel versions in
eq.\ (5.6) of ref.\ \cite{color}.

However, (\ref{eq:stou}) is not how $u$-channel states were defined
in ref.\ \cite{color}.  There, $t$-channel states were first defined
by%
\footnote{
  See, for example, eqs.\ (2.14) vs.\ (2.15) of ref.\ \cite{color}.
}
\begin {equation}
   2\leftrightarrow 3 \qquad \mbox{[for $s \to t$]} ,
\label {eq:stot}
\end {equation}
and then $u$-channel states were defined in terms of $t$-channel states
by%
\footnote{
  See eq.\ (2.9) of ref.\ \cite{color}.
}
\begin {equation}
   3 \leftrightarrow 4 \qquad \mbox{[for $t \to \bar u$]} ,
\label {eq:ttoubar}
\end {equation}
where we will use $\bar u$ to denote the $u$-channel conventions
of ref.\ \cite{color}.
Performing (\ref{eq:stot}) followed by (\ref{eq:ttoubar}) gives
\begin {equation}
   (1,2,3,4)  \longrightarrow  (1,4,2,3) \qquad \mbox{[for $s \to \bar u$]}
\label {eq:stoubar}
\end {equation}

The difference between our $u$-channel convention on the
right-hand side of (\ref{eq:stou}) and the convention on the
right-hand side of (\ref{eq:stoubar}) is
\begin {equation}
   1\leftrightarrow 4 \qquad \mbox{[for $u \to \bar u$]} ,
\label {eq:utoubar}
\end {equation}
The effect of this is to negate states that involves an anti-symmetric
combination of particles 1 and 4.  From (\ref{eq:AxA}), that's
$|{\rm A}_{aa}\rangle_u$ and $|\quote{\bm{10}+\overline{\bm{10}}}\rangle_u$
out of our $u$-channel states (\ref{eq:6dim}).  We can summarize this
as
\begin {equation}
  \begin{pmatrix}
    |{\bm 1}\rangle_{\bar u} \\
    |{\rm A}_{\rm aa}\rangle_{\bar u} \\
    |{\rm A}_{\rm ss}\rangle_{\bar u} \\
    |\quote{\bm{10}{+}\overline{\bm{10}}}\rangle_{\bar u} \\
    |\quote{\bm{27}}\rangle_{\bar u} \\
    |\quote{\bm 0}\rangle_{\bar u}
  \end{pmatrix}
  =
  \mathds{P}
  \begin{pmatrix}
    |{\bm 1}\rangle_u \\
    |{\rm A}_{\rm aa}\rangle_u \\
    |{\rm A}_{\rm ss}\rangle_u \\
    |\quote{\bm{10}{+}\overline{\bm{10}}}\rangle_u \\
    |\quote{\bm{27}}\rangle_u \\
    |\quote{\bm 0}\rangle_u
  \end{pmatrix}
\qquad\mbox{with}\qquad
  \mathds{P} \equiv
  \begin{pmatrix}
     +1 &&&&& \\
     & -1 &&&& \\
     && +1 &&& \\
     &&& -1 && \\
     &&&& +1 & \\
     &&&&& +1 \\
  \end {pmatrix} .
\label {eq:Pmatrix}
\end {equation}

The matrix that converts (for any $N$) between $s$-channel and $t$-channel
versions of the original basis states is given
by refs.\ \cite{color,NSZ6j,Sjodahl} as%
\footnote{
  This specific conversion is adapted from table IV of ref.\ \cite{color},
  which provides the entries of our
  (\ref{eq:Vconvert}) and whose last column provides the signs $\mathds{P}$
  in our (\ref{eq:convertsubar}).
  See footnote 13 of ref.\ \cite{color} for discussion
  of how those results are related
  to refs.\ \cite{NSZ6j,Sjodahl}.
}
\begin {equation}
  \begin{pmatrix}
    |{\bm 1}\rangle_s \\
    |{\rm A}_{\rm aa}\rangle_s \\
    |{\rm A}_{\rm ss}\rangle_s \\
    |\quote{\bm{10}{+}\overline{\bm{10}}}\rangle_s \\
    |\quote{\bm{27}}\rangle_s \\
    |\quote{\bm 0}\rangle_s
  \end{pmatrix}
  =
  \mathds{V}
  \begin{pmatrix}
    |{\bm 1}\rangle_t \\
    |{\rm A}_{\rm aa}\rangle_t \\
    |{\rm A}_{\rm ss}\rangle_t \\
    |\quote{\bm{10}{+}\overline{\bm{10}}}\rangle_t \\
    |\quote{\bm{27}}\rangle_t \\
    |\quote{\bm 0}\rangle_t
  \end{pmatrix}
\end {equation}
with
\begin {equation}
  \mathds{V} =
  \begin{pmatrix}
    \tfrac{1}{N^2-1}
    & \sqrt{\tfrac{1}{N^2-1}}
    & \sqrt{\tfrac{1}{N^2-1}}
    & \sqrt{\tfrac{N^2-4}{2(N^2-1)}}
    & \tfrac{N}{2(N+1)} \sqrt{\tfrac{N+3}{N-1}}
    & \tfrac{N}{2(N-1)} \sqrt{\tfrac{N-3}{N+1}}
  \\[5pt]
  
    & \tfrac12
    & \tfrac12
    & 0
    & -\tfrac12 \sqrt{\tfrac{N+3}{N+1}}
    &  \tfrac12 \sqrt{\tfrac{N-3}{N-1}}
  \\
  
    &
    & \tfrac{N^2-12}{2(N^2-4)}
    & -\sqrt{\tfrac{2}{N^2-4}}
    & \tfrac{N}{2(N+2)} \sqrt{\tfrac{N+3}{N+1}}
    & -\tfrac{N}{2(N-2)} \sqrt{\tfrac{N-3}{N-1}}
  \\[5pt]

    &
    &
    & \tfrac12
    & - \sqrt{\tfrac{(N-2)(N+3)}{8(N+1)(N+2)}}
    & - \sqrt{\tfrac{(N+2)(N-3)}{8(N-1)(N-2)}}
  \\[5pt]
    \multicolumn{3}{c}{\rm (symmetric) }
    &
    & \tfrac{N^2+N+2}{4(N+1)(N+2)}
    & \tfrac{1}{4} \sqrt{\tfrac{N^2-9}{N^2-1}}
  \\[5pt]
  
    &
    &
    &
    &
    & \tfrac{N^2-N+2}{4(N-1)(N-2)}
  \\
  \end{pmatrix} .
\label {eq:Vconvert}
\end {equation}
(Note that $\mathds{V} = \mathds{V}^\top = \mathds{V}^{-1}$.)
As explained in ref.\ \cite{color}, the conversion between
the $s$-channel and the $u$-channel basis of that paper is
correspondingly
\begin {equation}
  \begin{pmatrix}
    |{\bm 1}\rangle_s \\
    |{\rm A}_{\rm aa}\rangle_s \\
    |{\rm A}_{\rm ss}\rangle_s \\
    |\quote{\bm{10}{+}\overline{\bm{10}}}\rangle_s \\
    |\quote{\bm{27}}\rangle_s \\
    |\quote{\bm 0}\rangle_s
  \end{pmatrix}
  =
  \mathds{P}
  \mathds{V}
  \begin{pmatrix}
    |{\bm 1}\rangle_{\bar u} \\
    |{\rm A}_{\rm aa}\rangle_{\bar u} \\
    |{\rm A}_{\rm ss}\rangle_{\bar u} \\
    |\quote{\bm{10}{+}\overline{\bm{10}}}\rangle_{\bar u} \\
    |\quote{\bm{27}}\rangle_{\bar u} \\
    |\quote{\bm 0}\rangle_{\bar u}
  \end{pmatrix}
  .
\label {eq:convertsubar}
\end {equation}
Using (\ref{eq:Pmatrix}), the conversion between the $s$-basis
and $u$-basis in our paper here is given by (\ref{eq:su})
with $\mathds{U} = \mathds{P}\mathds{V}\mathds{P}$, which
equals (\ref{eq:Uconvert}).  For some purposes, it is useful to
have the $N{=}\infty$ limit of this matrix, which is
\begin {equation}
  \mathds{U}^\Ninf =
  \begin{pmatrix}
    0 & 0 & 0 & -\tfrac1{\sqrt2} & \tfrac12 & \tfrac12
      \\[3pt]
    0 & \tfrac12 & -\tfrac12 & 0 & \tfrac12 & -\tfrac12
      \\[3pt]
    0 & -\tfrac12 & \tfrac12 & 0 & \tfrac12 & -\tfrac12
      \\[3pt]
    -\tfrac1{\sqrt2} & 0 & 0 & \tfrac12 & \tfrac1{2\sqrt2} & \tfrac1{2\sqrt2}
      \\[3pt]
    \tfrac12 & \tfrac12 & \tfrac12 & \tfrac1{2\sqrt2} & \tfrac14 & \tfrac14
      \\[3pt]
    \tfrac12 & -\tfrac12 & -\tfrac12 & \tfrac1{2\sqrt2} & \tfrac14 & \tfrac14
  \end{pmatrix} .
\label {eq:Uinf}
\end {equation}


\subsection {Alternative descriptions for $N{=}\infty$}
\label {app:1234}

Here, we will justify the $N{=}\infty$
identifications made in (\ref{eq:1234}).
One relatively simple method is to start with a form where the
potential $V$ is written directly in terms of the transverse
positions $(\b_1,\b_2,\b_3,\b_4)$ of the four individual particles instead
of in terms of reduced variables such as the $(\C_{41},\C_{23})$ of
(\ref{eq:Veffective}).
This more direct expression is \cite{color}%
\footnote{
  Specifically, see eq.\ (3.10) of ref.\ \cite{color}.
}
\begin {align}
   \mtrx{V}(\b_1,\b_2,\b_3,\b_4) =
      \frac{i\qhatA}{4\CA} \biggl\{
         &
         \mtrx{\Tgen}_1\cdot\mtrx{\Tgen}_2
              \bigl[ (\b_1{-}\b_2)^2 + (\b_3{-}\b_4)^2 \bigr]
\nonumber\\ &
         +
         \mtrx{\Tgen}_1\cdot\mtrx{\Tgen}_3
              \bigl[ (\b_1{-}\b_3)^2 + (\b_2{-}\b_4)^2 \bigr]
\nonumber\\ &
         +
         \mtrx{\Tgen}_1\cdot\mtrx{\Tgen}_4
              \bigl[ (\b_1{-}\b_4)^2 + (\b_2{-}\b_3)^2 \bigr]
      \biggr\} .
\label {eq:V4Casimir}
\end {align}
(This expression only assumes the $\hat q$ approximation and not
that the medium is itself weakly coupled.)
Now we can use the expressions (\ref{eq:TdotT})%
\footnote{
  See also footnote \ref{foot:TdotT}.
}
for the
$\mtrx{\Tgen}_i\cdot\mtrx{\Tgen}_j$ in terms of
$\mtrx{S}_u$ and $\mtrx{T}_u$.  Then remember that for
$N{=}\infty$ our basis states
$(\one,\Ap,\Am,\Ax,\onexp,\onexm)$ are simultaneous eigenstates
of $\mtrx{S}_u$ and $\mtrx{T}_u$ with eigenvalues given by
the corresponding diagonal entries of (\ref{eq:ST0}).
All together, we can then write the explicit 4-particle potential
for each of our basis states.

For example, for $\sAm$, eq.\ (\ref{eq:ST0}) indicates that
$(S_u,T_u)=(\frac12,-\frac14)$ for $N{=}\infty$.  Using
(\ref{eq:TdotT}), the corresponding
potential (\ref{eq:V4Casimir}) is then
\begin {equation}
  V^\Ninf_{(\Am)}(\b_1,\b_2,\b_3,\b_4) =
  - \frac{i{\hat q}_{\rm A}}{8} \Bigl[
         (\b_1{-}\b_2)^2 + (\b_2{-}\b_3)^2 + (\b_3{-}\b_4)^2 + (\b_4{-}\b_1)^2
  \Bigr] .
\end {equation}
This is exactly the large-$N$ behavior that we characterized as
``(1234),'' where each particle can interact only with its neighbors
going around the cylinder.

Doing the same thing with $\sAx$, which has $(S_u,T_u)=(0,0)$ for $N{=}\infty$, 
gives
\begin {equation}
  V^\Ninf_{(\Ax)}(\b_1,\b_2,\b_3,\b_4) =
  - \frac{i{\hat q}_{\rm A}}{8} \Bigl[
         (\b_1{-}\b_2)^2 + (\b_2{-}\b_4)^2 + (\b_4{-}\b_3)^2 + (\b_3{-}\b_1)^2
  \Bigr] ,
\end {equation}
which corresponds to what we called (1243).

As another example, $\sonexm$ has $(S_u,T_u)=(0,-\frac12)$ for $N{=}\infty$,
which gives
\begin {equation}
  V^\Ninf_{(\onexm)}(\b_1,\b_2,\b_3,\b_4) =
  - \frac{i{\hat q}_{\rm A}}{4} \Bigl[
         (\b_1{-}\b_2)^2 + (\b_3{-}\b_4)^2
  \Bigr] ,
\end {equation}
where particles 1 and 2 interact only with each other, and
similarly particles 3 and 4 interact only with each other.
This corresponds to what we called $(12)(34)$.
One may similarly check all of the other identifications in (\ref{eq:1234}).


\section{Numerical method}
\label {app:nmethod}

Here, we give a few comments about our numerical method.  We have not
tried to figure out the most efficient method but have adopted
a brute-force approach implemented in Mathematica \cite{Mathematica}.
But there are issues, which we think may be useful to briefly
summarize.

First, because of round-off errors caused by subtractive cancellations,
we find that we typically need to do intermediate calculations with
much more than machine precision in order to succeed in a brute force
approach.  We therefore do our calculations with higher precision
arithmetic in Mathematica.

Some of our formulas, like (\ref{eq:xyxyresult}), involve
derivatives such as
$\partial_{j_1} \partial_{j_2} [\cdots] \Bigl|_{j_1=j_2=0}$.
After some experimentation, we decided to implement these
derivatives numerically as
\begin {equation}
  \partial_{j_1} \partial_{j_2} f(j_1,j_2) \Bigl|_{j_1=j_2=0}
  \simeq
  \frac{
    f({+}\eps,{+}\eps) - f({+}\eps,{-}\eps)
      - f({-}\eps,{+}\eps) + f({-}\eps,{-}\eps)
  }{ (2\eps)^2 }
  \qquad\mbox{($\eps$ small)}
\end {equation}
rather than doing the derivatives analytically (or using a more sophisticated
numerical estimate of the derivative).%
\footnote{
  If one wished to take the derivatives analytically, one could make use
  of small-$j$ expansions such as
  \[
    {\cal U}^{-1}
    =
    {\cal U}_0^{-1}
    + {\cal U}_0^{-1} (j_1 {\cal R}_1 {+} j_2 {\cal R}_2) {\cal U}_0^{-1}
    + j_1 j_2 {\cal U}_0^{-1}
      ( {\cal R}_1 {\cal U}_0^{-1} {\cal R}_2
                {+} {\cal R}_2 {\cal U}_0^{-1} {\cal R}_1 )
      {\cal U}_0^{-1}
    + O(j_1^2) + O(j_2^2)
  \] 
  and
  \[
    \det({\cal U}^{-1})
    =
    \det({\cal U}_0^{-1})
    \Bigl\{
      \bigl[
        1 + j_1 \tr({\cal U}_0^{-1} {\cal R}_1)
      \bigr]
      \bigl[
        1 + j_2 \tr({\cal U}_0^{-1} {\cal R}_2)
      \bigr]
      + j_1 j_2 \tr({\cal U}_0^{-1} {\cal R}_1 {\cal U}_0^{-1} {\cal R}_2)
    \Bigr\}
    + O(j_1^2) + O(j_2^2) ,
  \]
  where we have promoted the $2{\times}2$ matrices $R_1$ and $R_2$
  to $4{\times}4$ block-diagonal matrices by defining 
  ${\cal R}_1 \equiv
    \left( \begin{smallmatrix} R_1 & \\ & 0 \end{smallmatrix} \right)$
  and
  ${\cal R}_2 \equiv
    \left( \begin{smallmatrix} 0 & \\ & R_2 \end{smallmatrix} \right)$.
  However, this leads to more complicated formulas, which
  take extra CPU time to evaluate.
}

We let Mathematica handle the numerical evaluation of matrix inverses
and determinants in our formulas.

We found that naive use of canned, adaptive integration routines took too much
CPU time and caused a host of problems.  Rather than working out how to tweak
adaptive integration to do what we needed, we just did all of our
integrals as simply as possible by using midpoint Riemann sums.
For the $\Delta t$ integral (as opposed to the $t_1$ and $t_2$ integrals),
we found it convenient to change variables as
$
   \int_0^\infty d(\Delta t) \> f(\Delta t)
   =
   \int_{-\infty}^\infty dz \> e^z f(e^z) .
$
In practice, we then replace the infinite $z$ integration region
$(-\infty,+\infty)$
by a finite region $(z_{\rm min}, z_{\rm max})$ carefully chosen to
cover everywhere the integrand is non-negligible [a choice which must
be adjusted to study small values of $y$].

The cost of our brute-force method is that, because it is not adaptive, there
are an annoying number of numerical approximations one must check to
be sure that results are accurate, e.g.\
the size of $\eps$ in the numerical derivatives, the number of
Riemann intervals for the $(t_1,t_2,z)$ integrals, and the cut-offs
$(z_{\rm min},z_{\rm max})$.

Finally, we should mention the strategies we used to attempt to avoid human
error in our analysis and coding.  The result (\ref{eq:d2Gdet}) for
the $1/N^2$ correction $\delta^2 G$ to the 4-particle propagator was
initially derived independently, and implemented numerically, by each
of us in very different ways.  One way was the method presented in the
text.  The other way did not use any tricks for packaging the
transverse-position integrations into higher-dimensional vectors and
matrices like eqs.\ (\ref{eq:xidef}), (\ref{eq:Xidef}) and
(\ref{eq:calUABdef}) but instead did the Gaussian integrals separately
and explicitly, resulting in very long mathematical expressions for
$\delta^2 G$.  Once the two methods agreed numerically, we then
both switched to the method and code that most quickly produced
results for $\delta^2 G$, which was the method based on
(\ref{eq:d2Gdet}).  Now using the same code for $\delta^2 G$,
we each independently produced
or spot checked the various results in this paper.
Since we consulted with each other on general methods and
development [e.g.\ equations like (\ref{eq:XYZreadout})], and
helped each other to ferret out sources of numerical discrepancy,
our work was not completely independent, but the most error-prone
aspects of our numerical work were done independently.



\end {document}